\documentclass[twocolumn]{aastex61}

\usepackage{tikz}
\tikzset{
  treenode/.style = {shape=rectangle, rounded corners,
                     draw, align=center,
                     top color=white, bottom color=blue!20},
  root/.style     = {treenode, font=\large, bottom color=blue!20},
  env/.style      = {treenode, font=\ttfamily\normalsize},
  dummy/.style    = {circle,draw}
}
\usepackage{colordvi}
\usepackage{ulem}
\begin{document}

\title{Detection of low-energy breaks in gamma-ray burst prompt emission spectra}

\correspondingauthor{Gor Oganesyan}
\email{goganesy@sissa.it}

\author{Gor Oganesyan}
\affil{SISSA, via Bonomea 265, I--34136 Trieste, Italy}

\author{Lara Nava}
%\affil{Dipartimento di Fisica, Universit\`a di Trieste, via Valerio 2, I--34127 Trieste, Italy}
%\affiliation{INFN - Sezione di Trieste, via Valerio 2, I--34127 Trieste, Italy}
\affiliation{INAF -- Osservatorio Astronomico di Brera, via Bianchi 46, I--23807 Merate (LC), Italy}
\affiliation{INAF -- Osservatorio Astronomico di Trieste, via G.B. Tiepolo 11, I--34143 Trieste, Italy}

\author{Giancarlo Ghirlanda}
\affil{INAF -- Osservatorio Astronomico di Brera, via Bianchi 46, I--23807 Merate (LC), Italy}

\author{Annalisa Celotti}
\affil{SISSA, via Bonomea 265, I--34136 Trieste, Italy}
\affiliation{INFN - Sezione di Trieste, via Valerio 2, I--34127 Trieste, Italy}
\affiliation{INAF -- Osservatorio Astronomico di Brera, via Bianchi 46, I--23807 Merate (LC), Italy}

\begin{abstract} % LIMIT: 250 words. Number of words in the current version: 250!
The radiative process responsible for gamma-Ray Burst (GRB) prompt emission has not been identified yet. If dominated by fast-cooling synchrotron radiation, the part of the spectrum immediately below the $\nu F_\nu$ peak energy should display a power-law behavior with slope $\alpha_2=-3/2$, which breaks to a higher value $\alpha_1=-2/3$ (i.e. to a harder spectral shape) at lower energies. Prompt emission spectral data (usually available down to $\sim10-20\,$keV) are consistent with one single power-law behavior below the peak, with typical slope $\langle\alpha\rangle=-1$, higher than (and then inconsistent with) the expected value $\alpha_2=-3/2$. 
To better characterize the spectral shape at low energy, we analyzed 14 GRBs for which the {\it Swift} X-ray Telescope started observations during the prompt. 
When available, {\it Fermi}-GBM observations have been included in the analysis.
For 67\% of the spectra, models that usually give a satisfactory description of the prompt (e.g., the Band model) fail in reproducing the $0.5-1000\,$keV spectra: low-energy data outline the presence of a {\it spectral break around a few keV}.
We then introduce an empirical fitting function that includes a low-energy power law $\alpha_1$, a break energy $E_{\rm break}$, a second power law $\alpha_2$, and a peak energy $E_{\rm peak}$. 
We find $\langle\alpha_1\rangle=-0.66$  ($  \rm \sigma=0.35$), $\langle \log (E_{\rm break}/\rm keV)\rangle=0.63$ ($  \rm  \sigma=0.20$), $\langle\alpha_2\rangle=-1.46$ ($\rm \sigma=0.31$), and $\langle \log (E_{\rm peak}/\rm keV)\rangle=2.1$ ($  \rm \sigma=0.56$).
The values $\langle\alpha_1\rangle$ and $\langle\alpha_2\rangle$ are very close to expectations from synchrotron radiation. In this context, $E_{\rm break}$ corresponds to the cooling break frequency. The relatively small ratio $E_{\rm peak}/E_{\rm break}\sim30$ suggests a regime of moderately fast cooling, which might solve the long-lasting problem of the apparent inconsistency between measured and predicted low-energy spectral index.
\end{abstract}

%%%%%%%%%%%%%%%%%%%%%%%  INTRODUCTION  %%%%%%%%%%%%%%%%%%%%%%
\section{Introduction} \label{sec:intro}
The origin of prompt emission from gamma-ray bursts (GRBs) is still a mystery and represents one of the most pressing questions in GRB studies.
The nature of both the dissipation and radiative mechanisms has not been firmly identified yet.
This lack of knowledge on what is powering and shaping the prompt radiation is strictly related to a series of open questions about fundamental properties of GRBs, such as the jet composition, the location of the dissipation region, the efficiency and nature of the acceleration mechanism, and the strength and properties of the magnetic field in the emission region.
Even though the most natural radiative process expected to dominate the emission is synchrotron radiation \citep{katz94,rees94,sari96,sari98,tavani96}, the inconsistency between the observed spectral shape at low energies and predictions from the synchrotron theory represents a serious challenge for this interpretation.
As inferred from the spectral analysis, the photon index $\alpha$ describing the data at low energy (i.e. below the $\nu F_\nu$ peak energy) is distributed around a typical value $\alpha\sim-1$, higher than the value expected in the case of fast-cooling synchrotron radiation \citep{cohen97,crider97,preece98,Ghisellini_00}.
This result is independent of the spectral function adopted to fit the spectra (e.g., cutoff power law, smoothly broken power law, Band function), and it has been found to be similar from the analysis of the spectral data collected by different instruments \citep{preece98,frontera00,ghirlanda02,kaneko06,nava11,sakamoto11,goldstein12,gruber14,lien16,yu16}.

The problem has been widely discussed in the literature. The proposed solutions
can be classified into two types: models that invoke emission mechanisms different than synchrotron radiation, and models that propose modifications to the basic synchrotron scenario.
Among the first class of models, we recall scenarios invoking Comptonization and/or thermal components \citep{liang97,blinnikov99,ghisellini99,lazzati00,meszaros00,stern04,rees05,ryde09,guiriec11,guiriec15a,guiriec15b,guiriec16a,guiriec16b,ghirlanda13,burgess14}.
For the second class of models 
(studies that  consider synchrotron radiation) effects producing a hardening of the low-energy spectral index have been invoked, such as Klein-Nishina effects, marginally fast cooling regime, and anisotropic pitch angle distributions \citep{lloyd00,derishev01,derishev07,bosnjak09,nakar09,daigne11,uhm14}. In spite of all theoretical efforts, there is still no consensus on the origin of the prompt emission. The advantages and difficulties of some of these models have been recently reviewed by \cite{kumar15}. In spite of all theoretical efforts, there is still no consensus on the origin of the prompt emission.

%This work
Theoretical studies would benefit from a better characterization of prompt spectra, especially in the low-energy part, where observations are in contradiction with the synchrotron theory.
In this work, we collect a sample of 14 GRBs for which the X-Ray Telescope (XRT, 0.3-10\,keV), on board the {\it Swift} satellite,
started observations during the prompt emission, observed in the range 15-150\,keV by the Burst Alert Telescope (BAT). For these GRBs, we perform spectral analysis of the prompt emission from 0.5\,keV to 150\,keV, thanks to the joint analysis of XRT and BAT data, and from 0.5\,keV to $\gtrsim$\,1\,MeV when observations from the {\it Fermi} Gamma-Ray Burst Monitor (GBM) are available.
We find that the spectrum below $\sim10\,$keV does not lie on the extrapolation of the low-energy power law of the Band function (or similar functions that usually provide a satisfactory description of prompt spectra), but a spectral break around a few keV is required by the low-energy data.

The paper is organized as follows: In Section \S2 we describe the sample selection. In Section \S3 we summarize the method and procedure adopted to extract the data and perform the spectral analysis. The results are presented in Section \S4, and discussed in Section \S5. The main findings of this work are summarized in Section \S6.

%%%%%%%%%%%%%%%%%%%  SAMPLE  SELECTION  %%%%%%%%%%%%%%%%%%%%%%%%%%

\section{Sample selection}\label{sec:sample_selection}
%------------------------------------------------------------------------
\begin{figure*}[ht!]
%\begin{center}
\includegraphics[scale = 0.097]{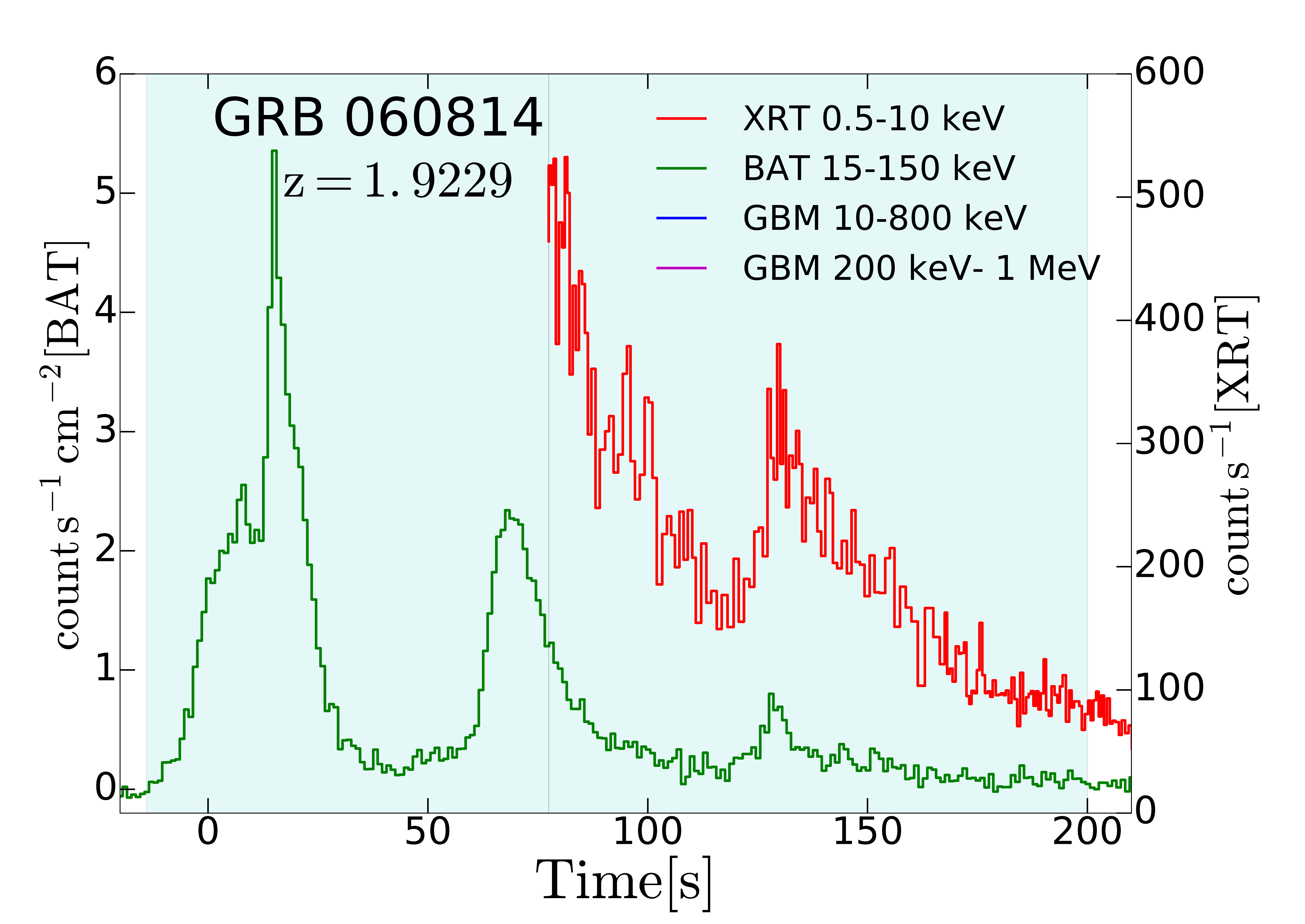} \hspace{0.06em}
\includegraphics[scale = 0.097]{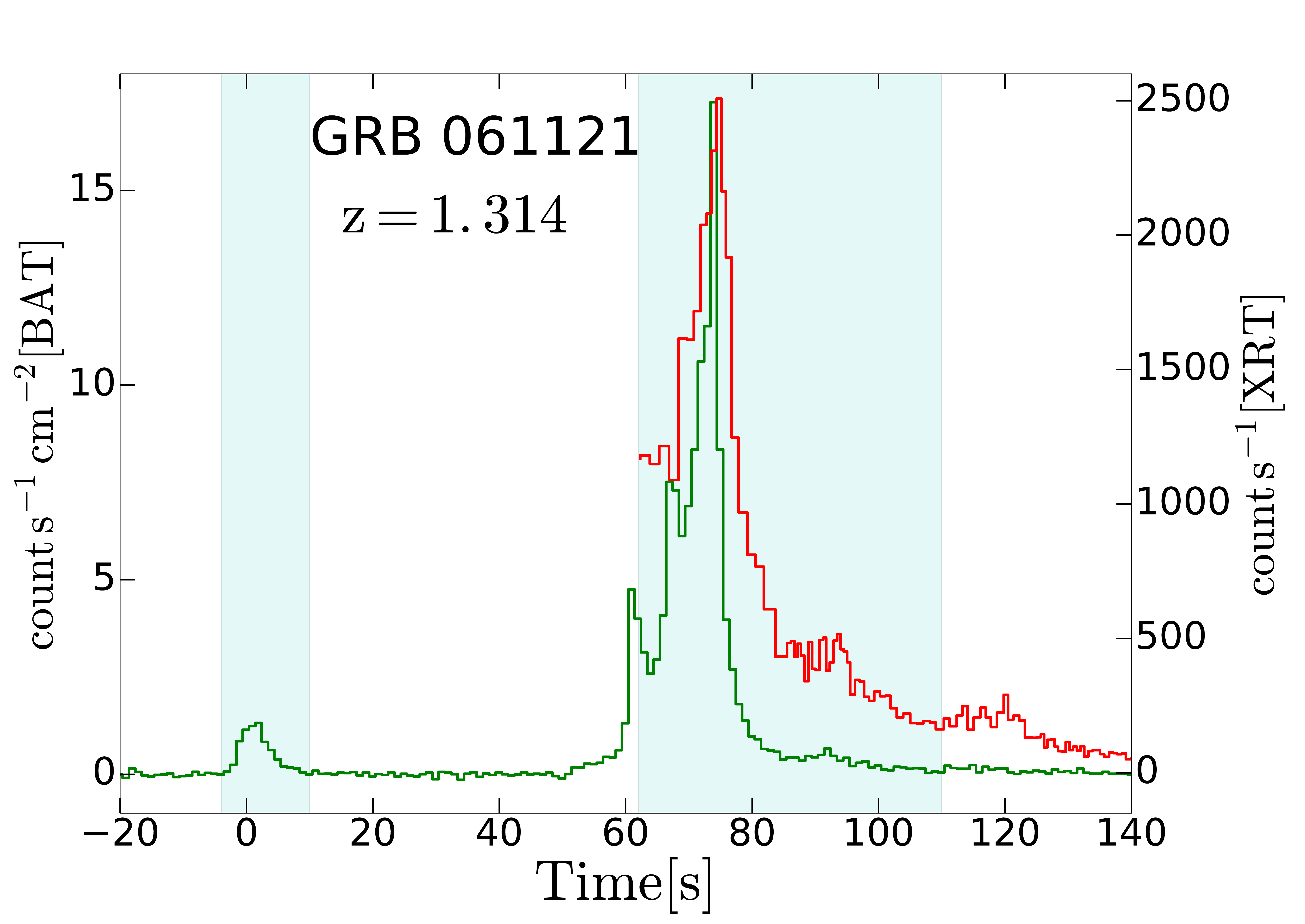} \hspace{0.06em}
\includegraphics[scale = 0.097]{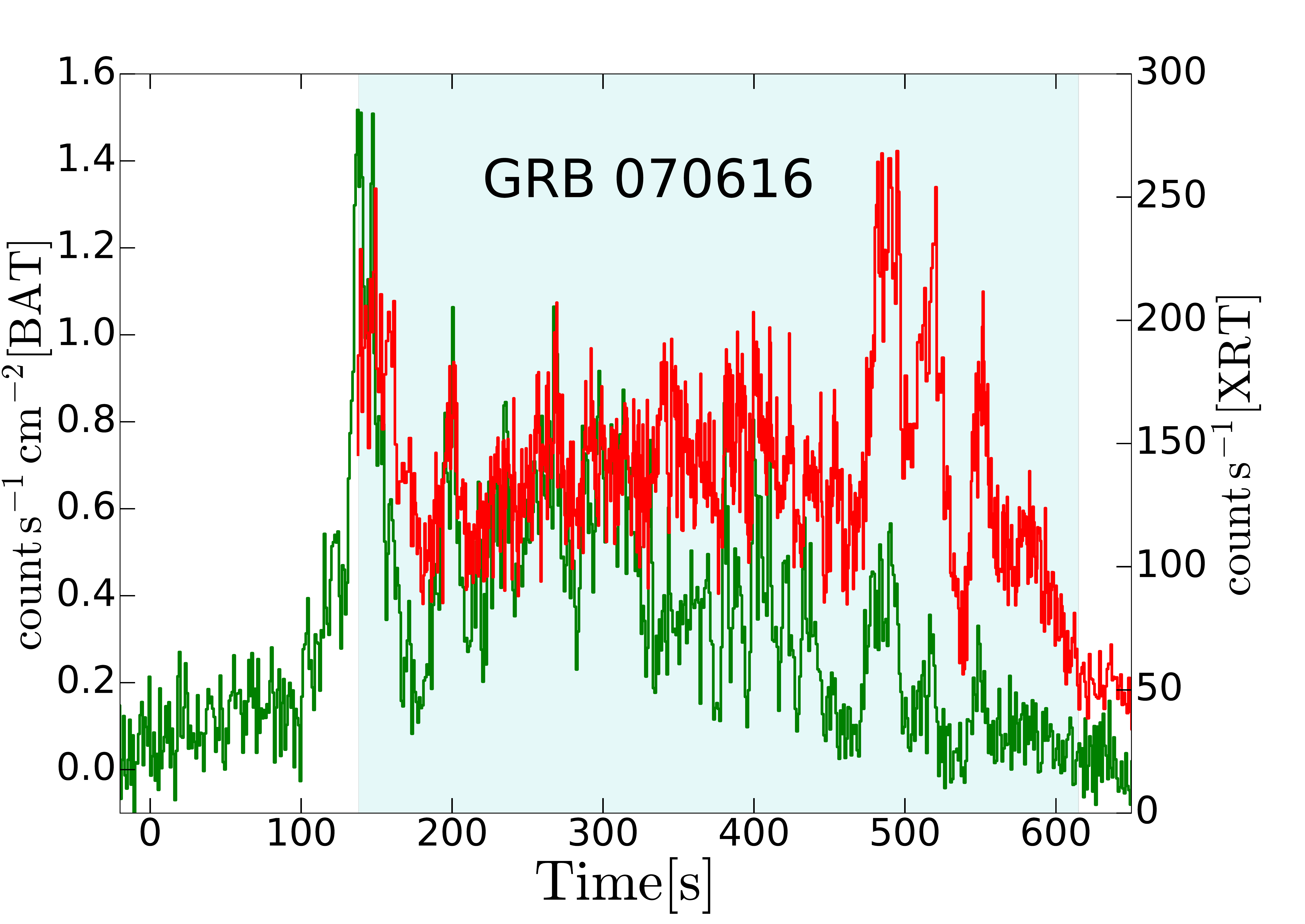} \\
\includegraphics[scale = 0.097]{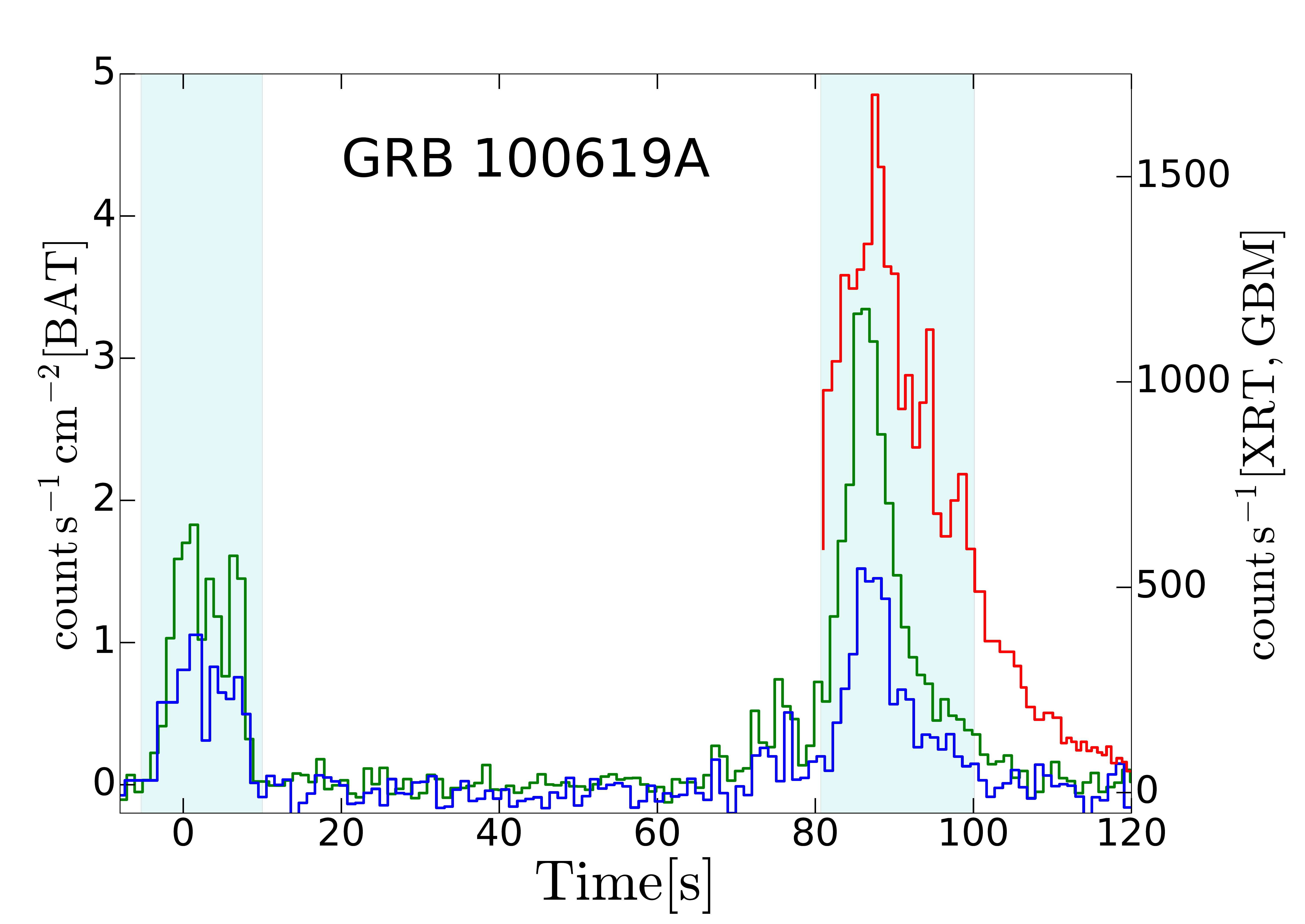} \hspace{0.06em} 
\includegraphics[scale = 0.097]{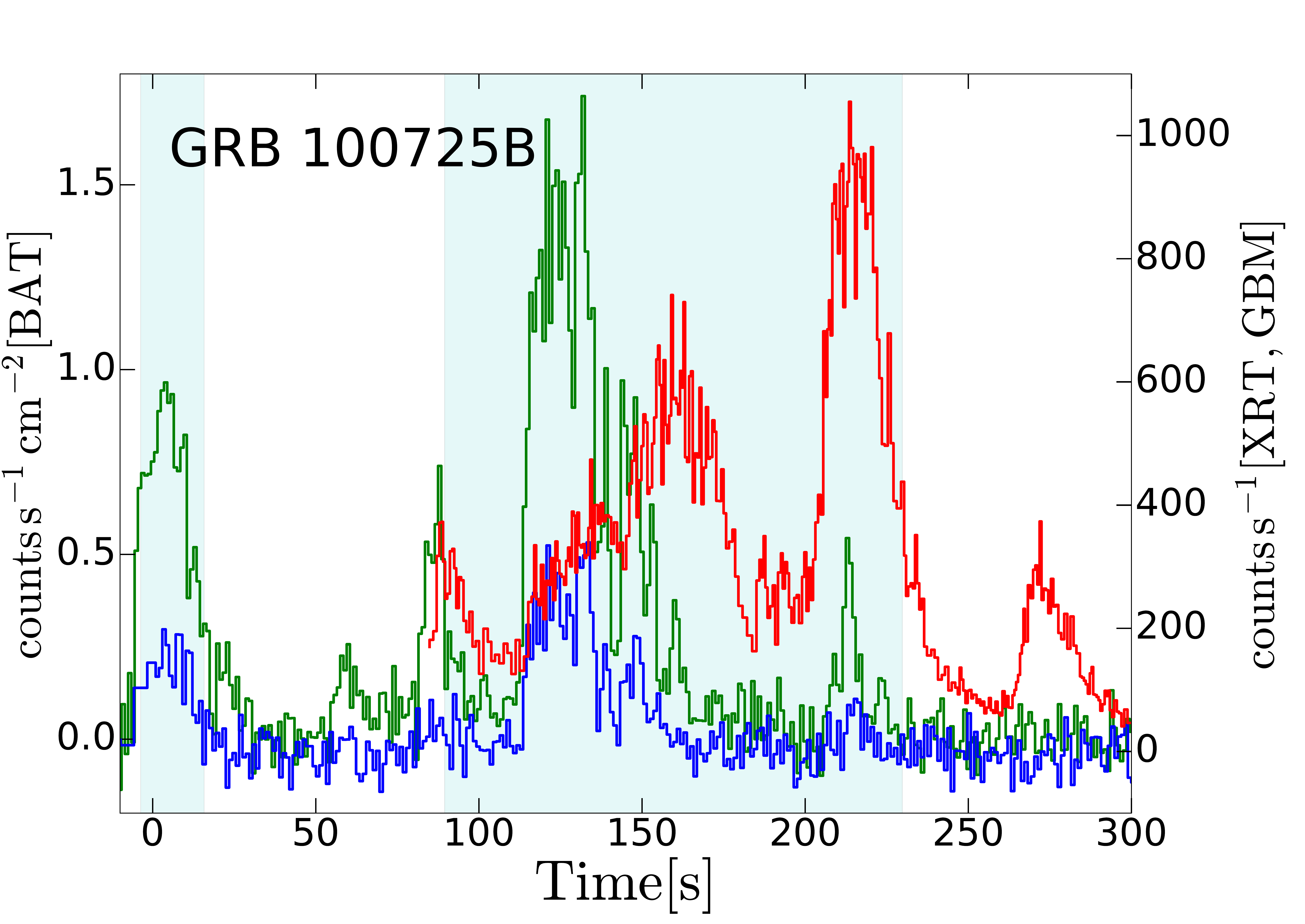} \hspace{0.06em}
\includegraphics[scale = 0.097]{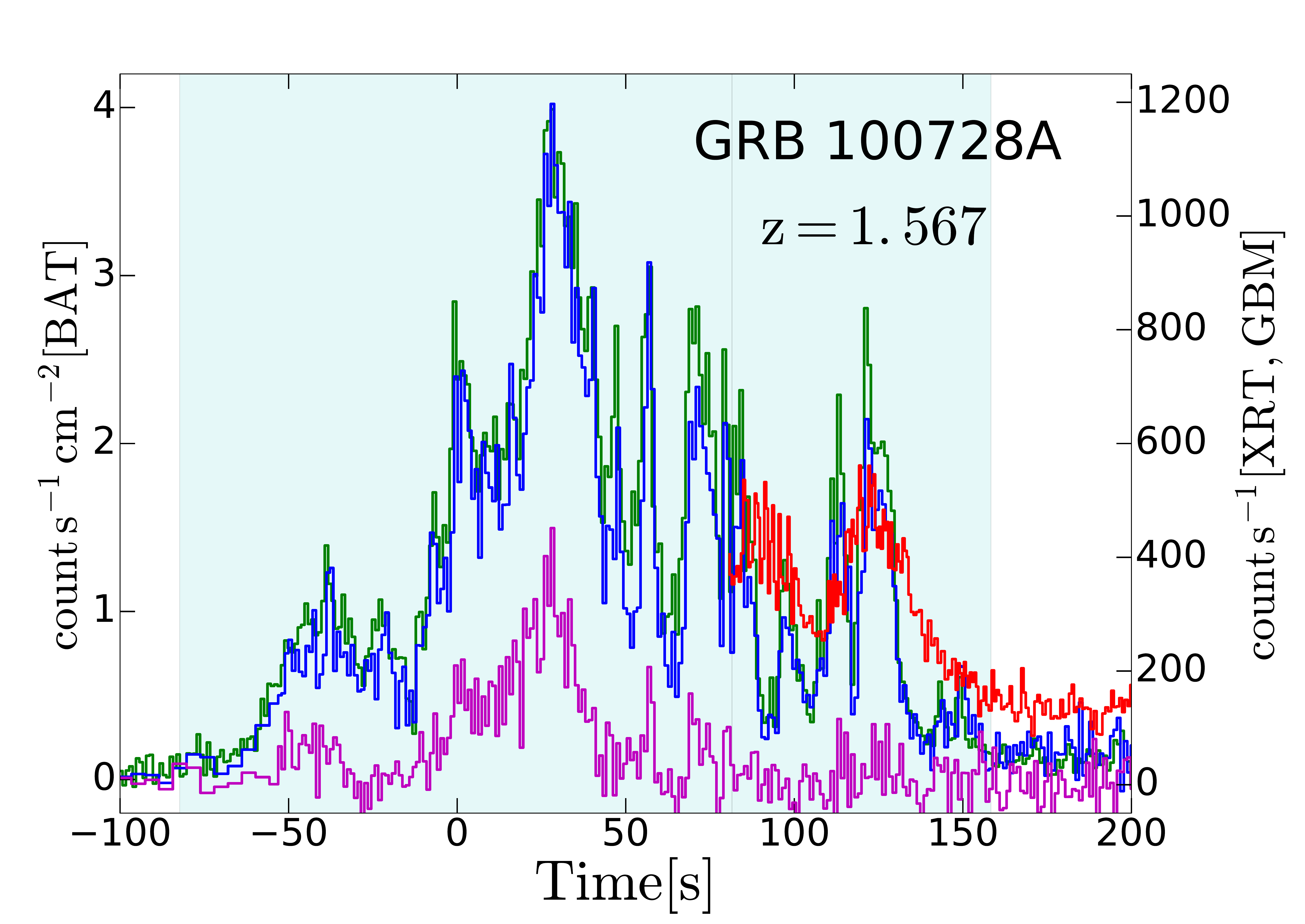} \\
\includegraphics[scale = 0.097]{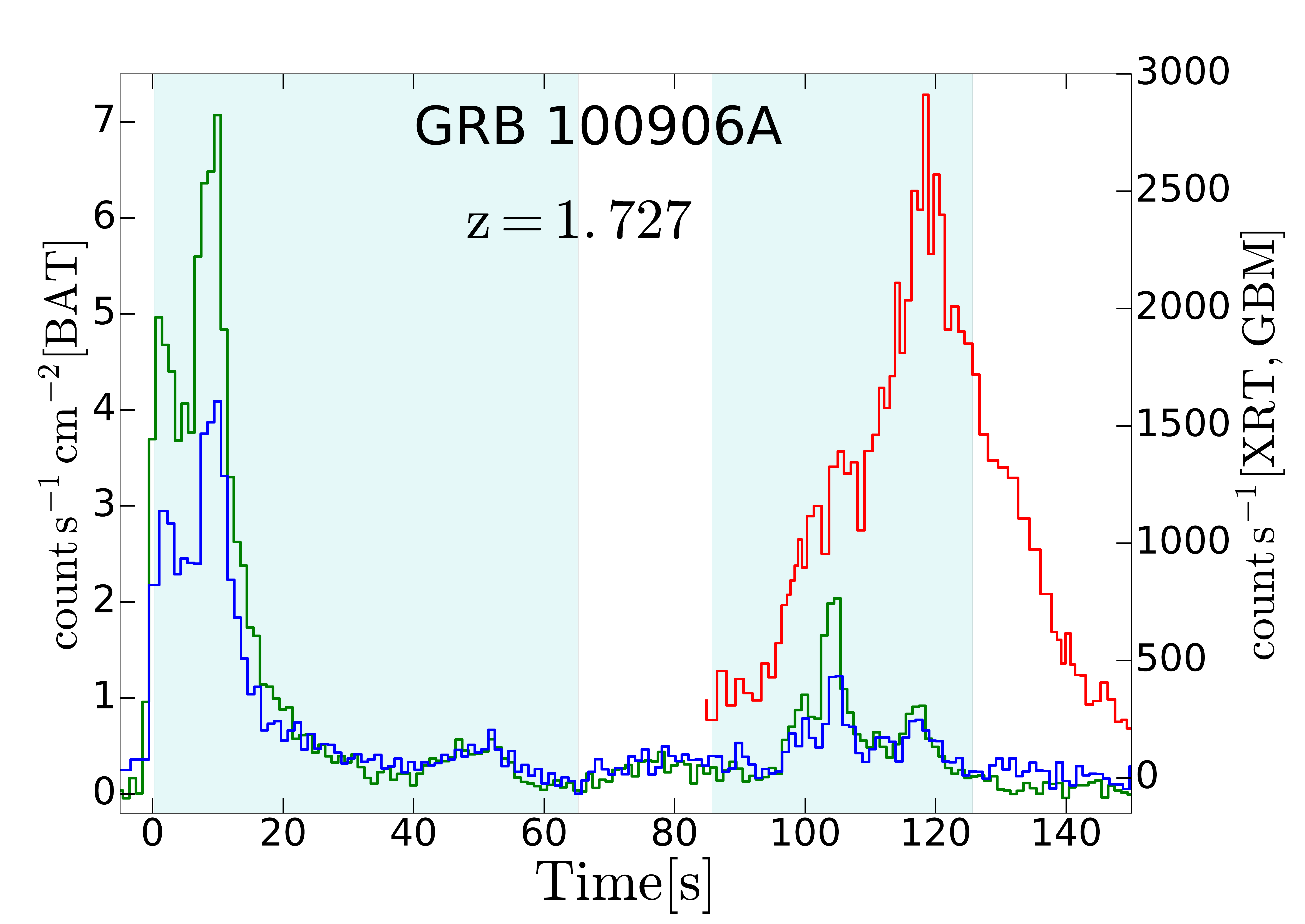} \hspace{0.06em}
\includegraphics[scale = 0.097]{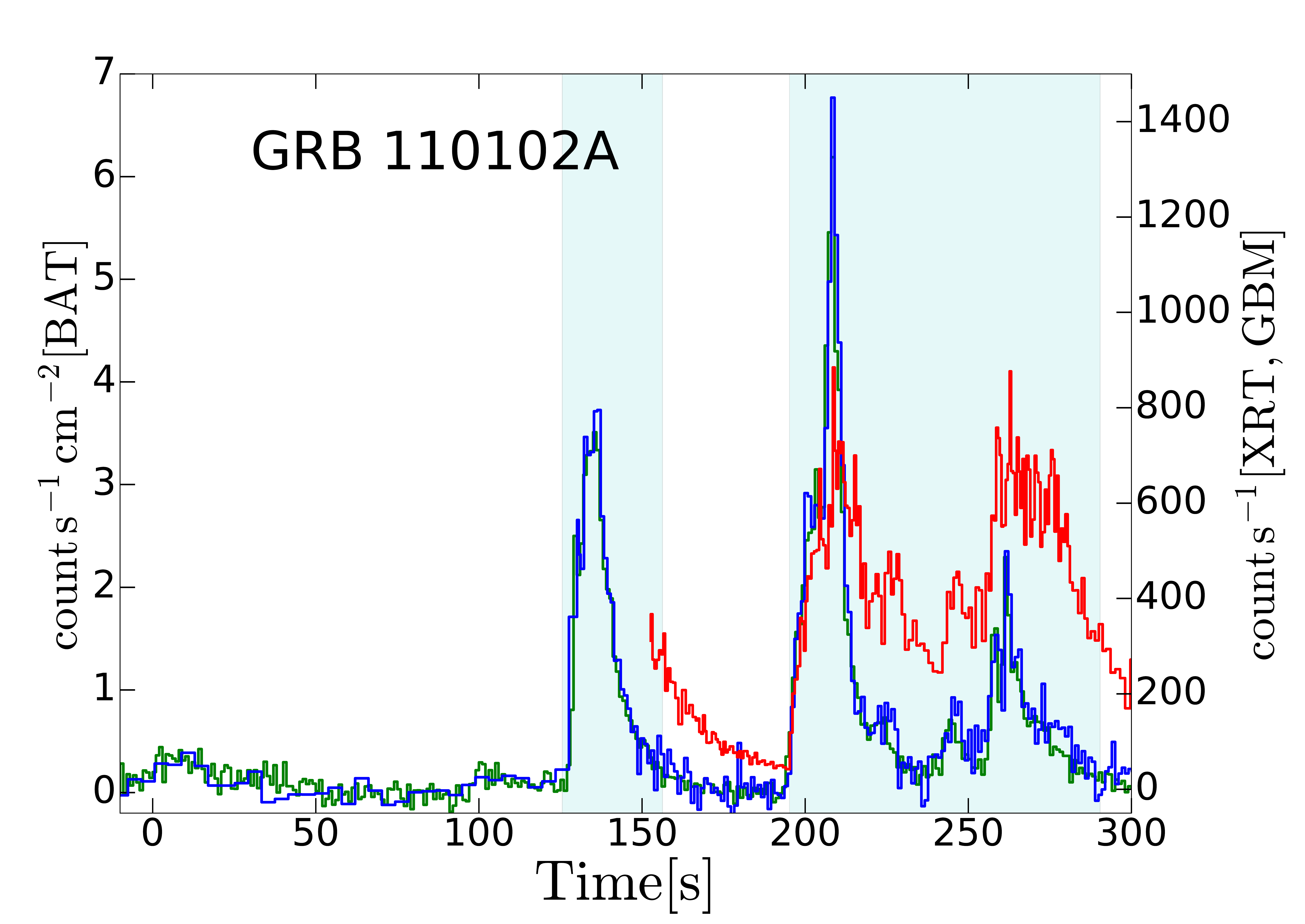} \hspace{0.06em}
\includegraphics[scale = 0.097]{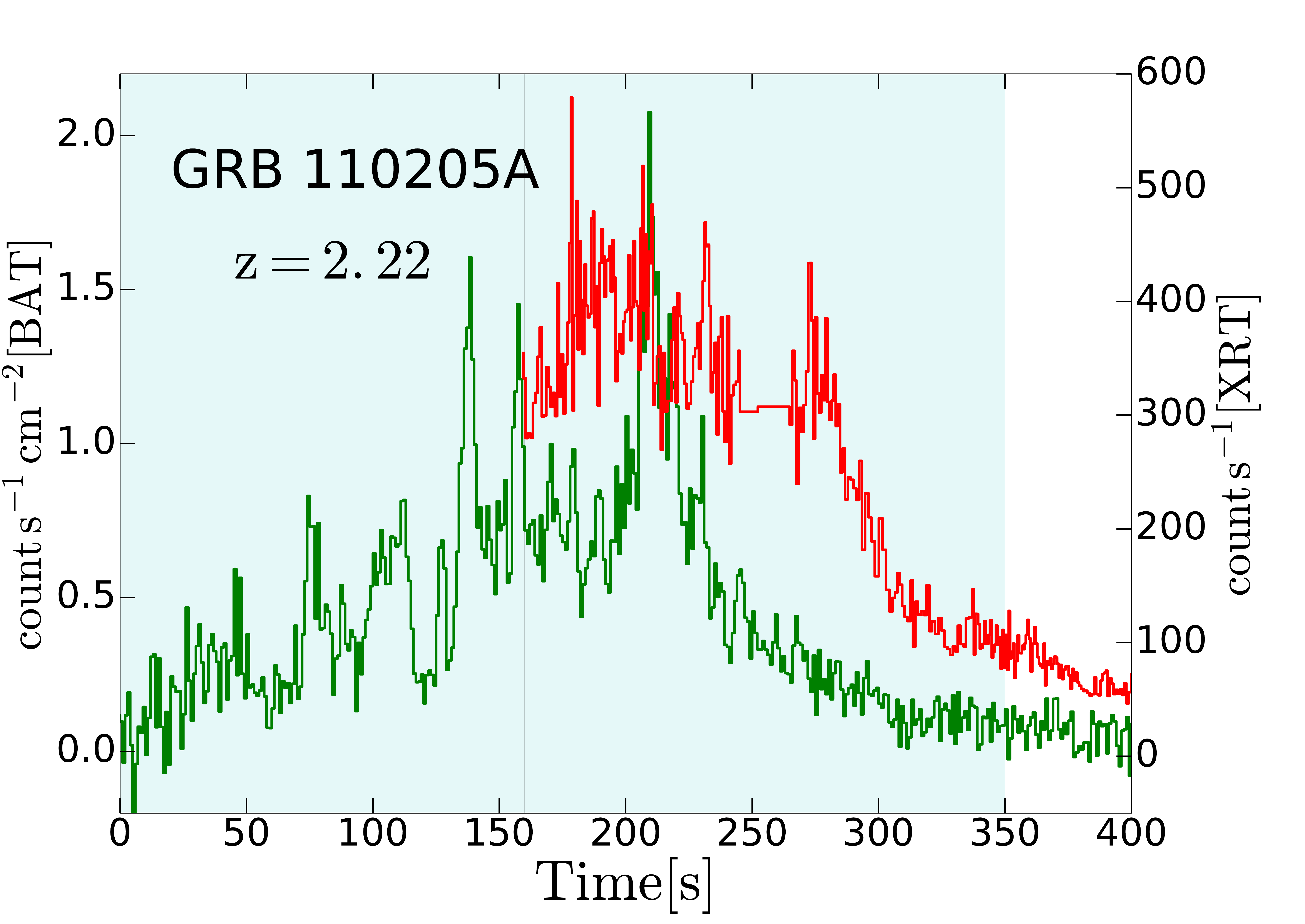} \\
\includegraphics[scale = 0.097]{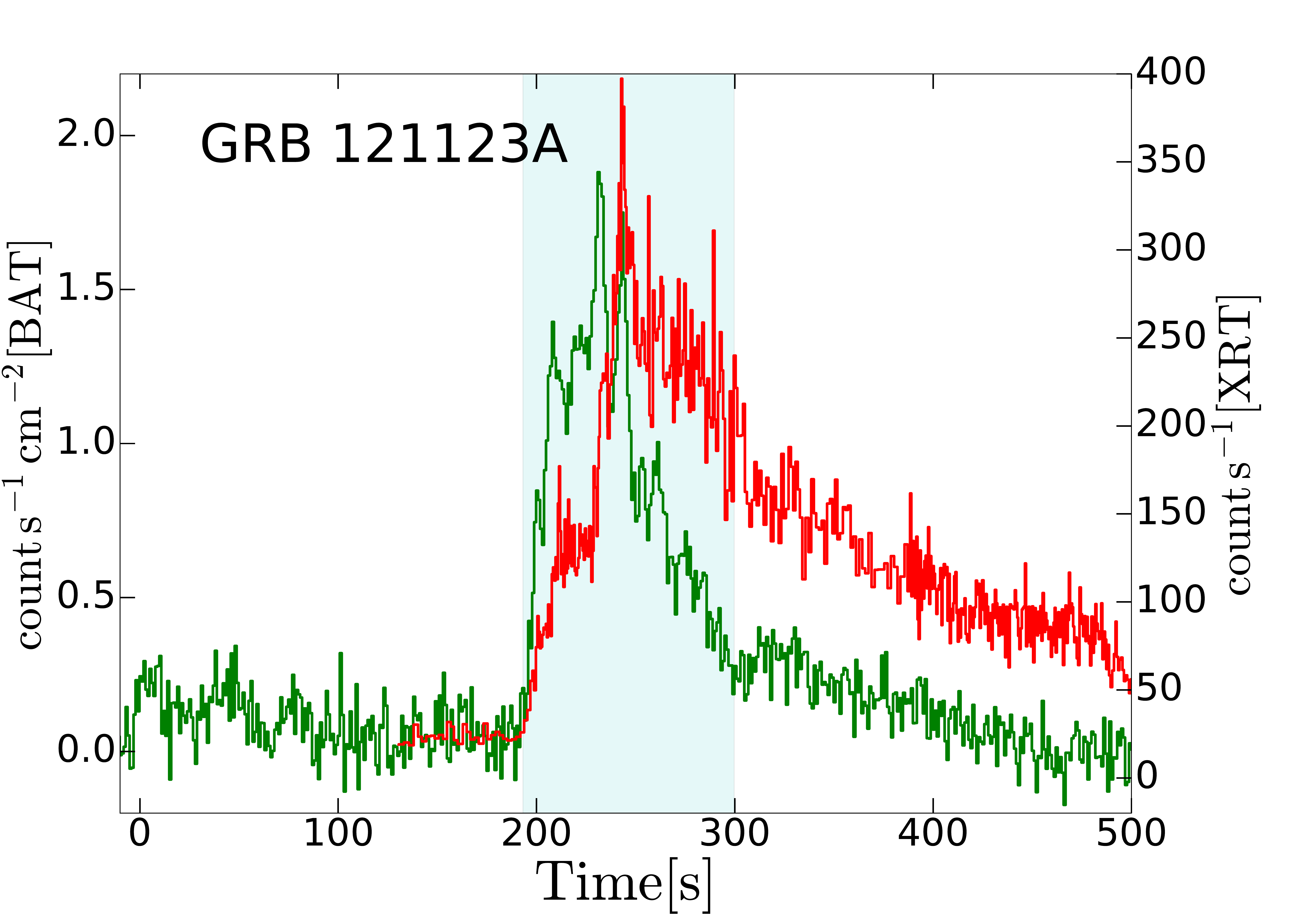} \hspace{0.06em}
\includegraphics[scale = 0.097]{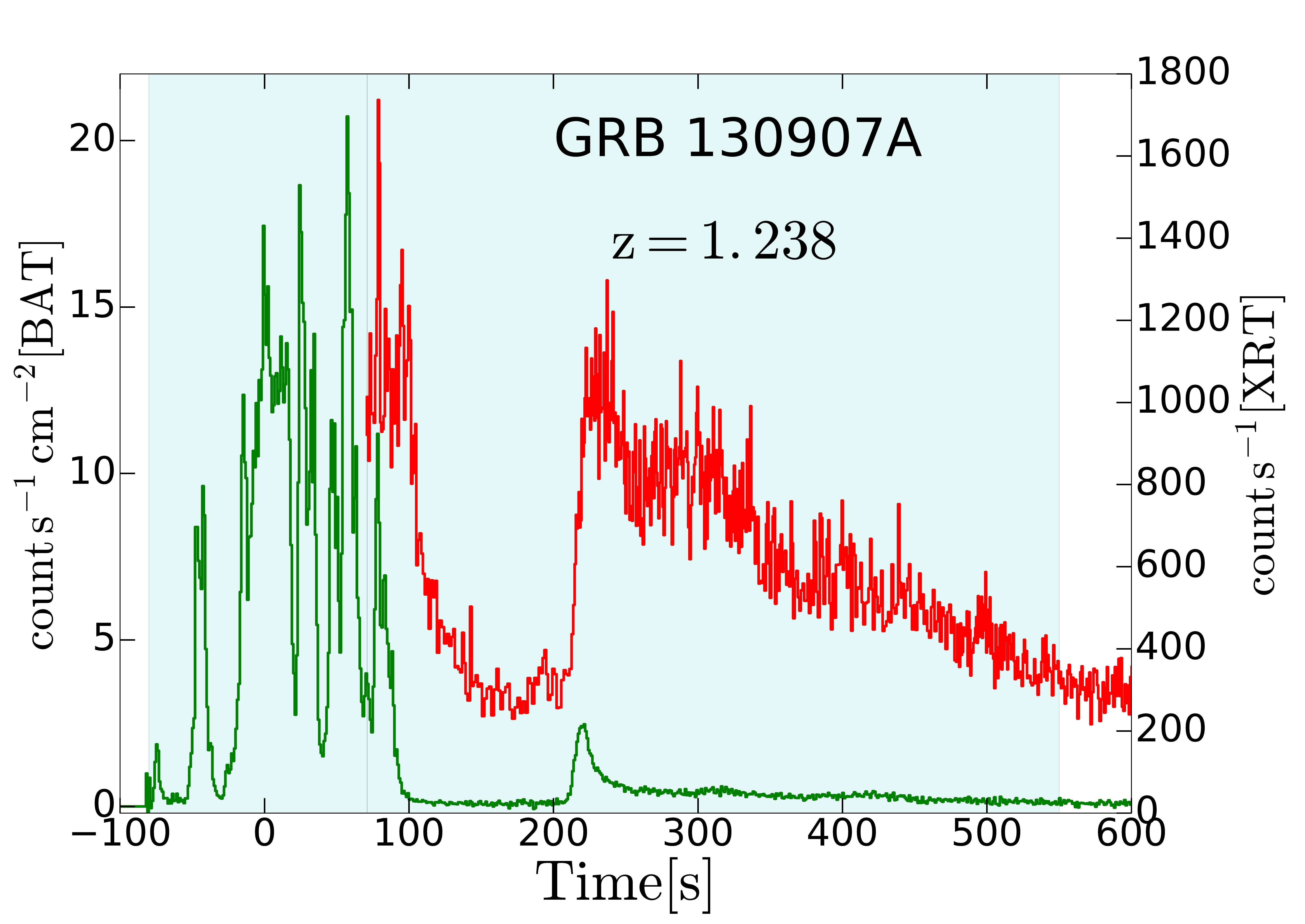} \hspace{0.06em}
\includegraphics[scale = 0.097]{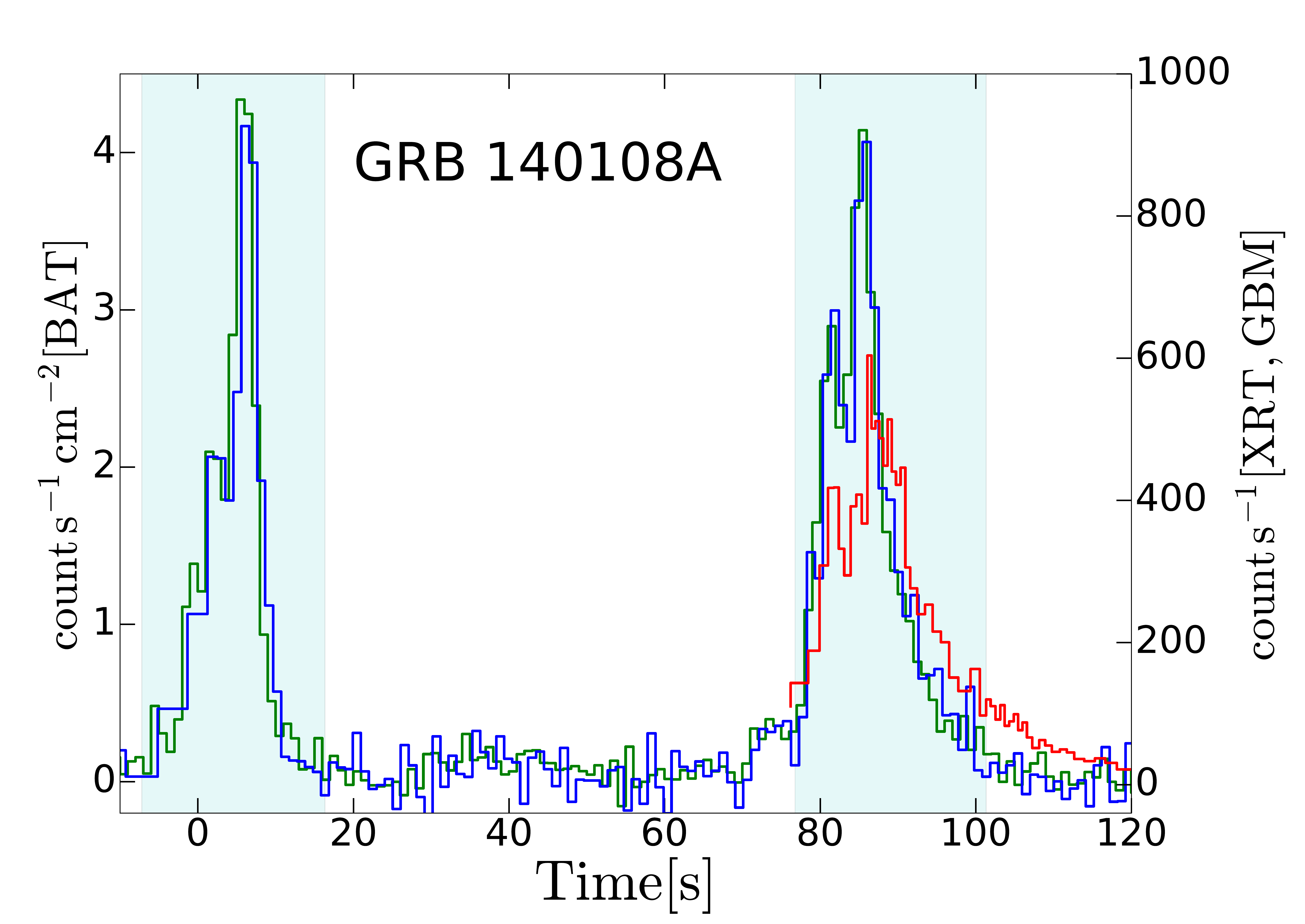} \\
\includegraphics[scale = 0.097]{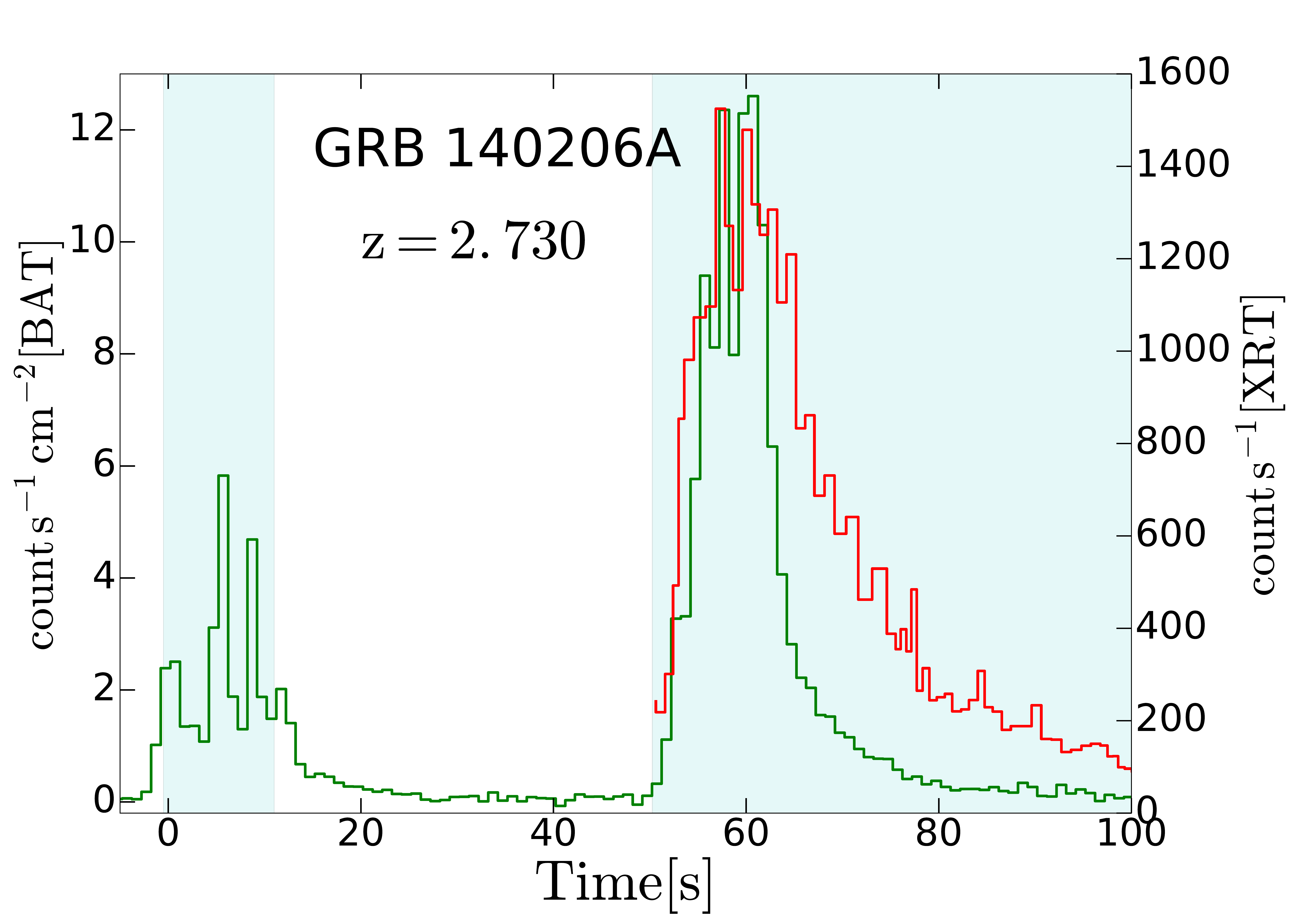} \hspace{0.06em}
\includegraphics[scale = 0.097]{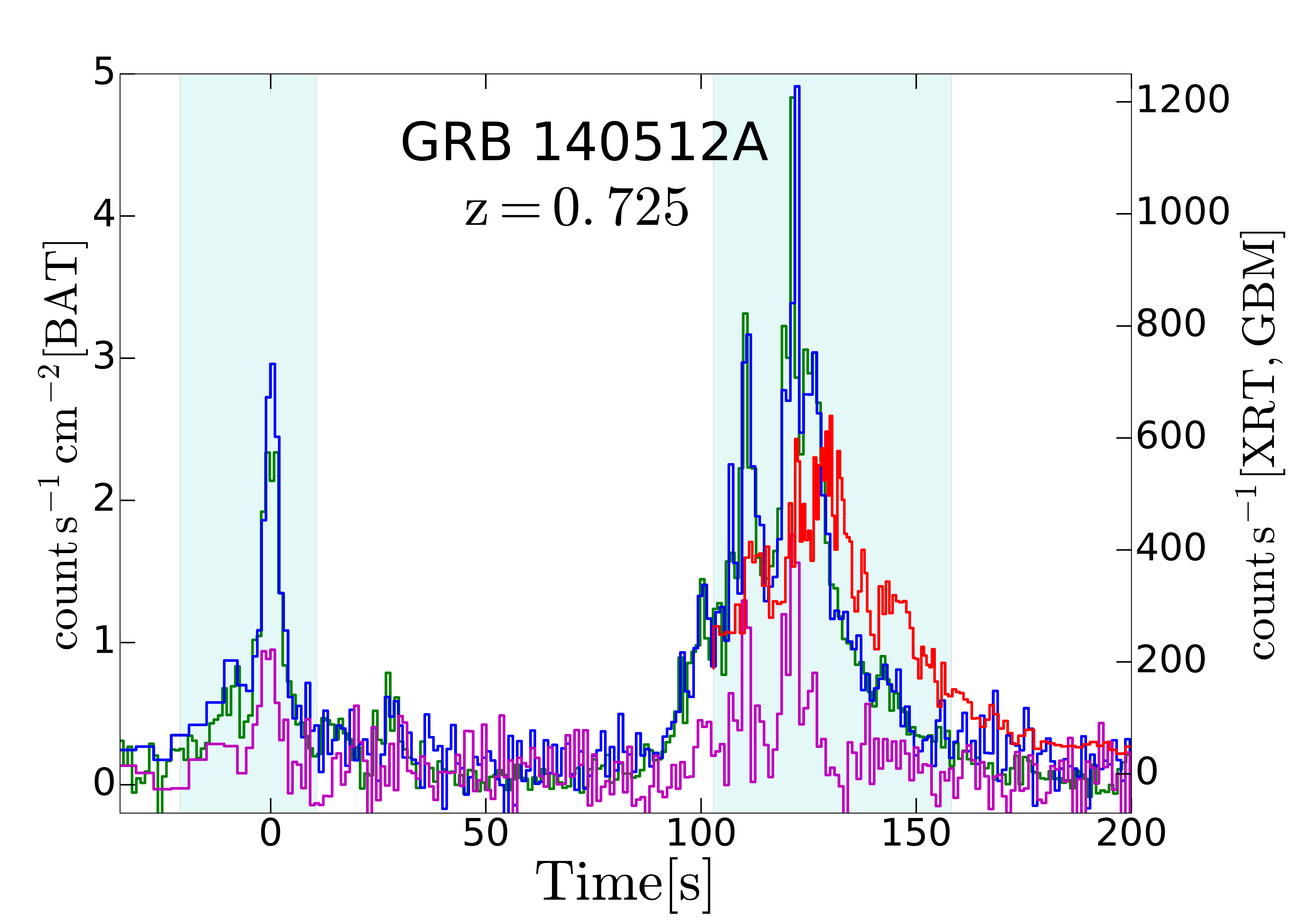} \hspace{0.06em}
%\end{center}
\caption{\label{fig:lc} Background-subtracted light curves of the 14 GRBs analyzed in this work. The time refers to the BAT trigger. \textit{Swift}-XRT light curves (in the range 0.5-10 keV) are shown in red, \textit{Swift}-BAT (15-150 keV) in green, \textit{Fermi}-GBM (8-800 keV) in blue, and \textit{Fermi}-GBM (200 keV-1 MeV) in purple. The time intervals where spectral analysis has been performed (light-blue shaded areas) have been determined on the basis of an $S/N$ criterion applied to BAT data. }
\end{figure*}

In order to extend the characterization of prompt spectra down to the soft X-ray band, we selected a sample of GRBs for which the prompt emission (or part of it) has been  
observed by the XRT in the 0.3-10\,keV range, in addition to the BAT in the 15-150 keV energy range.
To this aim, we inspected the XRT light curves of all events detected up to 2016 January reported in the online \textit{Swift}-XRT GRB catalog\footnote{\url{http://www.swift.ac.uk/xrt_live_cat}} \citep{Evans_09}.
The online tool automatically identifies the presence of pulses (defined as statistically significant positive deviations from an underlying power-law emission) and returns the time intervals where the pulses are present.   
This selection resulted in 329 GRBs with at least one significant pulse in X-rays.

The scope of the sample selection is to find GRBs with simultaneous signal in the XRT and BAT instruments that can be combined for a joint spectral analysis.
We then checked whether for these 329 GRBs the emission detected by the XRT was simultaneously observed also by the BAT.
To this aim, we extracted background-subtracted count rate BAT light curves in the energy range 15-150 keV. 
First, using the \textit{batgrbproduct} tool, we estimated the bust duration $T_{100}$, which corresponds to the duration that contains 100\% of the burst emission.
Then, we estimated the count rate outside the $T_{100}$ time interval and found that its value is always smaller than $\sim0.01\,$counts/s/detector, which is then chosen as reference value. 
Adopting as initial time the starting time of XRT observations, we applied the Bayesian block algorithm \citep{Scargle} to identify the possible presence of significant changes in the BAT signal during the XRT-detected emission, by requiring a BAT count rate higher than 0.01 counts/s/detector. 
This selection resulted in 77 GRBs with simultaneous signal detected by the BAT and the XRT.
Since the goal of our study is to perform reliable spectral analysis combining BAT and XRT data, we further limited the sample:
we required it to have at least four time bins with BAT signal-to-noise ratio $(S/N)$ larger than 30 during the time interval where the emission is simultaneously observed by the BAT and the XRT. The choice of the threshold value $(S/N>30)$ is based on the study of \cite{savchenko09}, where they show that BAT spectra with $S/N>30$ return photon indices similar to those of the complete catalog of BATSE time-resolved spectra \citep{kaneko06}.
After applying all these selection criteria, we ended up with a sample of 15 GRBs. Among these, we excluded GRB~130427A because 
%From the 15 GRBs fulfilling these requirements, we excluded GRB~130427A, 
data extraction for this GRB requires a nonstandard pipeline processing \citep{maselli14}.
The final sample includes 14 GRBs (Table~\ref{tab:nh}).  
In seven cases, {\it Fermi}-GBM observations are also available and have been included in the spectral analysis. 
The light curves of all 14 GRBs are shown in Figure~\ref{fig:lc}. In each panel we show the XRT ($0.5-10\,$keV; red curve), BAT ($15-150\,$keV; green), and when available also the GBM ($8-800\,$keV in blue and 200\,keV$-1\,$MeV in purple) count light curves. 
Note that in most cases XRT observations are available during the brightest part of the prompt emission, while in the remaining few cases they cover the less intense part of the prompt phase.
The redshift (available for eight GRBs) ranges between $z=0.725$ and $z=2.73$ (Table~\ref{tab:nh}).
%%%%%%%%%%%%%%%%%%%%%%%    DATA EXTRACTION AND ANALYSIS  %%%%%%%%%%%%%%%%%%%%%%%%%%%% 
\section{Data extraction and spectral analysis}\label{sec:data}
%======================  SWIFT ======================
\subsection{Swift-BAT and Swift-XRT data extraction}
The \textit{Swift} data have been processed using standard procedures, which we briefly summarize in the following.
We downloaded the BAT event files from the Swift data archive\footnote{\url{http://heasarc.gsfc.nasa.gov/cgi-bin/W3Browse/swift.pl}}. 
We extracted the \textit{Swift}-BAT spectra and light curves using the latest version of the {\sc heasoft} package (v6.17). 
The background-subtracted mask-weighted BAT light curves have been extracted in the energy range 15-150 keV using the \textit{batmaskwtevt} and \textit{batbinevt} tasks in FTOOLS. 
BAT spectral files have been produced using the \textit{batbinevt} task and have been corrected through the  \textit{batupdatephakw} and  \textit{batphasyserr} tasks to include systematic errors. 
Using \textit{batdrmgen}, we generated different response matrices for intervals before, during, and after the satellite slew. 
The latest calibration files (CALDB release 2015 November 13) have been adopted.

The XRT light curves have been retrieved from the \textit{Swift} Science Data Center, provided by the University of Leicester\footnote{\url{http://www.swift.ac.uk/xrt_curves/}}
\citep{Evans_09}.  
To extract the spectra, we downloaded the XRT event files from the \textit{Swift}-XRT archive\footnote{\url{http://www.swift.ac.uk/archive/}}. 
Since for all our GRBs XRT data are heavily piled up, data from the central region\footnote{By ``central region" we mean the 
circular region centred on the pixel with the largest number of counts detected within the time of interest  \citep{Romano_06}.} 
have been excluded \citep{Romano_06}. 
For each GRB, the size of the exclusion region has been determined so that the maximum count rate in the time interval of interest does not exceed 150 counts s$^{-1}$.
We extracted source and background spectra in each time bin using the \textit{xselect} tool. 
For each time bin, the ancillary response file has been generated using the task \textit{xrtmkarf}. 
We excluded from the spectral analysis all channels below 0.5 keV.
In order to use $\chi^2$ statistics, energy channels have been grouped together using the {\it grppha} tool by requiring at least 20 counts per bin.

%======================  FERMI  ========================
\subsection{Fermi-GBM data extraction}
The GBM is composed of 12 sodium iodide (NaI) and two bismuth germanate (BGO) scintillation detectors \citep{meegan09}.
We considered the data from the two NaI and the BGO detectors with the highest signal. 
For each detector, we retrieved the data and the detector response matrices from the \textit{Fermi} website\footnote{\url{http://fermi.gsfc.nasa.gov/}}. 
We selected CSPEC data, i.e. time sequences of 128 energy-channel spectra with integration time of 1024 ms each. 
%characterised by full spectral resolution (128 energy channels).
Channels with energies in the range $8-800\,$keV and 200\,keV$-1\,$MeV were selected for the NaI and BGO detectors, respectively.
The extraction of spectra and light curves has been performed using \textsc{RMFIT} (v4.3.2). 
We selected pre- and post-burst data to model the background and fit an energy- and time-dependent polynomial.
Spectra and background files have been exported from \textsc{RMFIT} to \textsc{XSPEC}(v12.7.1) format in order to fit GBM spectra jointly with BAT and XRT spectra. 
The extraction of GBM spectra is compliant with the standard procedures adopted in the literature (e.g.  \citealt{goldstein12,gruber14}). 
Energy channels have been grouped together using the {\it grppha} tool by requiring at least 20 counts per bin.

%====================   SPECTRAL ANALYSIS ======================
\subsection{Spectral analysis}
The spectral analysis has been performed using \textsc{XSPEC}(v12.7.1).
To account for intercalibration uncertainties between the different instruments, we introduced multiplicative factors in the fitting models.
In particular, when GBM data are not available, we multiplied the XRT model by a factor left free to vary between 0.9 and 1.1. When GBM data are available, we froze to 1 the factor between XRT and BAT and multiplied the GBM model by a free factor.
Inspecting the results inferred from the best-fit models, we found that in all cases the calibrations between the GBM and XRT/BAT agree within $\rm 15 \%$.

The time intervals for the temporally resolved analysis have been defined so that in each bin the BAT $\rm S/N$ is larger than 30.
Moreover, when possible, we redefined the time bins (provided that the criterion on the BAT $\rm S/N$ is always satisfied) in order not to mix the rising and decaying parts of a pulse, or, if a pulse is composed by the superposition of many spikes, in order not to mix different spikes.
The analysis was applied also to the initial part of the emission, before XRT observations started. Time intervals selected for the analysis are outlined in Figure~\ref{fig:lc} with gray-shaded areas.
The total number of time-resolved spectra analyzed is 128. For 86 of these, XRT data are available.

In the following, we explain in detail how the spectral analysis has been performed.
First, we discuss the method adopted to account for absorption in the soft X-ray band.
Then, we introduce the spectral models and the criteria adopted for the selection of the best-fit model.

%-------------------------------------
\subsubsection{Absorption model}
For GRBs with known redshift, we accounted for both Galactic and intrinsic metal absorption using the XSPEC models $tbabs$ and $ztbabs$, respectively \citep{wilms00}. 
The Galactic contribution to absorption in the direction of the burst has been estimated from \cite{Kalberia_05}. 
The intrinsic absorption has been fixed to the value estimated from spectral analysis of late-time ($\gtrsim10^4\,$s) XRT data \citep{Butler_07}.
During XRT pulses, indeed, if the intrinsic $N_{\rm H}$ is left as a free parameter, a dramatic variation (even by a factor of 10) of its value is often observed. 
While an increase of $N_{\rm H}$ could be induced by photoionization effects of the circumburst medium by the prompt radiation (e.g. \citealt{perna02,lazzati03,perna03,frontera04}), a fast decrease of $N_{\rm H}$ is more difficult to explain. This could hide a temporal 
evolution of the spectrum, e.g. the passage of any spectral break across the XRT energy band \citep{Butler_07}. Therefore, the best estimate of $N_{\rm H}$ could be obtained when there is no strong spectral evolution and the light curve is well described by a simple power-law decay.
We chose the latest-available XRT time interval (provided that no spectral evolution is apparent and the light curve is well described by a power-law decay) and modeled the  extracted spectrum with an absorbed power law.
When extracting the late-time spectrum, we considered an integration time large enough to constrain the intrinsic $N_{\rm H}$. 
This value of $N_{\rm H}$ has then been used as an input (fixed) parameter for the early-time spectral analysis.

For GRBs with unknown redshift, the late-time X-ray spectrum has been fitted by applying the $tbabs$ model only.
We verified that in all cases the best-fit value of $N_{\rm H}$ derived from this fit was larger than the Galactic value estimated from \cite{Kalberia_05}.
This value of $N_{\rm H}$ has then be used as a fixed input parameter for early-time spectral analysis, where this time only the $tbabs$ model was applied.

For each GRB, the value of the intrinsic $N_{\rm H}$ inferred from late-time data and the late-time interval (LTI) chosen for the analysis are listed in Table~\ref{tab:nh}.

%-------------------------------
\subsubsection{Spectral models}
\begin{figure}
\centering
\includegraphics[scale=0.5]{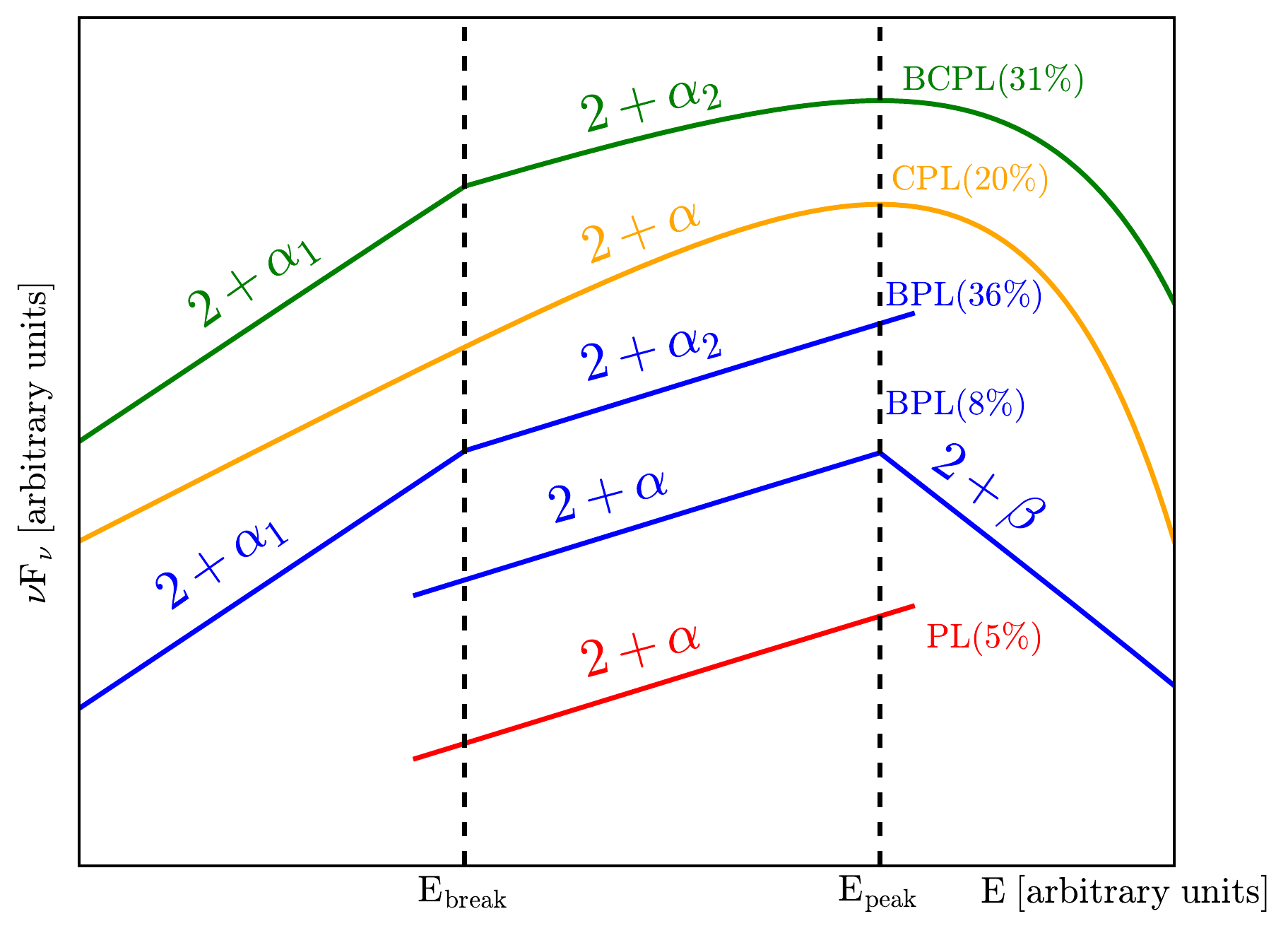}
\caption{Summary of the four spectral models adopted in this work, and definition of the adopted notation for the spectral indices and characteristic energies. Each one of the analyzed spectra is well described by one of these models. From top to bottom: a broken power law with a high-energy exponential cutoff (BCPL; green), a cutoff power law (CPL; orange), a broken power law (BPL; blue), and a single power law (PL; red). For the BPL model, we found two different cases in our analysis: both indices are larger than -2 (in this case they are called $\alpha_1$ and $\alpha_2$), or the first index is $>-2$ and the second one is $<-2$ (in this case they are called $\alpha$ and $\beta$, respectively). The percentages quoted next to each model name refer to time-resolved analysis for periods where XRT data are available.}
\label{fig:sketch}
\end{figure}

Spectral models commonly applied to GRB prompt spectra include a single power law (PL), a power law with an exponential cutoff (CPL), and a Band function \citep{preece00,kaneko06,nava11,goldstein12,gruber14,bhat16,lien16}. 
These empirical models usually return a satisfactory fit to most spectra.
However, as we will show in the following, all these models are in most cases inadequate when the energy range available for the analysis is extended down to 0.5\,keV by the inclusion of XRT data.
Stated differently, XRT spectra do not lie on the low-energy extrapolation of the spectral shape defined by $>10\,$keV data. 
A spectral break in the soft X-ray band must be introduced in order to fit with one single spectral component the prompt spectrum from $\sim$0.5 keV to $\sim$1 MeV. 
We then extend the standard models (PL, CPL, and Band) to include a low-energy break. 
This leads us to introduce three additional models: a broken PL (BPL), a CPL with a break at low energies (BCPL), and a Band model with a break at low energies.
However, the high-energy spectral index of the Band model and that of the Band model with a low-energy break are always unconstrained.
This is due to the fact that for half of the sample, GBM data are not available.
Moreover, even when they are available, the relatively small $\rm S/N$ of the time-resolved analysis makes it difficult to constrain the value of $\beta$.
The value of $\rm \beta$ is constrained in a few cases where a BPL is the best-fit model. We also tried to apply a smoothly broken power-law model 
with a high-energy cutoff, but we did not succeed in constraining the smoothness parameter and/or the shape of the spectrum below the break energy.
Summarizing, we found that all the spectra analyzed in this work are well described (i.e., the best-fit model gives a reduced chi-square $\chi^2_{\rm red}<1.15$, except for one case, where $\chi^2_{\rm red}=1.3$) by one of the following four models: PL, CPL, BPL, or BCPL (see Figure~\ref{fig:sketch}).

We use the following conventions.
A photon index is called $\alpha$ if its value is larger than $-2$ in the notation $dN(\nu)/d\nu\propto \nu^{\alpha}$, where $dN$ represents the photon number (i.e., $\alpha$ identifies a part of the spectrum that is rising in the $\nu F_\nu = \nu^2 N_\nu$ representation). If there are two (consecutive) segments where the spectrum is rising (which is a common case in our analysis), we call them $\alpha_1$ and $\alpha_2$. The break energy that separates these two rising power-law segments is called $E_{\rm break}$. 
Following the traditional notation, when the spectrum has a peak in $\nu F_\nu$, we refer to it as the peak energy $E_{\rm peak}$.
Finally, we use the letter $\beta$ when the photon index is lower than $-2$ (i.e., describing a part of the spectrum that is decreasing in $\nu F_\nu$).
We found a few cases where the photon spectral index has a value around $-2$. In these cases, we refer to it as $\beta$ if it is smaller than -2 within 1$\sigma$ error.

A schematic representation of all the models, the notation, and different cases found in our analysis is shown in Figure~\ref{fig:sketch}. As can be seen in this plot, a BPL model can describe two different situations: either both indices are $>-2$ ($\alpha_1$ and $\alpha_2$, separated by a break energy $E_{\rm break}$), or the first index is $>-2$ and the second one is $<-2$ (in this case we call them $\alpha$ and $\beta$, and they are separated by the peak energy $E_{\rm peak}$). For PL, CPL, and BCPL models, instead, we find only cases where the photon indices are $>-2$.

%-----------------------------------------------
%-----------------------------------------------

We fitted all the time-integrated and time-resolved spectra to all the models, and for each spectrum we identified the best-fit model.
In general, the $F-test$ is used to compare different models and choose the best one, but only when the models to be compared are nested \citep{Protassov_02}. 
Since we are testing the existence of a new feature (i.e., a spectral break), we decided to perform a conservative analysis and set at 3$\sigma$ the significance level of the $F-test$ required to select a more complex model.
In Appendix (Figure \ref{fig:scheme}) we provide a scheme of the method applied to determine the best-fit model. 
We start with the simplest function (PL) and consider progressively more complex functions.
A single PL can be generalized in two different ways: by adding a break or by adding a high-energy exponential cutoff.
In both cases the fit obtained with the resulting model (BPL and CPL, respectively) can be compared with the PL fit through an $F-test$.
Depending on the result of the comparison, different cases are possible:
\begin{itemize}
\item Neither of the two models significantly (at more than $3\sigma$) improves the fit. In this case the best-fit model is a PL;
\item Only one of the two models improves the PL fit. We then select this model (either a CPL or BPL) and compare it to the fit performed with a BCPL, through an $F-test$. A BCPL model is chosen only if the improvement is significant at more than $3\sigma$;
\item Both models (CPL and BPL) improve the PL fit. First, we compare them one to each other. Note that they are not nested, and the $F-test$ cannot be performed.
Since the number of parameters is different (3 for the CPL and 4 for the BPL), if the total $\chi^2$ of the BPL is the largest between the two, then a CPL is preferred and is compared to the BCPL. In the opposite case ($\chi^2_{\rm CPL}>\chi^2_{\rm BPL}$), we separately compare each of them to the BCPL fit.
If the BCPL significantly improves both of them, then we choose the BCPL.
If the improvement over a CPL is significant, but the improvement over a BPL is not, it means that the spectrum has a significant break, but not a significant exponential cutoff, and a BPL is then chosen.
If the opposite case is verified (BCPL is better than a BPL but not better than a CPL), it means that a high-energy cutoff is clearly present, while the low-energy break is not significant. A CPL is then chosen. 
The validity of this method is confirmed by the inspection, case by case, of the shape of the residuals.
A peculiar situation (which is realized only in six spectra) is provided by the case when a BCPL is not improving either the CPL or the BPL fit, and one of the latter models must be chosen.
In these cases, we inspect the residuals and choose the model fit for which the residuals do not show evidence of systematic trends. 
\end{itemize}
We verified that for all spectra the selected best-fit model gives a reduced chi-square $\chi^2_{\rm red}<1.15$, except for one case, where $\chi^2_{\rm red}=1.3$ (see Figure~\ref{fig:BB}).

%%%%%%%%%%%%%%%%%%%%%%%    RESULTS  %%%%%%%%%%%%%%%%%%%%%%%%

\section{Results}\label{sec:results}
We first present and discuss in detail the results of our analysis applied to one event, GRB~140512A, as an example. 
In the second part of this section, we present the results obtained by applying the same analysis to all GRBs in our sample.
For all time-integrated and time-resolved spectra, the results (best-fit models, parameters, and fluxes)  are reported in Table~\ref{tab:table}.

%=========================  GRB 140512A  ==========================
\subsection{GRB 140512A}
%------------------------------------------------------------------
\begin{figure*}
   {\centering
  \includegraphics[width=0.43\textwidth]{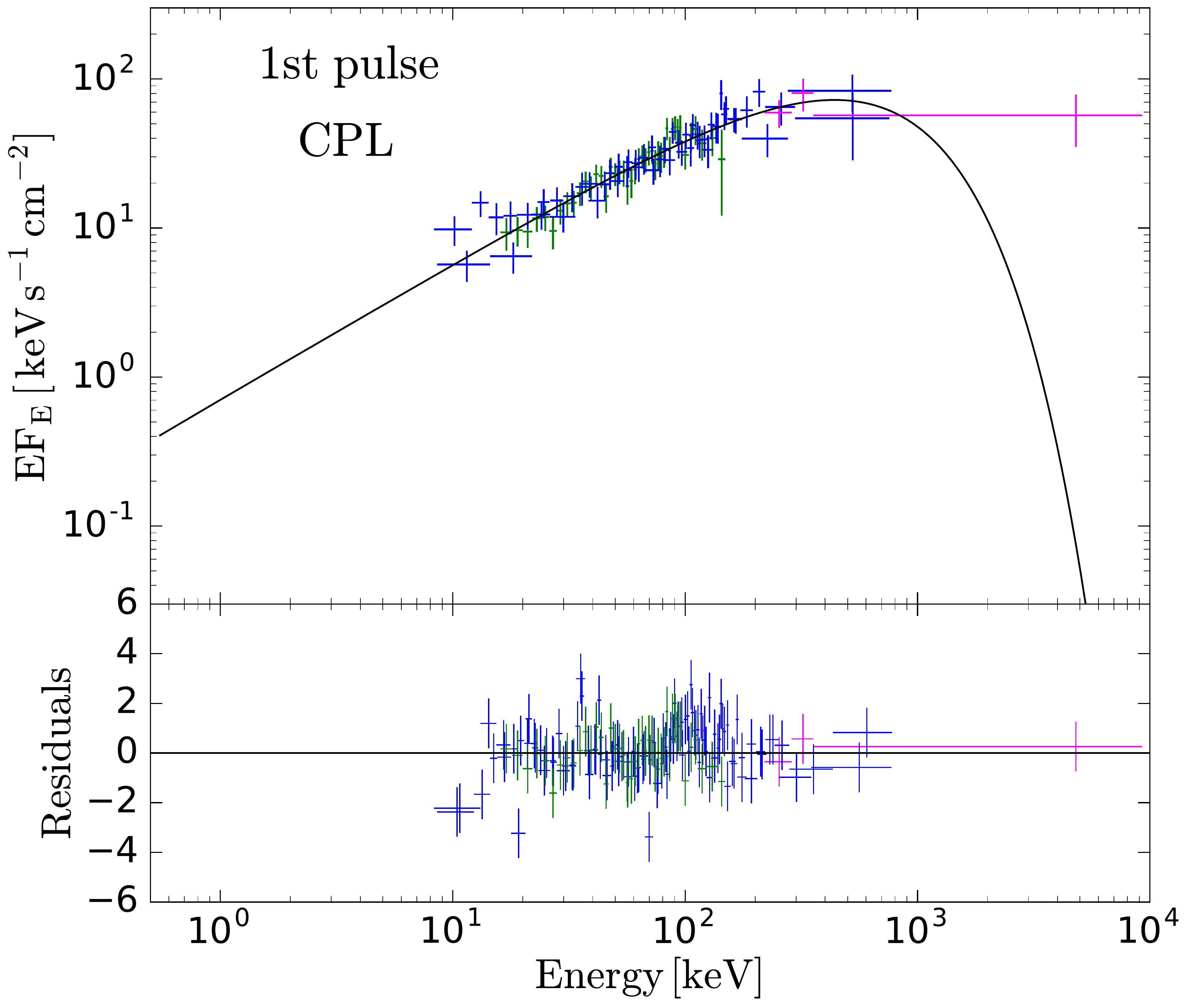}
   \includegraphics[width=0.43\textwidth]{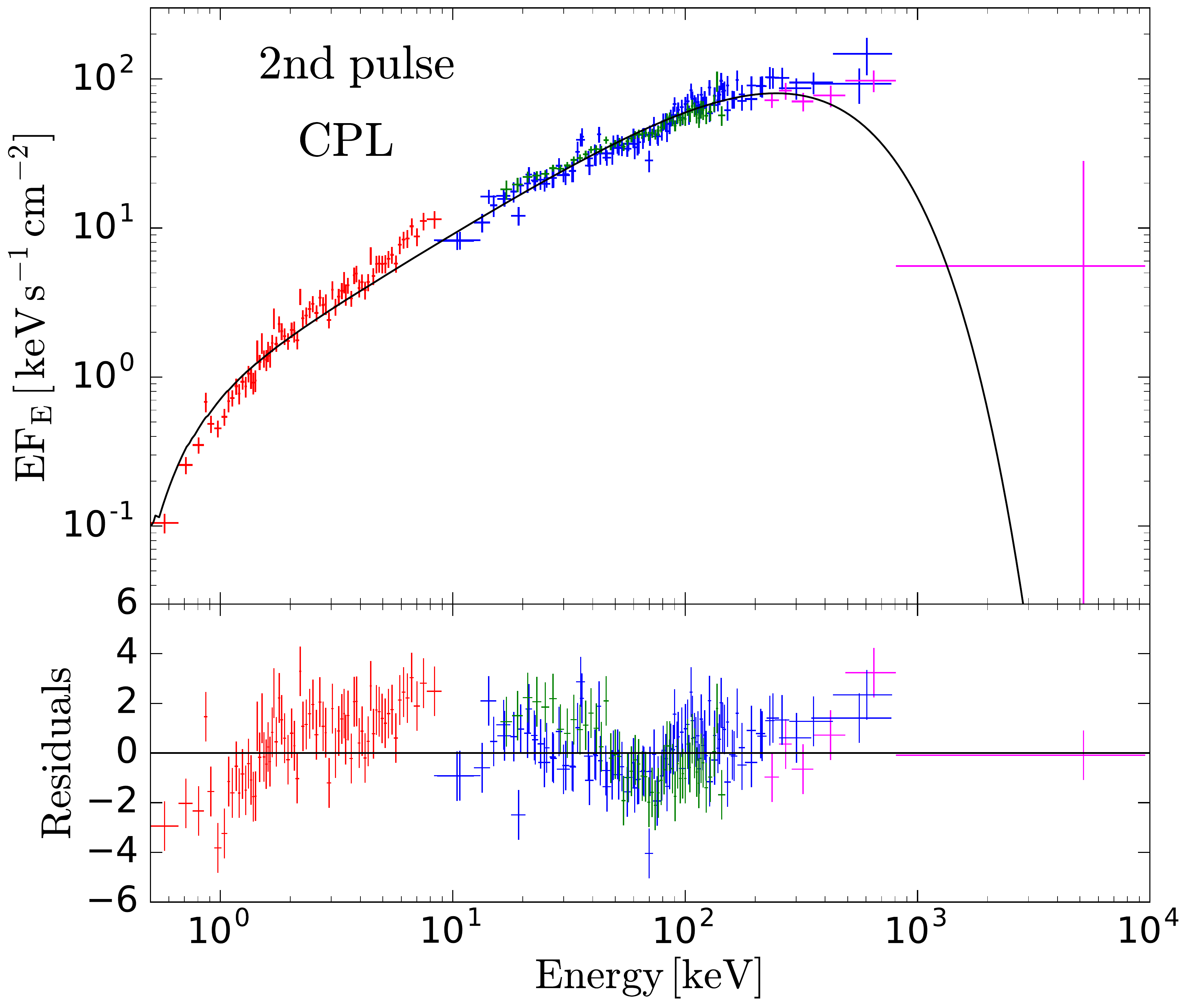}
   \includegraphics[width=0.43\textwidth]{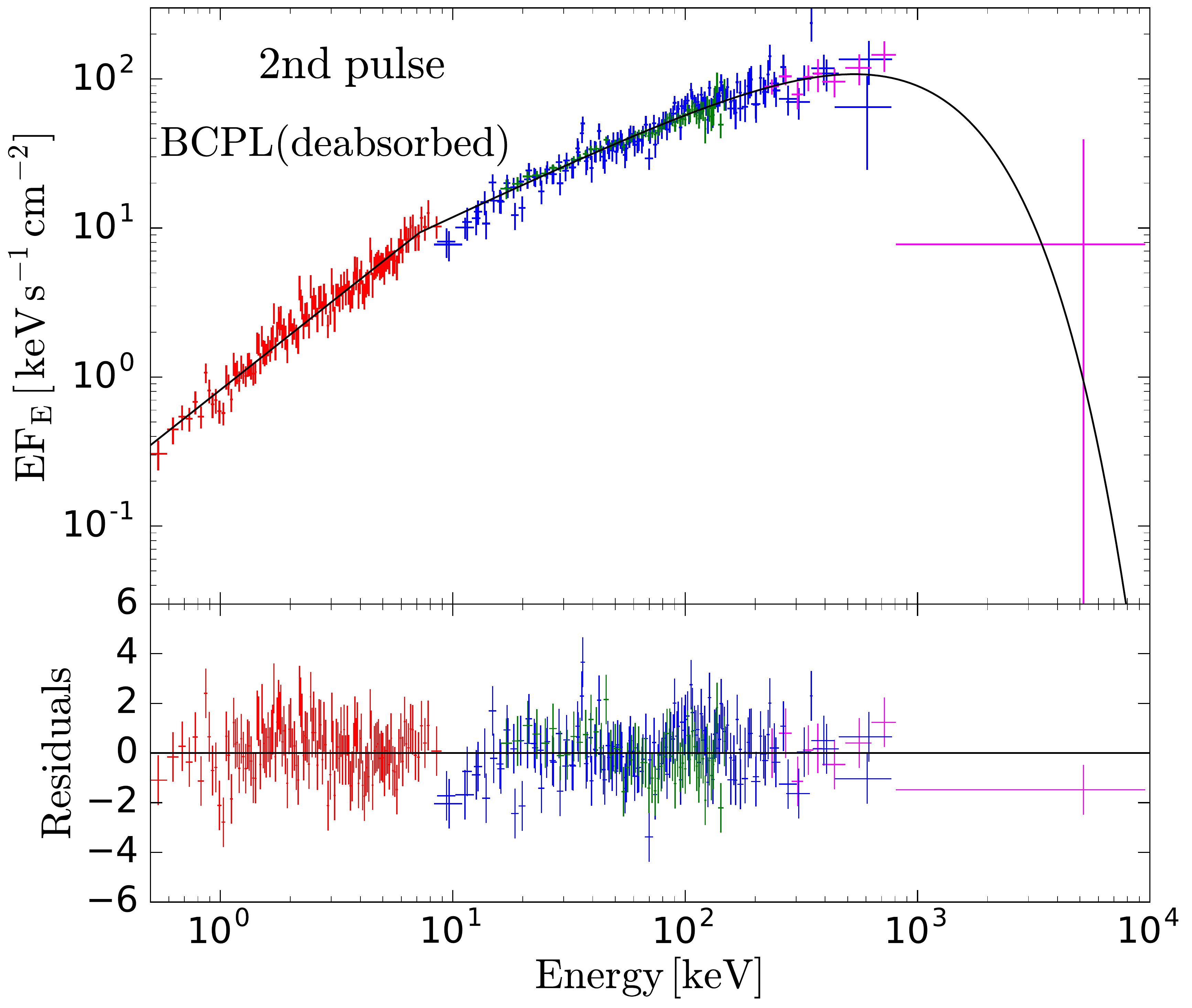}
   \includegraphics[width=0.43\textwidth]{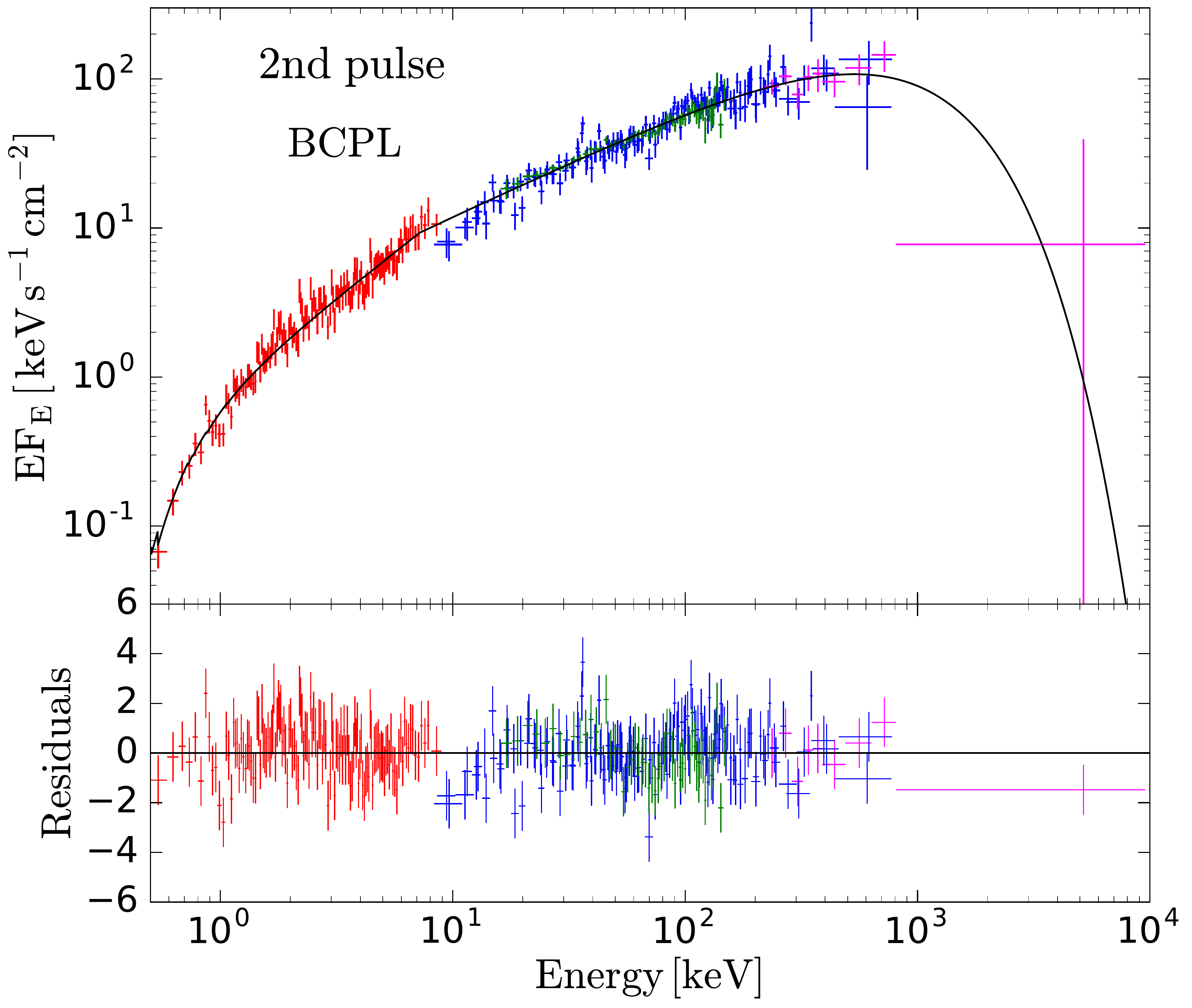}\\
   }
\caption{\label{fig:TI_spectra_140512A} Time-integrated spectral analysis performed separately on the first and second pulses of GRB~140512A (see the light curve in Figure~\ref{fig:140512A}). 
The bottom sections of each panel show the fit residuals (see text).  
The joint BAT (green) and GBM (blue and purple, corresponding to the NaI and BGO detectors, respectively) spectrum of the first pulse (integrated from $t=-21.05\,$s to $t=10.70\,$s since BAT trigger time) is shown in the top left panel. During this temporal window, XRT data are not available. 
The spectrum is well modeled by a CPL (black line).
The other three panels show the spectrum integrated during the second emission episode (i.e., from $t=102.86\,$s to $t=158.16\,$s). 
In this time interval, XRT observations are available and are included in the analysis (red data points).
The fits with a CPL (top right panel) and with a BCPL (bottom right panel) are shown.
For this last fit, the de-absorbed model and data are shown in the bottom left panel.}
\end{figure*}
%------------------------------------------------------------------

%------------------------------------------------------------------
\begin{figure*}
   {\centering
   \includegraphics[width=0.7\textwidth]{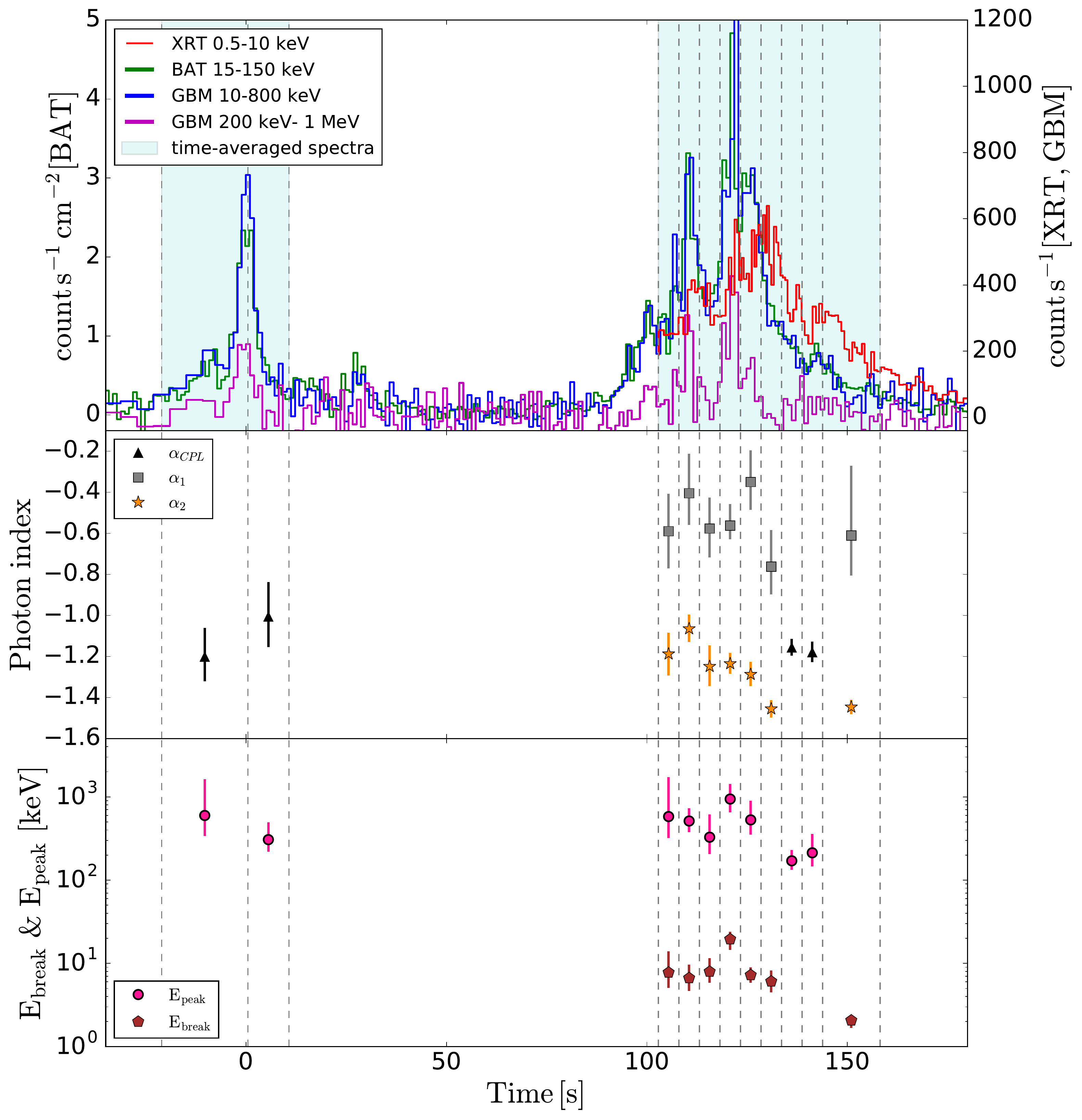}\\
   }
\caption{\label{fig:140512A} Results of the time-resolved spectral analysis of GRB~140512A. 
The top panel shows the XRT (red), BAT (green), and GBM (blue and purple) light curves.
The shaded vertical stripes 
show the time intervals selected for the time-average 
spectral analysis of the first and second pulses (the 
corresponding spectra are shown in Fig.~\ref{fig:TI_spectra_140512A}).
The dashed vertical lines show the time bins selected for the time-resolved spectral analysis. In the first interval, XRT data are not available.
The middle and bottom panels show the best-fit parameters (photon indices and break/peak energies, respectively)  with $\rm 1 \sigma$ level errors.}
\end{figure*}
%------------------------------------------------------------------
The light curve of GRB 140512A is composed of two separated emission episodes (see the top panel of Figure~\ref{fig:140512A}), which we call the first and second pulses.
During the first pulse, only BAT and GBM observations are available. For the second episode (where most of the radiation is emitted) there are also XRT data. 
First, we discuss the time-integrated spectral analysis, which has been performed on the two pulses separately.
The time intervals chosen for the analysis are shown by the cyan-shaded regions in Figure~\ref{fig:140512A} (top panel).  
The time-integrated spectra of each pulse are shown in Figure~\ref{fig:TI_spectra_140512A}.
The spectrum of the first pulse (top left panel) is well fitted by a CPL model (solid line, $\rm  \chi^{2}=196.9$, for $317$ degrees of freedom  [dof]), which according to the $F-test$ improves the PL fit ($\rm  \chi^{2}=235.0$, for $318$ dof) with a $\rm 3 \sigma$ significance. We note that both the PL model and CPL model overfit the data, since they result in a $\rm \chi_{red}^{2} < 1$. A Band model does not improve the CPL fit (i.e., a high-energy power law is not required by the data).
The best-fit parameters are $\rm \alpha=-1.09_{-0.11}^{+0.12}$ and $E_{\rm peak}=439_{-134}^{+293}\,$ keV.

In the second pulse, a CPL model (top right panel in Figure~\ref{fig:TI_spectra_140512A}) appears adequate for the description of $>8\,$keV data, but cannot account for the harder spectral shape characterizing the XRT band. The CPL model returns $\rm \chi^{2}=613.6$ ($\rm d.o.f.=480$) and shows a systematic trend in the residuals (defined as the difference between the data and the model, divided by the error, and shown in the bottom sections of each spectrum). 
We then allow for a spectral break at low energies and verify that a BCPL model (bottom right panel) gives a significantly better description of the data. 
For this model $\chi^{2}=442.8$ ($\rm d.o.f.=478$), corresponding to an improvement (with respect to the CPL one) of $8.4\sigma$ significance.
The best-fit parameters are $E_{\rm break}=7.18_{-1.0}^{+1.12}\,$keV, $\rm \alpha_1=-0.76_{-0.04}^{+0.05}$, $\alpha_2=-1.26 \pm 0.04$, and $E_{\rm peak}=532_{-123}^{+190}\,$keV. 
The curvature below $\sim3$\,keV visible in the data and in the model is due to the absorption, which we inferred to correspond to $N_{\rm H}=4.4\times 10^{21}\,$cm$^{-2}$ from the spectral analysis of the data accumulated between $2.8\times 10^4\,$s and  $3.3\times 10^5\,$s (see Table~\ref{tab:nh}).
For convenience, for the BCPL fit we also show (bottom left panel) the de-absorbed model and data, so that the intrinsic shape of the spectrum can be better appreciated.

The results of time-resolved spectral analysis performed on each pulse are shown in Figure~\ref{fig:140512A} (middle and bottom panels, respectively).
The first pulse is divided into two time bins. In both bins, the spectra are best fitted by a CPL. 
The second pulse is divided into nine time bins. 
In seven cases, the best-fit model is a BCPL. In the remaining two cases, a CPL model is chosen, because the addition of a low-energy break improves the fit with a 2$\sigma$ significance, which, according to our 3$\sigma$ requirement, is not sufficient to claim the presence of a break.
The spectral indices as a function of time are plotted in the middle panel of Figure~\ref{fig:140512A}.
When the best-fit model is a BCPL, the spectral index $\alpha_2$ (stars), representing the spectral shape just below the peak energy, is softer as compared to the standard value $\alpha\simeq-1$, i.e., we find $-1.5<\alpha_2<-1$.
At lower energies, below the break energy, the spectral slope (squares) is higher and spans the 
 range $-0.9<\alpha_1<-0.2$ (this range includes the $1\sigma$ statistical uncertainty on the smallest and largest measured values of $\alpha_1$).
The break energy $E_{\rm break}$ (bottom panel, pentagon symbols) assumes values between 2 and 20\,keV, while for the peak energy $E_{\rm peak}$ (circles) we found standard values, between 200\,keV and 1\,MeV.
For the first six time bins of the second pulse, the spectra and their modeling with different spectral models are shown in the Appendix (Figure~\ref{fig:TR_spectra_140512A}). The six different rows refer to the six different time bins. For each time bin, the three panels show the fits and residuals obtained with a CPL (first panel), BPL (second), and BCPL (third) model. In these six time bins, the best model is always the BCPL.

%===============================  WHOLE SAMPLE  ============================
\subsection{Whole sample}
The results of the spectral analysis on time-integrated and time-resolved spectra for the entire sample (14 GRBs) are reported in Table~\ref{tab:table}.
For each spectrum, we report the time interval, the name of the best-fit model, the best-fit parameters, the flux, and the instruments included in the analysis.

We first comment on the results for time-integrated spectra. For almost all GRBs, two integration windows can be defined: a first one where XRT has not started observations yet (and only BAT and eventually GBM data are available), and a second one where also XRT observations are available and have been included in the analysis.
We note that all time-integrated spectra accumulated over epochs lacking XRT observations have best-fit functions represented by one of the standard models (PL or CPL).
Conversely (with only two exceptions represented by GRB~100906A and GRB~121123A), in spectra integrated over times where XRT observations are available, a break energy $E_{\rm break}$ is firmly  identified (i.e., the best model is either a BCPL or a BPL with both indices $>-2$, and the significance of the improvement as compared to models without a break is larger than 3$\sigma$).

We also performed time-resolved analysis and found that for time bins with XRT the best-fit model is a PL in 4 cases, a CPL in 17 cases, a peaked BPL in 7 cases, a BPL (with $\alpha_1,\alpha_2>-2$) in 31 cases, and a BCPL in the remaining 27 cases.
This means that in $67\%$ of the time-resolved spectra that take advantage of the presence of XRT observations, a break energy $E_{\rm break}$ is found and is firmly constrained. 
The significance of the break is higher than 5 $\sigma$ in 
65\% of cases, while in the remaining cases it is between 3$\sigma$ and 5$\sigma$.
For all GRBs except one (GRB~100906A) we can constrain the break energy at least in one time-resolved spectrum.
Conversely, when XRT is not available, a break energy is never found, and the best-fit model is either a PL (15 cases) or a CPL (27 cases). 

\begin{figure*}
\hskip 0.3truecm
\includegraphics[scale=0.26]{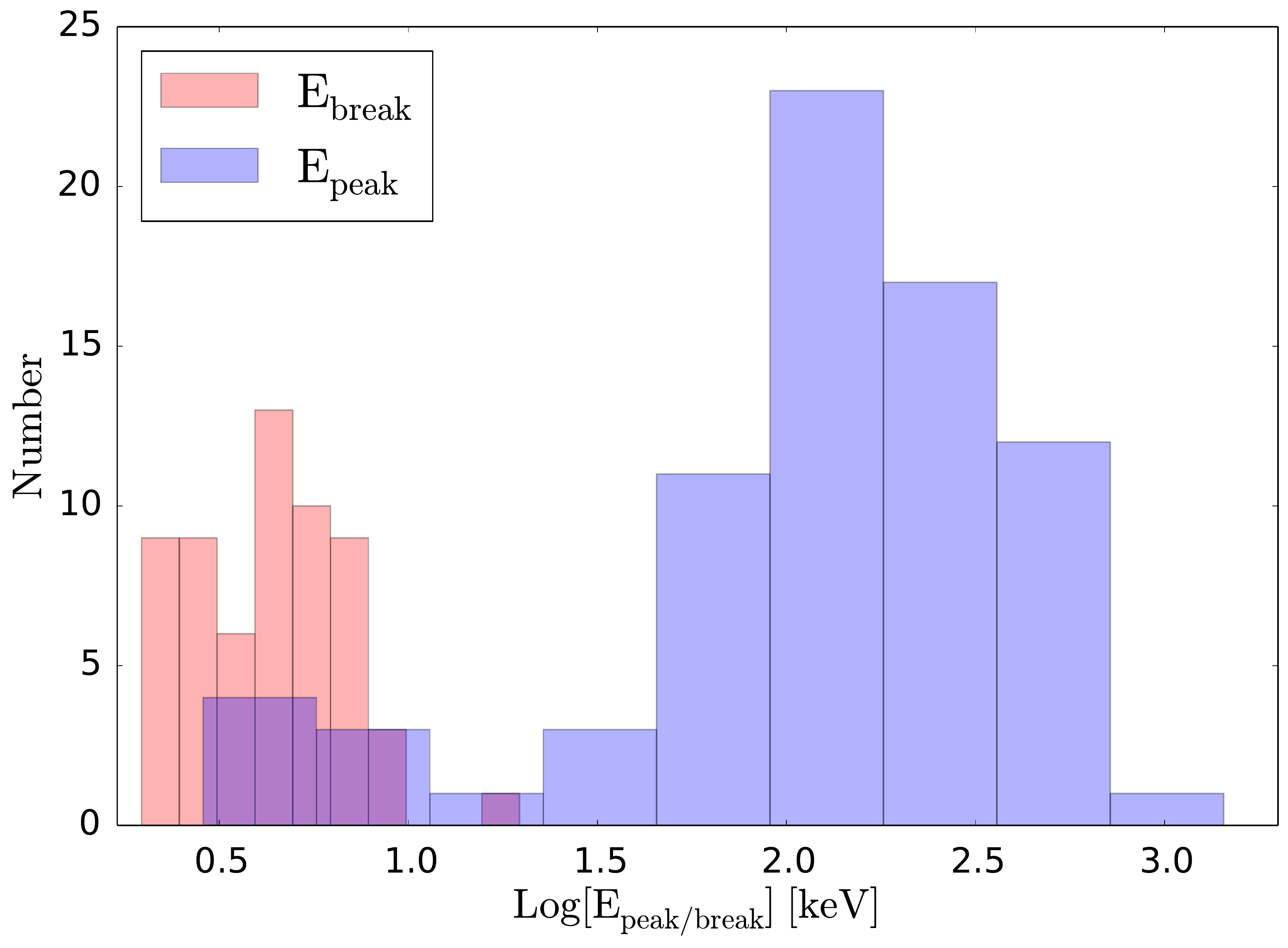}
\includegraphics[scale=0.263]{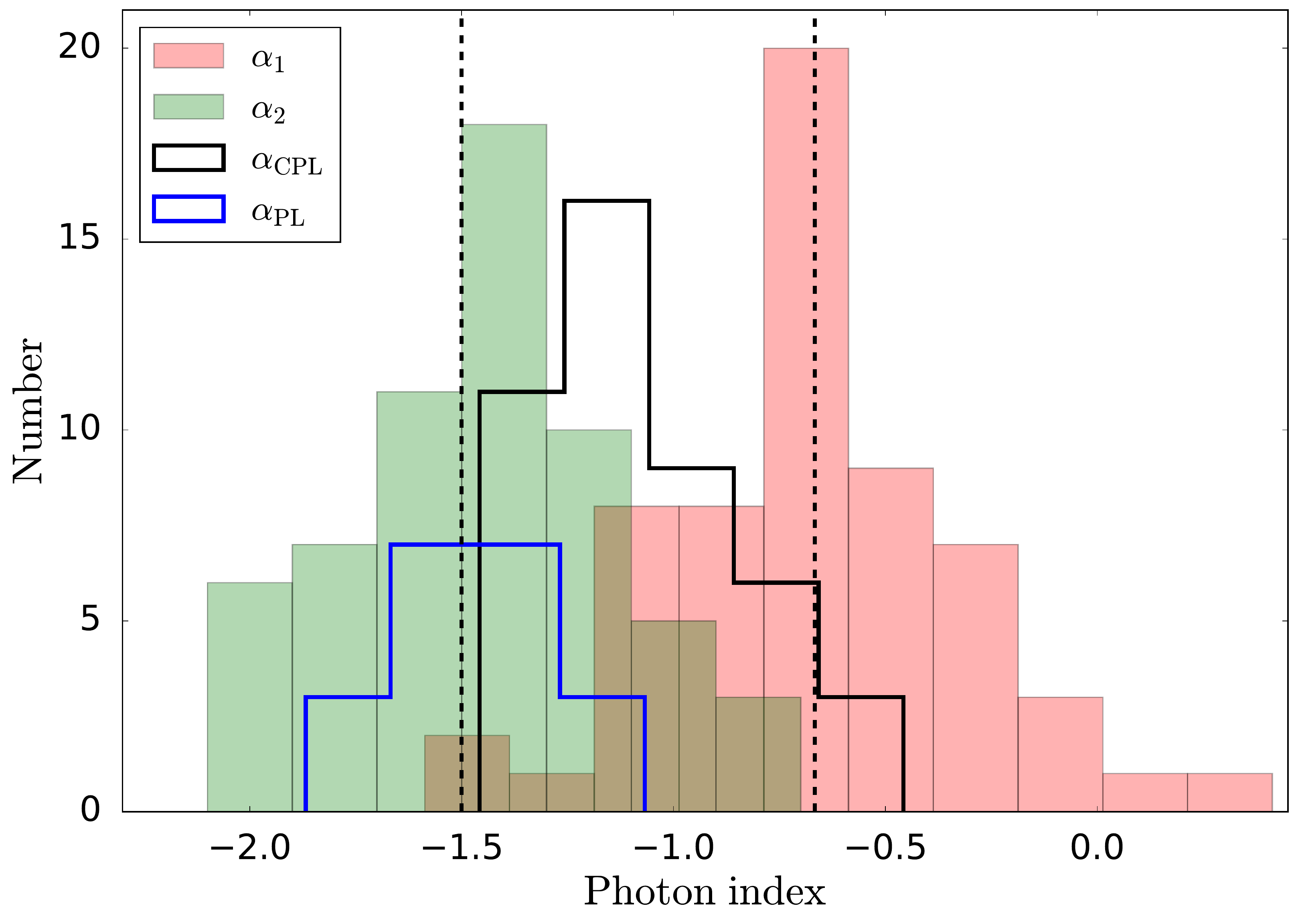}
\caption{\label{fig:histo} Distributions of the best-fit parameters for all time-resolved spectra. Left panel: the distributions of the peak and break energies are shown by the blue and red histograms, respectively. Right panel: photon index distributions. Different histograms refer to different models, according to the legend (refer to Figure~\ref{fig:sketch} for the notation). Dashed black lines show the values of the spectral slopes theoretically expected below and above the cooling break energy in fast-cooling synchrotron emission.}
\end{figure*}

For the time-resolved spectra of all GRBs included in the sample, we show in Figure~\ref{fig:histo} the distributions of the best-fit parameters.
We fit the distributions with gaussian functions and report the mean values and $1\sigma$ widths in Table~\ref{tab:gaussian_fits}.
The $E_{\rm peak}$ distribution (blue histogram in the left panel of Figure~\ref{fig:histo}) peaks at $\sim 120\,$keV, a value larger than that found in the BAT catalog ($\sim80\,$keV, \citealt{lien16}), reflecting the fact that for half of the GRBs included in the sample GBM data are available, allowing the determination of $E_{\rm peak}$ even when its value is above the BAT high-energy threshold. 
The inclusion of XRT data allowed us to find low value $E_{\rm peak}<20\,$keV, whose measure is usually precluded by analysis of GBM, BATSE, or BAT data alone, but whose existence has been already proven by the analysis of HETE data \citep{Sakamoto05} and X-ray flares \citep{Butler_07,Margutti_10}.

The pink histogram (left panel of Figure~\ref{fig:histo}) shows the distribution of the break energy $E_{\rm break}$.
We find that its logarithmic mean value corresponds to $\langle E_{\rm break}\rangle\sim4\,$keV, and its distribution spans one order of magnitude, from 2 to 20\,keV. The largest value found for $E_{\rm break}$ is then at the bottom edge of the BAT sensitivity range. This implies that BAT or GBM observations alone would not be sufficient to firmly reveal the presence of the break. The $E_{\rm break}$ distribution covers the whole XRT energy range, down to $\lesssim2\,$keV. Values smaller than $\sim$2\,keV cannot be recovered.

The right panel of Figure~\ref{fig:histo} shows the distribution of the spectral indices. We distinguish between $\alpha_{\rm CPL}$, $\alpha_{\rm PL}$, $\alpha_1$, and $\alpha_2$ (see their definition in Figure~\ref{fig:sketch}). The distribution of $\beta$ (not shown) is flat and ranges from -2 to -3.
Consistently with previous spectral catalogs, the mean value for $\alpha_{\rm CPL}$ is around -1, and the mean value of $\alpha_{\rm PL}$ is softer, around -1.5 (see Table~\ref{tab:gaussian_fits}).
When $E_{\rm break}$ is identified in the spectrum, the slope below and above the break ($\alpha_1$ and $\alpha_2$, respectively) can be defined. Their mean values are $\langle\alpha_1\rangle=-0.66$ 
($\rm \sigma=0.35$) and $\langle\alpha_2\rangle=-1.46$  ($\rm \sigma=0.31$).
Remarkably, these mean values are very close to expectations from synchrotron emission in a regime of fast cooling: $\alpha_1^{\rm syn}=-0.67$ and $\alpha_2^{\rm syn}=-1.5$ (vertical dashed lines). This naturally leads us to identify the peak energy $E_{\rm peak}$ with the characteristic synchrotron frequency corresponding to the minimum frequency $\nu_{\rm m}$ of the nonthermal accelerated electrons, and the break energy $E_{\rm break}$ with the cooling break frequency $\nu_{\rm c}$.
However, we note that the distributions are wide and there are 14 cases with $\rm\alpha_1>-0.67$ (at more than $\rm 1 \sigma$), which cannot be interpreted as nonthermal emission spectra.

\begin{figure*}
\epsscale{1.05}
\plottwo{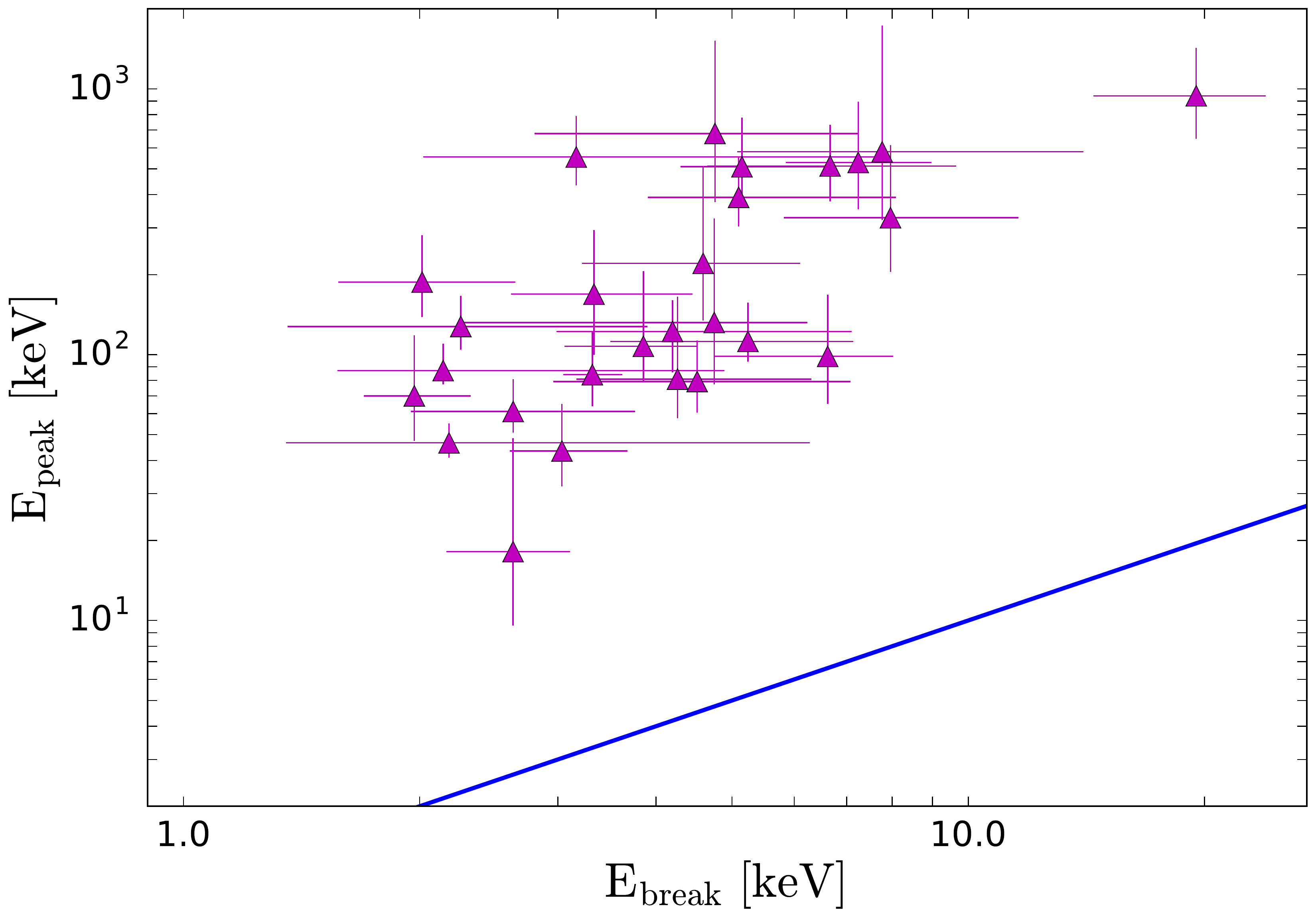}{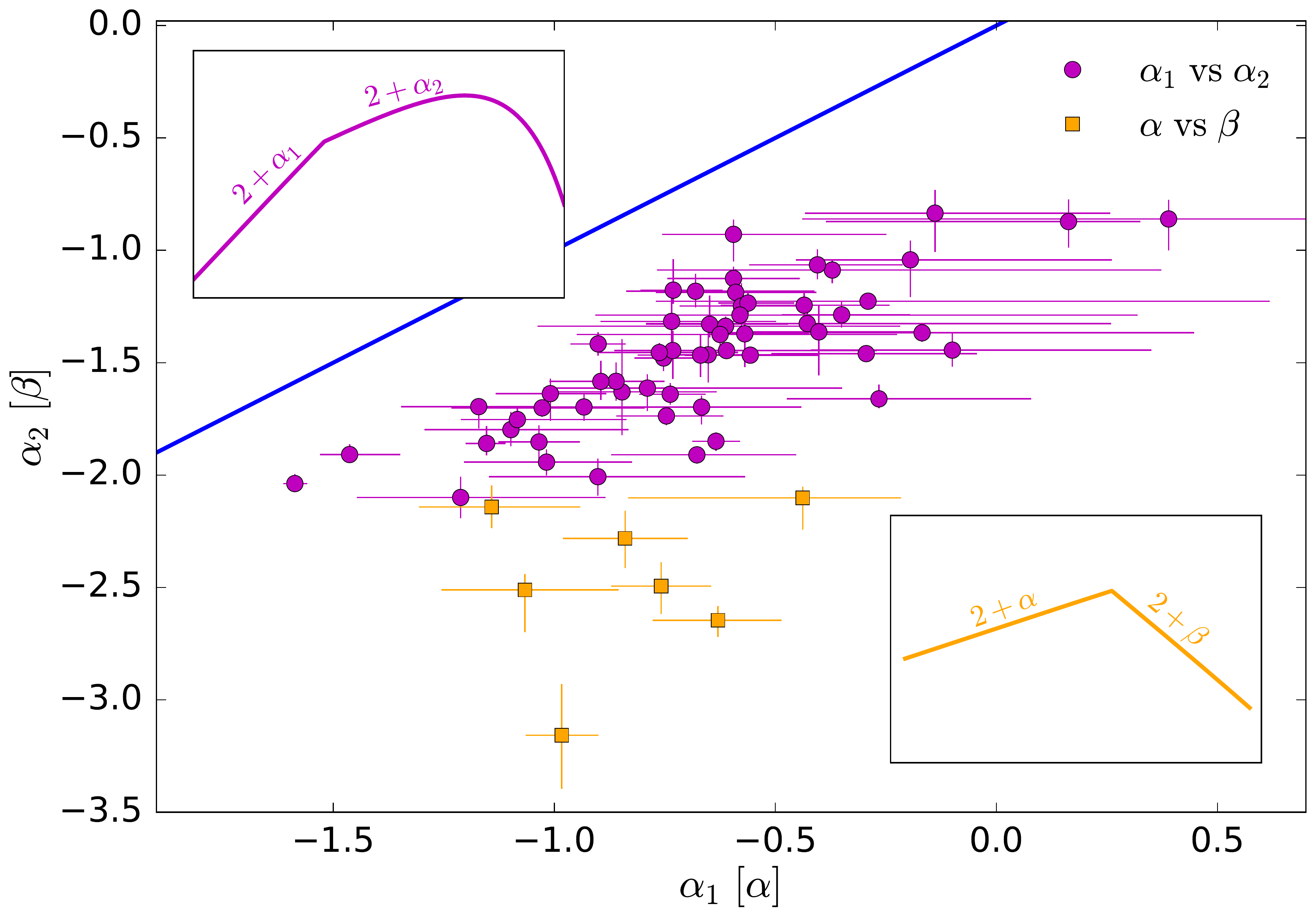}
\caption{\label{fig:scatter_plots} Correlations among best-fit parameters for time-resolved spectra. Left panel: peak energy $E_{\rm peak}$ vs.  break energy $E_{\rm break}$ for spectra in which both spectral features are constrained (i.e., spectra fitted by a BCPL). 
Right panel: $\alpha_1$ vs. $\alpha_2$ (circles)  for those spectra best modeled by either a BPL or a BCPL. Squares symbols show instead $\alpha$ vs. $\beta$ for cases where the best fit is a BPL with high-energy index $<-2$. In both panels, the solid blue line shows the identity line.}
\end{figure*}

We note that the distributions of data points in both panels of Fig.~\ref{fig:scatter_plots} lie far from the equality line. The gap between the points and the line could be, in principle, filled with points, but if $E_{\rm peak}$ and $E_{\rm break}$ are very close to each other, it is hard to distinguish them and find $\alpha_1$ and $\alpha_2$ from spectral analysis.

%------------------------------------------------------------------
%------------------------------------------------------------------

\floattable
\begin{deluxetable}{ccc}
\tablecaption{Summary of the mean values and $1\sigma$ width of the gaussian fits to the best-fit parameters of interest.   \label{tab:gaussian_fits}}
\startdata \\[-0.1cm]
Parameter &  Mean Value    & $\sigma$ \\[0.1cm]
\hline
$\log E_{\rm peak}$ &     2.07       &     0.56      \\ 
$\log E_{\rm break}$  &   0.63   &   0.20  \\[0.1 cm]
\hline 
$\alpha_1$ &   -0.66  &  0.35 \\
$\alpha_2$&    -1.46    &  0.31 \\
$\alpha_{\rm PL}$ &  -1.47  &  0.20 \\
$\alpha_{\rm CPL}$ &  -1.08 & 0.23\\ 
\enddata
\end{deluxetable}

%------------------------------------------------------------------
%------------------------------------------------------------------
 
Correlations among model parameters are investigated in Figure~\ref{fig:scatter_plots}. For those spectra where both $E_{\rm break}$ and $E_{\rm peak}$ are constrained, the two quantities are plotted one versus the other in the left panel. 
Note that $E_{\rm peak}$ spans over two decades, while $E_{\rm break}$ is confined to a narrower range (one order of magnitude).
This narrow range is clearly limited by the instrument energy threshold: values smaller than $\sim1\,$keV cannot be recovered. An upper bound to $E_{\rm break}$ in principle is not present. The lack of break energies in excess of 20\,keV might then suggest that these values are intrinsically not present, which would also explain why these breaks have not been identified so far, with instruments sensitive at energies from 8\,keV up.

In the right panel of Figure~\ref{fig:scatter_plots}, circles show the relation between $\alpha_1$ and $\alpha_2$ for spectra modeled either a BCPL or a BPL with both indices larger than -2. Cases where the best-fit model is a BPL with a high-energy index smaller than -2 are shown with squares, and refer to $\alpha$ versus $\beta$.

 %-------------------------------------------------------
\begin{figure*}
\epsscale{1.08}
\plottwo{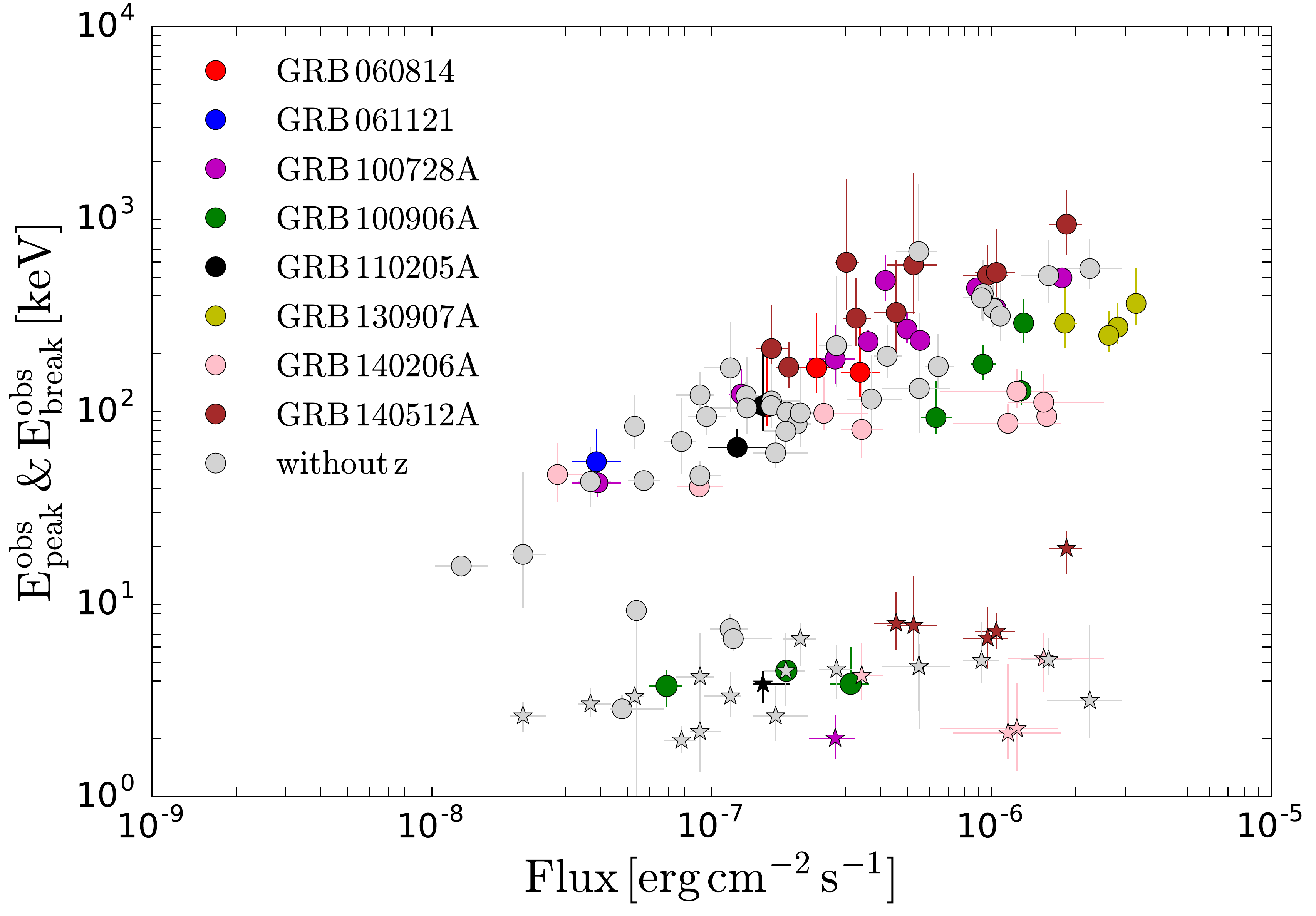}{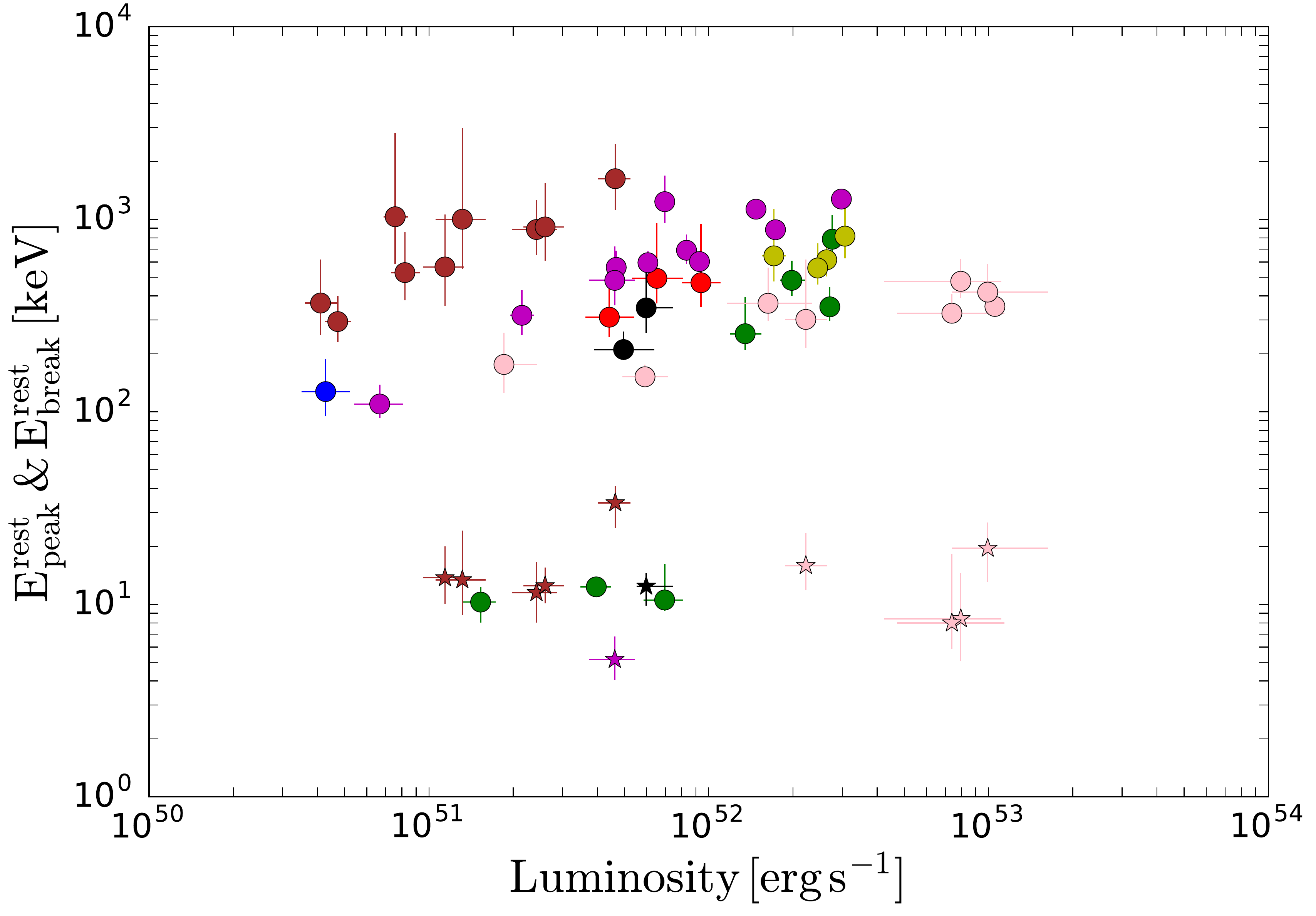}
\caption{\label{fig:flux} Left panel: time-resolved peak (circles) and break (stars) energies are plotted as a function of the flux (estimated in the energy range 0.5\,keV-10\,MeV). 
Right panel: for those GRBs with measured redshift, the rest-frame peak and break energies are plotted as a function of the luminosity.
In both panels, different colours are used for different GRBs.}
\end{figure*}
%-------------------------------------------------------

For each time-resolved spectrum, we also estimate the unabsorbed flux, in the energy range 0.5\,keV - 10\,MeV.
When the GBM data and/or XRT data are not available, this requires an extrapolation of the best-fit model up to 10\,MeV and/or down to 0.5\,keV. 
If the best-fit model is a peaked (in $\nu F_\nu$) function, we perform the extrapolation. The value obtained (and its error) is reported in Table~\ref{tab:table}.
When the peak energy is not constrained, we estimate a lower limit and an upper limit to the 0.5\,keV - 10\,MeV flux, and report their values in Table~\ref{tab:table} within square brackets.
The lower limit is estimated by integrating the best-fit model only in the energy range where data are actually available (i.e., no extrapolation is performed). 
The upper limit is instead estimated by extrapolating the best-fit model up to 10\,MeV and/or down to 0.5\,keV.
Figure~\ref{fig:flux} (left panel) shows the peak and break energies as a function of the flux.
It has been shown in several studies \citep{Ghirlanda_10,Ghirlanda_11a,Ghirlanda_11b} that in single GRBs a correlation between the time-resolved $E_{\rm peak}$ and the instantaneous flux is present. We mark different GRBs with different colors and verify that such a correlation is present also in our sample (circles). The investigation of the existence of a similar correlation also for $E_{\rm break}$ (stars) is more difficult, given the small range spanned by $E_{\rm break}$ and the smaller number of points. 
For the subsample of GRBs with measured redshift, we estimate the rest-frame characteristic energies ($E_{\rm peak}^{\rm rest}$ and $E_{\rm break}^{\rm rest}$) and plot them as a function of the luminosity (Figure~\ref{fig:flux}, right panel).
A standard flat $\Lambda\rm CDM$ cosmological model with $\Omega_\Lambda=0.69$ and $H_0=68\,$km\,s$^{-1}$Mpc$^{-1}$ has been adopted for the estimate of the luminosity distance.
%%%%%%%%%%%%%%%%%%%%%%%   DISCUSSION  %%%%%%%%%%%%%%%%%%%%%%
\section{Discussion}\label{sec:discussion}
\subsection{Reliability of the analysis}
The necessity to introduce low-energy breaks in the spectral models is motivated by a hardening of the spectra in the XRT energy range.
One might wonder if such a hardening can be the result of an incorrect estimate of the $N_{\rm H}$ and/or an insufficient correction for pileup.
The fact that the value of $E_{\rm break}$ ranges from 2 to 20 keV and varies with time during a single GRB might suggest that these breaks are intrinsic features. 
In any case, to test how robust are our results against possible absorption and pileup effects, we performed a series of tests, which confirmed the solidity of our claim on the spectral hardening at low energy. 
We briefly summarize here the tests performed and the results obtained; see Appendix~\ref{sec:tests} for more details.

A hardening in the observed soft X-ray spectrum can be caused by pileup effects when two or more low-energy photons are detected as one single photon of higher energy.
To avoid this effect, we excluded a region at the center of the PSF, large enough to lower the maximum count rate down to 150\,counts s$^{-1}$, so that effects of pileup are negligible \citep{Romano_06}.
A simple test consists in further lowering the maximum count rate and verifying that the results of spectral analysis do not change.
We applied this test to one bright event (GRB~140206A) and two fainter events (GRB~110102A and GRB~140512A) and found that the results are unchanged: by progressively decreasing the maximum count rate down to 70 counts s$^{-1}$, the presence of a spectral break in the XRT band is always significant. Moreover, we find that, within the errors, its location is unaffected. 
In Appendix~\ref{sec:tests}, we show as an example the results of this test applied to one time-resolved spectrum taken from GRB~140512A (Table~\ref{tab:pileup}).

A spurious hardening in the soft X-ray band can also be caused by an incorrect estimate of the amount of absorption by neutral hydrogen. 
Absorption is energy dependent and can then produce a curvature in the observed spectrum below a few keV, depending also on the redshift of the source.
There is then a degeneracy between the amount of absorption and the intrinsic spectral curvature.
If absorption is underestimated, a curvature in the spectral model must be introduced in order to model the data.
Conversely, to fail in recognizing the presence of a spectral break/curvature and/or spectral evolution in the intrinsic spectrum leads to overestimating the value of $N_{\rm H}$ \citep{Butler_07}.
In our analysis, we have estimated the column density from late-time X-ray spectra, selecting a region where the light curve decays in time as a power law and the photon index is roughly constant. 
The derived value has then been used as a fixed input parameter for the joint XRT+BAT spectral analysis. 

In order to test whether an underestimate of the $N_{\rm H}$ is at the origin of the spectral breaks we found, we propose two different tests, which we have applied to GRB~140206A, GRB~110102A, and GRB~140512A. As an example, we show in Appendix~\ref{sec:tests} the results of these tests applied to the time-averaged spectrum of GRB~140512A integrated from 102.86 to 158.16\,s (corresponding to the second pulse).
In the first test, we considered the intrinsic absorption as a free parameter, rather than fixing its value to the one found from spectral analysis of late-time data. 
We modeled the spectrum with both CPL and BCPL models and we found that adding a low-energy break significantly improves the fit at more than $\rm 8\sigma$ (see Figure~\ref{fig:Nh_test1}). 
In the second test, we compared the CPL and BCPL fits obtained after excluding the X-ray data below $\rm 3\,$keV. Also in this case, an $F-test$ reveals that the presence of the break is statistically significant ($\rm 6 \sigma$; see Figure~\ref{fig:Nh_test2}).

\subsection{Comparison with previous studies}
XRT+BAT joint spectral analysis of simultaneous observations has been already performed in several studies.
However, to our knowledge, this is the first time that break energies in the XRT energy range are \sout{found and are} identified as a common feature.
In this section we address the question why this X-ray hardening - which, according to our investigation, appears to be a common feature - has never been reported before.

For GRBs in our sample for which the XRT+BAT joint spectral analysis has been already performed and published in previous papers, we report in Appendix~\ref{app:comparison} a detailed comparison between previous model fits and the fits proposed in this work. 
Here, we summarize the results of such a comparison.
We found that in some cases breaks have been indeed identified in studies focusing on single GRBs \citep{Page_07,starling08,zheng12}.
Systematic studies of larger samples of GRBs whose flaring activity was detected simultaneously by BAT and XRT have been performed by \citep{Friis_13} and \citep{Peng_14}. These analyses proposed, for most of the spectra, a two-component model including a blackbody (BB) and a nonthermal component (see also a similar model proposed by \citealt{guiriec16b} to explain the spectrum of GRB~110205A).
The reason why two completely different models can both account for the same data can be understood from Figure~\ref{fig:peng}, showing the same spectrum fitted with a BPL (left panel) and with a BB+PL (right panel).
In general, the role of the BB is to contribute to the flux at intermediates energies, while the PL segment of the nonthermal component dominates at low and high energies. The total spectrum then mimics the shape of a BPL.

To better compare the two different interpretations (one-component models invoking breaks [B(C)PL] and two-component models including a BB and an unbroken nonthermal component [BB+(C)PL]), in Appendix~\ref{app:comparison} we consider all time-resolved spectra for which we found the presence of a spectral break and refit them with a BB+(C)PL. 
 
%In the sample studied in this work, an interpretation in terms of BB + non-thermal component requires a BB that is never too strong to produce a visible bump in the spectrum, and never too weak to be completely negligible. 
We find that in these fits the role of the BB is to contribute to the flux at low energies, modifying the low-energy PL behavior of the nonthermal component producing an overall 
change in the spectral slope. The empirical fitting function proposed in this work suggests an alternative description of the data, where the overall spectrum can be modeled with one single component (nonthermal with a low-energy break), with a spectral shape resembling the one predicted by the synchrotron model.
A simple comparison of the reduced chi-square values (Figure~\ref{fig:BB}) shows that both models return acceptable fits, with a tendency of single-component models proposed in this work to give a smaller chi-square.
We stress that a completely different case is represented by GRBs where a BB component has been clearly identified \citep{ghirlanda03,ghirlanda13,ryde10,guiriec16a}, and dominates the emission (typically in the initial phase of the prompt). 
The presence of a thermal component in a small fraction of GRBs is not called into question by our findings. Conversely, however, we suggest that the addition of a blackbody  component when not explicitly required can hide important features, such as spectral breaks, which might shed light on the nature of the dominant emission mechanism in GRB prompt radiation.

%Moreover, an interpretation in terms of BB+(C)PL requires a BB component that is never too strong to produce a visible bump in the spectrum, and never too weak to be completely negligible. 

In general, we conclude that similar studies on the same GRBs have failed in recognizing that GRB spectra at low energy are characterized by a change in slope consistent with the synchrotron fast-cooling model for several reasons. First, a peak and break feature have rarely been introduced in the fitting model at the same time. 
Moreover, even when a BPL or a Band model with $\beta>-2$ has been identified as a best fit model, the feature has been often referred to as peak energy \citep{Peng_14}. 
In other cases, the change in slope at a few keV has been interpreted as being caused by the contribution of an additional, thermal component with a temperature at $\sim1$\,keV.
 More importantly, even 
%Again, also 
in analyses recognizing the break feature, the study has been performed on one single GRB \citep{Page_07,starling08,zheng12}.

Finally, several studies have focused on joint XRT+BAT spectra with the aim of investigating X-ray flare spectral properties, focusing on the question of the evolution of the peak energy down to the XRT energy range \citep{Butler_07,Margutti_10}. Our requirement to have bright signal in BAT probably excluded these cases and selected cases where the spectral peak is still in the BAT energy range and where XRT is observing a large part of the prompt emission, rather than the late-time flaring activity.

\subsection{Interpretation}
The hard photon index ($N_{\nu}\propto \nu^{-1}$) describing prompt emission spectra at low energies represents a serious challenge for an interpretation in terms of synchrotron radiation.
In the standard synchrotron fast-cooling model, the spectrum below the $\nu F_\nu$ peak is expected to have a softer index (-3/2), which hardens only at even lower energies, reaching the limiting value -2/3 below the cooling break frequency \citep{preece98,Ghisellini_00}.
A marginally fast cooling regime (i.e. a situation where $\nu_{\rm c}\lesssim\nu_{\rm m}$ rather than $\nu_{\rm c}<<\nu_{\rm m}$) has been considered as a possible solution to the inconsistency between the expected and measured photon index \citep{derishev07,daigne11,beniamini13,uhm14}.
In the context of prompt emission, a theoretical prediction of the location of the cooling break frequency and of the ratio $\nu_{\rm c}/\nu_{\rm m}$ is difficult to make, given the large uncertainties on the properties of the emitting region, such as dissipation radius, bulk Lorentz factor, magnetic field, and particle acceleration mechanism and efficiency. 
\cite{daigne11} showed that if $0.01<\nu_{\rm c}/\nu_{\rm m}<1$, the spectrum displays a continuous curvature toward the value -2/3. In this case, between $\nu_{\rm c}$ and $\nu_{\rm m}$ a PL behavior with index -3/2 provides a satisfactory description of the spectrum only in a very narrow range of energies. 
Only well below $\nu_{\rm c}$ will the spectrum be satisfactorily approximated by a PL, with index -2/3.
In this case, the spectral index inferred from spectral analysis does not necessarily need to be equal to -1.5:
its value will depend on the relative location of $\nu_{\rm c}, \nu_{\rm m}$, the low-energy threshold of the detector, and also the empirical fitting function adopted to model the data.

Even though such a situation can in principle explain why we do not typically observe the value -3/2 and why the inferred slope is higher than this expected value, the question now is what are the physical conditions required to attain a regime of marginally fast cooling and whether such a conditions are realistic.
\cite{daigne11} addressed this question and found that a regime of marginally fast cooling can be obtained for small radii, and/or large Lorentz factors, and/or small magnetic fields.
A similar study on physical conditions leading to $\nu_{\rm c}\lesssim\nu_{\rm m}$ is discussed also in \cite{beniamini13}, and in \cite{beniamini14} in the context of magnetically dominated jets.
These studies have assumed a homogeneous magnetic field and an instantaneous, one-shot acceleration. Other scenarios leading to a similar spectral shape invoke a magnetic field that decays downstream with a strength that depends on the distance from the shock front \citep{derishev07,uhm14}, or continuous electron acceleration \citep{kumar08,asano09}.

%%%%%%%%%%%%%%%%%%%%%%%   CONCLUSIONS  %%%%%%%%%%%%%%%%%%%%%%
\section{Conclusions}\label{sec:conclusions}
To more properly characterize the shape of the prompt spectra at low energy, where observations are in tension with the theory, 
it would be very beneficial to dispose of observations extending well below the low-energy threshold of instruments dedicated to prompt emission studies (typically $\sim10-20\,$keV).
This can be done in several fortunate cases thanks to XRT observations of prompt emission.
With the aim of improving our knowledge on the shape of the low-energy part of the prompt spectrum, we looked for cases where the XRT started observations during the prompt emission. 
For these GRBs, simultaneous XRT and BAT spectral data allowed us to study the prompt emission (or part of it; see Figure~\ref{fig:lc}) down to 0.5\,keV.
We selected events where the emission in the BAT is bright enough to allow reliable time-resolved spectral analysis in at least four temporal bins. 
Fourteen long GRBs satisfy the selection criteria. 
In 12 cases, we found robust evidence for a change in the spectral slope around a few keV.
{\it Fermi}-GBM observations, available for seven GRBs, have been included in the spectral analysis. 
The list of GRBs and their redshift (available for eight events) is reported in Table~\ref{tab:nh}. 
Their BAT and XRT (and, if available, also GBM) lightcurves can be found in Figure~\ref{fig:lc}. In 10 cases, the XRT is observing the main emission episode, while in the remaining four GRBs, the XRT is observing secondary peaks.

For all 14 GRBs in our sample, we have performed time-integrated (26 time bins) and time-resolved (128 time bins) spectral analysis, covering the entire prompt emission. 
For time bins where XRT observations are not available, we found standard results: the spectra are well modeled by a single PL or a CPL.
The peak energy and spectral index distributions (Figure~\ref{fig:histo}) are consistent with those derived from spectral analysis of larger samples of BAT and GBM GRBs. 
In particular, when the peak energy is constrained, the low-energy index $\alpha$ has a distribution peaked around $-1$ (see $\alpha_{\rm CPL}$ in the right panel of Figure~\ref{fig:histo} and in Table~\ref{tab:gaussian_fits}). 
The value of the spectral index is instead softer when the best-fit model is a single PL: $\langle\alpha_{\rm PL}\rangle\simeq -1.5$. 
Both results perfectly agree with spectral index distributions derived in spectral catalogs of BAT \citep{lien16} and GBM \citep{gruber14} long GRBs.

The situation is different for temporal bins where XRT observations are available. 
The spectra in the whole energy range ($0.5-150\,$keV or $0.5-1000\,$keV) can still be fitted by one single spectral component, but in the 67\% of the cases a low-energy break must be added to the empirical fitting function, resulting in a significant (more than 3$\sigma$) improvement of the fit (see an example in Figure~\ref{fig:TI_spectra_140512A}, right panels). 
This led us to introduce two additional spectral models: a cutoff PL with a low-energy break (BCPL), describing cases where both the low-energy break $E_{\rm break}$ and the peak energy $E_{\rm peak}$ are constrained (31\% of time-resolved spectra with XRT data), and a BPL with both indices $\alpha_1, \alpha_2>-2$ (36\%), describing cases where $E_{\rm break}$ is constrained, while $E_{\rm peak}$ falls near or above the high-energy threshold, and cannot be determined.
A summary of the models and notation chosen for the model parameters can be found in Figure~\ref{fig:sketch}.

This systematic difference between best-fit models in spectra with and without XRT observations suggests that our  knowledge of the prompt emission spectral shape is usually limited (and possibly biased) by the lack of low-energy observations.
The results of spectral analysis down to 0.5\,keV revealed that the typical GRB spectrum has two characteristic energies ($E_{\rm break}$ and $E_{\rm peak}$, with $E_{\rm break}< E_{\rm peak}$) and three power-law segments ($\alpha_1, \alpha_2>-2$, and $\beta<-2$). 
We speculate that this result might be quite general: the sample investigated in this work has been selected based on the main requirement of simultaneous XRT and BAT observations of the prompt emission (and relatively bright BAT emission). The selected GRBs have fluences and energies in the range $7\times10^{-6}- 8\times10^{-4}$erg~cm$^{-2}$ and $6\times10^{52}-3\times10^{54}$erg, respectively, indicating that these are not necessarily the brightest events. Their light curves differ in morphology one from the other, and the redshift spans the range $z=0.725$ to $z=2.73$.
From the point of view of temporal properties, energetics, and redshift, these GRBs do not seem to belong to a subclass of peculiar events. 

In the sample studied in this work, the break energy $E_{\rm break}$ has a distribution peaked around 4\,keV in the observer frame (10\,keV in the rest frame), and the peak energy $E_{\rm peak}$ has a distribution peaked around 120\,keV in the observer frame (300\,keV in the rest frame).
The typical ratio $E_{\rm break}/E_{\rm peak}$ is around $0.03$.
It is very likely that the observed distribution of $E_{\rm break}$ is significantly biased by the fact that values smaller than $\sim2\,$keV cannot be constrained. 
It is very tempting to associate these characteristic energies with the synchrotron cooling and typical frequencies $\nu_{\rm c}$ and $\nu_{\rm m}$.
This is supported by the average values of the photon indices. In a synchrotron context, the expected values are $\alpha_1=-2/3$ below $\nu_{\rm c}$ and $\alpha_2=-3/2$ between $\nu_{\rm c}$ and $\nu_{\rm m}$.
From spectral analysis we found $\langle\alpha_1\rangle=-0.66$ ($  \rm \sigma=0.35$) and $\langle\alpha_2\rangle=-1.46$ ($\rm \sigma=0.31$).

In the synchrotron-prompt emission scenario, the physical parameters of the emitting region have not been constrained yet.
Observations of typical prompt fluxes, peak energies, and timescales are not enough to constrain all the unknown parameters governing the physics of acceleration, dissipation, and emission.
Studies that use observations to constrain the theory can only identify an allowed parameter space \citep{kumar08,daigne11,beniamini13,beniamini14}.
These studies can now take advantage of an additional, important constraint: the location of the cooling break frequency. 
Further constraints on the properties of the emission region (dissipation radius, strength of the magnetic field, Lorentz factor, particle acceleration) can be derived. Implications for physical models coming from the location of the cooling break at $\sim10\,$keV will be treated in an upcoming work in preparation. 

Even though the spectra are qualitatively consistent with synchrotron radiation, additional studies are required to firmly assess the consistency of data with theoretical expectations from the synchrotron process.
Recent studies have pointed out the importance of reproducing also the narrowness of the spectral shape \citep{beloborodov13,axelsson15,yu15,vurm16}, arguing that most prompt emission spectra are too narrow to be reproduced by synchrotron radiation, even in the limiting case of a Maxwellian electron distribution.
Moreover, it is unclear how spectra with a low-energy photon index higher than -0.67 (present both in this sample and in the BATSE and GBM GRB catalogs) can be reconciled with the synchrotron scenario.
While the results found in this work clearly show that a spectral break is present in the keV range, the interpretation of the spectral shape in terms of synchrotron radiation (although encouraged by the average values of the photon indices) demands a more quantitative investigation.
 
\acknowledgements
L.N. and G.O. thank  INAF-Osservatorio Astronomico di Brera for kind hospitality during the completion of this work.
G.O. is grateful to Alessio Pescalli and Sergio Campana for their help with the analysis of GBM and XRT data, respectively.
This work made use of public {\it Fermi}-GBM data and data supplied by the UK \textit{Swift} Science Data Centre at the University of Leicester.
This research has made use of data and software provided by the High Energy Astrophysics Science Archive Research Center (HEASARC), which is a service of the Astrophysics Science Division at NASA/GSFC and the High Energy Astrophysics Division of the Smithsonian Astrophysical Observatory.

%%%%%%%%%%%%%%%%%%%%%%%%%%%%%%%%%%%%%%%%%%%
\newpage

\appendix

\section{Tables}

%\floattable
\begin{deluxetable}{cccc}[ht!]
\tablecaption{\label{tab:nh} List of GRBs analyzed in this work. The name and redshift are reported in the first and second column. 
The third column lists the values of the $N_{\rm H}$, derived from spectral analysis of late time XRT observations. The late time interval (LTI, from BAT trigger time) chosen for the derivation of $N_{\rm H}$ can be found in the last column.}
\startdata \\[-0.3cm]
GRB & redshift     &  $N_{\rm H}$ &  LTI  \\
&  & ${\rm 10^{22} \ cm^{-2}}$& ${\rm 10^{4} \ s }$ \\
\hline\\[-0.3cm]
060814   & $1.92$    & $3.05$ &  $16.83-137.78$\\
061121   & $1.314$   & $0.72$ &  $3.46-9.25$\\ 
070616   & ...           &  $0.49$ & $0.46-37.11$\\
100619A & ...           & $0.76$ & $5.34-100.59$\\
100725B & ...          &  $0.59$ & $2.18-80.35$\\
100728A & $1.567$  &  $3.25$ & $0.50-68.29$\\
100906A & $1.727$  & $1.32$ &  $1.06-46.86$\\
110102A & ...            & $0.20$ & $1.04-24.32$ \\
110205A & $2.22$   &  $0.59$ & $0.14-38.29$\\
121123A & ...           &   $0.12$ & $1.66-13.91$\\
130907A & $1.238$  & $1.15$  &  $0.76-238.41$\\
140108A & ...           &   $0.71$ & $1.05-43.16$ \\
140206A & $2.73$   & $1.40$ &  $2.12-8.71$\\
140512A & $0.725$  & $0.44$ &  $2.79-32.94$\\
\enddata
\end{deluxetable}

%%%%%%%%%%%%%%%%%%%%%%%%%%%%%%%%%%%%%%%%%%%%%%
\LTcapwidth=0.94\textwidth
%\begin{center}
%\onecolumn
%\longtab{
\begin{longtable}{cccccccccc}
%\centering
%\begin{threeparttable}
\caption{ \label{tab:table} Best-fit parameters for time-integrated and time-resolved spectra. The table lists the time interval (since the BAT trigger time), the best fit model (PL=power-law, CPL=cutoff power-law, BPL= broken power-law, BCPL=broken power-law with a high energy cutoff), the best fit parameters (columns 3 to 7, for a definition see Figure~\ref{fig:sketch}),  the flux $F$ (or its lower and upper limits, in square brackets), integrated in the energy range 0.5\,keV - 10\,MeV,  the total chi-square $\rm \chi^{2}$, and the degrees of freedom (d.o.f.). The last column reports the instruments included in the spectral analysis: X=XRT, B=BAT, G=GBM.  Time bins marked with a bold font identify time-integrated spectra.}\\
\hline\hline
 Time bin & Model & ${\rm \alpha_{1}}$ & ${E_{\rm break}}$ & ${\rm  \alpha / \alpha_{2} }$ &   ${E_{\rm peak}}$  & ${\rm \beta}$ &  ${\rm F_{-7}}$ & $\rm \chi^{2}$ (dof) & Instr.\\
          ${\rm s}$                   &       &                  &    ${\rm keV}$               &                  &   ${\rm keV}$   &  & ${\rm [erg \ cm^{-2} s^{-1} ]}$  & &   \\ [0.05 cm]\hline 
\endfirsthead 
\caption{continued.}\\
\hline\hline
 Time bin & Model & ${\rm \alpha_{1}}$ & ${E_{\rm break}}$ & ${\rm  \alpha / \alpha_{2} }$ & ${E_{\rm peak}}$  & ${\rm \beta}$ & ${\rm F_{-7}}$ & $\rm \chi^{2}$ (dof) & Instr.\\
          ${\rm s}$                   &       &                  &    ${\rm keV}$               &                  &   ${\rm keV}$   & &  ${\rm [erg \ cm^{-2} s^{-1} ]}$ &  & \\ [0.05 cm]\hline 
\endhead

%%%%%%
\hline
\endfoot    

%%%%%% 
\\
\multicolumn{10}{c}{${ \bf GRB~060814,\, z=1.92 }$ } \\   [0.1 cm]

 $\bf{[-14.00 , 77.50]}$ & CPL &  &  & $-1.26_{-0.13}^{+0.13}$ &   $157_{-32}^{+80}$ &   & $2.49_{-0.28}^{+0.33}$ & $27.1 (55)$ & B\\
 
 $\bf{[77.50 , 200.00]}$ & BPL & $-0.98_{-0.09}^{+0.13}$ & $2.83_{-0.45}^{+0.39}$  & $-1.71_{-0.06}^{+0.04}$  &  &  & $[0.39-6.42]$ & $172.1 (186)$ & X,B\\

 $[-14.00 , 11.00]$ & CPL&  &  & $-0.99_{-0.21}^{+0.21}$ & $169_{-44}^{+159}$ &   &$2.37_{-0.44}^{+0.56}$ & $46.8 (55)$ & B \\

 $[11.00 , 15.00]$ & PL&  &  & $-1.42_{-0.06}^{+0.06}$ &  & &  $[2.54-481.03]$  & $55.3 (56)$ &  B\\

 $[15.00 , 40.00]$ & CPL&  &  & $-1.29_{-0.16}^{+0.17}$ & $160_{-41}^{+162}$ & &  $3.39_{-0.49}^{+0.60}$  & $34.2 (55)$ & B \\ 

 $[40.00 , 77.50]$ & CPL&  &  & $-1.42_{-0.19}^{+0.20}$ & $106_{-22}^{+87}$ &  &$1.58_{-0.28}^{+0.36}$  & $35.6 (55)$ & B \\        

 $[77.50 , 97.00]$ & BPL& $-0.27_{-0.21}^{+0.34}$ & $2.22_{-0.33}^{+0.26}$  & $-1.66_{-0.04}^{+0.06}$  & &  & $[0.88-21.35]$ & $124.6 (131)$ & X,B \\ 

 $[97.00 , 120.00]$ & BPL& $-1.10_{-0.20}^{+0.27}$ & $2.28_{-0.50}^{+0.87}$  & $-1.80_{-0.07}^{+0.08}$  & &  & $[0.33-2.83]$ & $102.0 (106)$ & X,B \\

 $[120.00 , 130.00]$ & BPL& $-1.01_{-0.12}^{+0.13}$ & $4.78_{-1.10}^{+1.03}$  & $-1.64_{-0.12}^{+0.07}$  & &  & $[0.53-15.73]$ & $80.6 (89)$ & X,B \\

 $[130.00 , 200.00]$ & BPL& $-1.17_{-0.18}^{+0.14}$ & $2.71_{-0.33}^{+2.40}$  & $-1.70_{-0.10}^{+0.04}$  & &  & $[0.24-4.37]$ & $174.4 (152)$  & X,B \\
\\
 \multicolumn{10}{c}{${ \bf GRB~061121,\,z=1.314 }$ } \\    [0.1 cm]      
 
 $\bf{[-4.00 , 10.00]}$ & PL &  &  & $-1.66_{-0.10}^{+0.09}$ &  &  &  $[0.48-18.04]$ & $58.0 (56)$ & B \\

 $\bf{[62.00 , 110.00]}$ & BPL & $-0.78_{-0.10}^{+0.10}$ & $4.97_{-0.84}^{+0.96}$  & $-1.43_{-0.03}^{+0.02}$  & & & $[3.18-529.42]$ & $137.8 (150)$ & X,B \\ 

  $[-4.00 , 10.00]$ & PL&  &  & $-1.66_{-0.10}^{+0.09}$ &  & &  $[0.48-18.04]$ & $58.0 (56)$ & B \\

 $[50.00 , 62.00]$ & PL&  &  & $-1.64_{-0.10}^{+0.10}$ & & & $[0.50-21.60]$ & $51.7 (56)$ & B \\

 $[62.00 , 68.00]$ & BPL& $-0.56_{-0.15}^{+0.16}$ & $4.53_{-0.64}^{+1.01}$  & $-1.47_{-0.03}^{+0.03}$  & & & $[4.73-574.02]$ & $87.0 (85)$ & X,B \\

 $[68.00 , 74.00]$ & BPL& $-0.29_{-0.21}^{+0.25}$ & $4.83_{-0.91}^{+2.18}$  & $-1.46_{-0.03}^{+0.03}$  & &  & $[9.98-1297.70]$ & $48.8 (77)$ & X,B \\ 

 $[74.00 , 78.00]$ & BPL& $-0.29_{-0.48}^{+0.91}$ & $2.52_{-0.74}^{+1.71}$  & $-1.23_{-0.03}^{+0.02}$  &  & & $[11.20-10385.00]$ & $49.1 (73)$ & X,B \\  

 $[78.00 , 90.00]$ & BPL& $-0.93_{-0.24}^{+0.23}$ & $2.82_{-0.57}^{+1.60}$  & $-1.70_{-0.06}^{+0.06}$  & &  & $[1.12-20.48]$ & $92.0 (86)$ & X,B \\

 $[90.00 , 110.00]$ & CPL&  &  & $-1.42_{-0.08}^{+0.06}$ & $55_{-14}^{+26}$  & &  $0.39_{-0.07}^{+0.09}$ & $99.4 (104)$ & X,B \\

\\   
 \multicolumn{10}{c}{${ \bf GRB~070616}$ } \\    [0.1 cm]  
 $\bf{[138.00 , 615.00]}$ & BCPL & $-0.84_{-0.04}^{+0.05}$ & $3.22_{-0.35}^{+0.32}$  & $-1.29_{-0.07}^{+0.01}$ & $102_{-14}^{+25}$ & & $0.57_{-0.01}^{+0.06}$ & $228.7 (234)$ & X,B \\
 
  $[-10.00 , 95.00]$ & PL&  &  & $-1.66_{-0.19}^{+0.18}$ & &  & $[0.11-3.98]$ & $50.2 (56)$ & B \\

  $[95.00 , 138.00]$ & PL&  &  & $-1.38_{-0.07}^{+0.07}$ & &  & $[0.43-135.90]$ & $43.8 (56)$ & B \\

 $[138.00 , 175.00]$ & BPL& $-0.90_{-0.06}^{+0.06}$ & $8.10_{-0.96}^{+4.88}$  & $-1.42_{-0.05}^{+0.05}$  & &  & $[0.87-162.03]$ & $164.8 (153)$ & X,B \\

 $[175.00 , 210.00]$ & BPL& $-0.89_{-0.08}^{+0.14}$ & $7.21_{-0.74}^{+2.97}$  & $-1.58_{-0.08}^{+0.09}$  & & & $[0.44-25.37]$ & $128.0 (140)$ & X,B \\

 $[210.00 , 282.00]$ & BCPL & $-0.73_{-0.07}^{+0.11}$ & $4.20_{-1.21}^{+2.90}$  & $-1.18_{-0.06}^{+0.14}$ & $122_{-37}^{+38}$ &  & $0.91_{-0.16}^{+0.11}$ & $188.8 (186)$ & X,B\\ 

 $[282.00 , 330.00]$ & BCPL & $-0.65_{-0.14}^{+0.18}$ & $3.34_{-0.72}^{+1.12}$  & $-1.33_{-0.06}^{+0.13}$ & $169_{-69}^{+125}$ &  & $1.17_{-0.22}^{+0.11}$ & $138.3 (174)$ & X,B \\ 

 $[330.00 , 460.00]$ & BCPL & $-0.75_{-0.07}^{+0.07}$ & $3.32_{-0.27}^{+0.30}$  & $-1.48_{-0.06}^{+0.05}$ & $84_{-20}^{+38}$ &  & $0.53_{-0.03}^{+0.04}$ & $185.4 (217)$ & X,B \\

 $[460.00 , 500.00]$ & BCPL & $-0.73_{-0.13}^{+0.16}$ & $3.04_{-0.43}^{+0.64}$  & $-1.45_{-0.13}^{+0.09}$ & $43_{-11}^{+22}$  & &  $0.37_{-0.03}^{+0.07}$  & $123.0 (126)$ & X,B \\ 

 $[500.00 , 530.00]$ & BCPL & $-0.85_{-0.16}^{+0.21}$ & $2.63_{-0.47}^{+0.48}$  & $-1.63_{-0.19}^{+0.23}$ & $18_{-9}^{+30}$ &  & $0.21_{-0.02}^{+0.04}$ & $128.8  (112)$ & X,B \\ 

 $[530.00 , 615.00]$ & CPL&  &  & $-1.33_{-0.06}^{+0.06}$ & $16_{-2}^{+2}$  &  & $0.13_{-0.02}^{+0.03}$    & $159.0 (163)$ & X,B    \\

\\     
  \multicolumn{10}{c}{${ \bf GRB~100619A}$ } \\    [0.1 cm]
  
 $\bf{[-5.34 , 10.02]}$ & CPL &  &  & $-1.23_{-0.19}^{+0.21}$ & $110_{-21}^{+46}$ & &  $1.62_{-12.34}^{+12.85}$ & $184.8 (204)$ & B,G \\ 
 
$\bf{[80.68 , 100.13]}$ & BPL & $-1.01_{-0.12}^{+0.12}$ & $5.13_{-0.60}^{+1.02}$  & $-1.93_{-0.04}^{+0.04}$  & & & $[3.08-8.79]$ & $218.3 (271)$ & X,B,G \\
         
$[80.68 , 86.82]$ & BPL & $-1.02_{-0.19}^{+0.19}$ & $6.33_{-1.15}^{+2.08}$  & $-1.94_{-0.06}^{+0.06}$  & &   & $[3.26-9.25]$ & $169.3 (163)$ & X,B,G \\
 
$[86.82 , 89.89]$ & BCPL & $-0.79_{-0.23}^{+0.44}$ & $4.75_{-2.50}^{+1.49}$  & $-1.61_{-0.10}^{+0.06}$ & $132_{-55}^{+193}$ &  & $5.51_{-1.45}^{+1.58}$ & $161.1 (159)$ & X,B,G \\
 
$[89.89 , 92.97]$ & BPL& $-0.90_{-0.25}^{+0.33}$ & $5.41_{-1.54}^{+1.80}$  & $-2.01_{-0.08}^{+0.08}$  & &  & $[2.81-6.58]$ & $123.4 (130)$ & X,B,G \\
 
$[92.97 , 100.13]$ & BPL& $-1.21_{-0.24}^{+0.33}$ & $3.92_{-1.18}^{+1.46}$  & $-2.10_{-0.09}^{+0.10}$  & &  & $[1.35-2.13]$ & $157.5 (182)$ & X,B,G  \\  
 
 \\      
\multicolumn{10}{c}{${ \bf GRB~100725B}$ } \\    [0.1 cm]

$\bf{[-3.70 , 15.76]}$ & PL &  &  & $-1.56_{-0.07}^{+0.07}$ &  &  & $[2.12-57.27]$ & $152.4 (144)$ & B,G \\

$\bf{[89.49 , 229.78]}$ & BPL & $-1.25_{-0.06}^{+0.09}$ & $5.19_{-1.23}^{+0.89}$  & $-2.06_{-0.05}^{+0.06}$  & & & $1.47_{-0.08}^{+0.08}$ & $348.2 (310)$ & X,B,G \\

$[109.97 , 120.21]$ & CPL &  &  & $-1.15_{-0.05}^{+0.06}$ & $94_{-19}^{+33}$ & &  $0.96_{-0.14}^{+0.16}$ & $255.0 (226)$ & X,B,G \\ 

$[120.21 , 129.43]$ & BCPL & $-0.65_{-0.11}^{+0.15}$ & $6.62_{-1.88}^{+1.40}$  & $-1.46_{-0.12}^{+0.15}$ & $99_{-33}^{+70}$ & &  $2.07_{-0.27}^{+0.29}$ & $295.9 (227)$ & X,B,G  \\ 

$[129.43 , 136.59]$ & BCPL & $-0.73_{-0.16}^{+0.24}$ & $4.51_{-1.55}^{+2.57}$  & $-1.32_{-0.12}^{+0.10}$ & $79_{-19}^{+34}$ &  & $1.84_{-0.30}^{+0.31}$ & $222.0 (193)$  & X,B,G \\

$[136.59 , 143.76]$ & BPL& & & $-0.84_{-0.14}^{+0.14}$ & $7.47_{-1.19}^{+1.45}$  & $-2.28_{-0.13}^{+0.12}$  &   $1.16_{-0.18}^{+0.19}$ & $167.8 (185)$ & X,B,G \\      

$[143.76 , 155.03]$ & BPL& & & $-0.76_{-0.11}^{+0.11}$ & $9.29_{-9.31}^{+0.56}$  & $-2.49_{-0.12}^{+0.11}$  &   $1.19_{-0.12}^{+0.45}$ & $265.6 (235)$ & X,B,G \\ 

$[155.03 , 170.39]$ & BPL& & & $-0.98_{-0.08}^{+0.08}$ & $6.63_{-0.94}^{+0.78}$  & $-3.16_{-0.24}^{+0.23}$  &   $0.54_{-0.05}^{+0.05}$ & $128.7 (127)$ & X,B,G \\   

$[205.20 , 229.78]$ & BPL& & & $-1.07_{-0.19}^{+0.21}$ & $2.86_{-0.33}^{+0.52}$  & $-2.51_{-0.19}^{+0.07}$  &  $0.48_{-0.04}^{+0.20}$ & $160.0 (148)$ & X,B,G \\ 

\\
 \multicolumn{10}{c}{${ \bf GRB~100728A,\,z=1.567 }$ } \\    [0.1 cm]      
$\bf{[-82.31 , 81.53]}$  & CPL &  &  & $-0.69_{-0.03}^{+0.03}$ & $342_{-19}^{+21}$ & & $5.85_{-0.11}^{+0.11}$ & $370.1 (363)$ & B,G \\ 
 
$\bf{[81.53 , 158.33]}$ & BCPL & $-0.97_{-0.12}^{+0.19}$ & $2.24_{-0.51}^{+0.54}$  & $-1.34_{-0.02}^{+0.02}$ & $186_{-25}^{+33}$ & & $2.00_{-0.14}^{+0.13}$ & $360.3 (350)$ & X,B,G \\
 
$[-82.31 , -48.52]$ & PL&  &  & $-1.17_{-0.07}^{+0.07}$ & & & $[11.09-502.29]$ & $348.4 (324)$ & B,G  \\
 
$[-48.52 , -13.70]$ & CPL&  &  & $-0.98_{-0.07}^{+0.07}$ & $481_{-107}^{+176}$ & &  $4.17_{-0.17}^{+0.20}$ & $370.3 (321)$ & B,G \\
 
$[-13.70 , 14.97]$ & CPL&  &  & $-0.74_{-0.04}^{+0.04}$ & $439_{-42}^{+50}$ & &  $8.84_{-0.22}^{+0.24}$ & $326.8 (323)$ & B,G  \\ 
 
$[14.97 , 28.29]$ & CPL&  &  & $-0.60_{-0.04}^{+0.04}$ & $496_{-38}^{+44}$ & &  $17.85_{-0.40}^{+0.41}$ & $330.1 (290)$ & B,G \\ 
 
$[28.29 , 52.86]$ & CPL&  &  & $-0.74_{-0.04}^{+0.04}$ & $344_{-23}^{+26}$ & &  $10.38_{-0.23}^{+0.25}$ & $348.9 (317)$ & B,G \\
 
$[52.86 , 65.15]$ & CPL&  &  & $-0.90_{-0.08}^{+0.08}$ & $269_{-40}^{+56}$  & &  $4.99_{-0.27}^{+0.32}$ & $284.3 (270)$ & B,G \\ 
 
$[65.15 , 81.53]$ & CPL&  &  & $-0.76_{-0.07}^{+0.07}$ & $235_{-23}^{+28}$ & &  $5.55_{-0.26}^{+0.29}$ & $325.5 (287)$  & B,G \\
 
$[81.53 , 92.79]$ & BCPL & $-0.43_{-0.36}^{+0.69}$ & $2.01_{-0.44}^{+0.63}$  & $-1.33_{-0.05}^{+0.05}$ & $188_{-49}^{+94}$ & & $2.76_{-0.53}^{+0.50}$ & $167.2 (166)$ & X,B,G \\

$[92.79 , 106.11]$ & CPL &  &  & $-1.35_{-0.04}^{+0.04}$ & $123_{-26}^{+44}$ & &  $1.27_{-0.12}^{+0.14}$ & $188.7 (164)$ & X,B,G \\  

$[106.11 , 118.39]$ & CPL &  &  & $-1.20_{-0.03}^{+0.03}$ & $219_{-35}^{+49}$ &  & $2.79_{-0.15}^{+0.16}$ & $206.5 (236)$ & X,B,G \\ 

$[118.39 , 135.80]$ & CPL &  &  & $-1.19_{-0.02}^{+0.02}$ & $232_{-27}^{+35}$ &  & $3.62_{-0.14}^{+0.15}$ & $319.3 (280)$ & X,B,G \\

$[135.80 , 158.33]$ & CPL &  &  & $-1.31_{-0.07}^{+0.06}$ & $43_{-7}^{+11}$ &  &  $0.39_{-0.07}^{+0.08}$ & $114.0 (106)$ & X,B,G \\

\\       
\multicolumn{10}{c}{${ \bf GRB~100906A,\,z=1.727 }$ } \\    [0.1 cm]
$\bf{[0.22 , 65.24]}$ & CPL &  &  & $-1.42_{-0.08}^{+0.09}$ & $182_{-36}^{+66}$ & & $3.23_{-0.21}^{+0.26}$ & $231.1 (342)$ & B,G \\
 
$\bf{[85.72 , 125.65]}$ & BPL & &  & $-0.56_{-0.17}^{+0.13}$ & $3.84_{-0.27}^{+0.58}$  & $-2.45_{-0.07}^{+0.04}$  &  $1.58_{-0.15}^{+0.16}$ & $394.8 (378)$ & X,B,G \\
 
$[0.22 , 2.77]$ & CPL&  &  & $-1.04_{-0.09}^{+0.10}$ & $289_{-60}^{+98}$ & &  $13.02_{-0.89}^{+1.08}$ & $159.3 (171)$ & B,G \\ 
 
$[2.77 , 5.84]$ & CPL&  &  & $-1.09_{-0.10}^{+0.11}$ & $177_{-30}^{+47}$ &  & $9.32_{-0.85}^{+1.06}$ & $182.6 (185)$ & B,G  \\ 
 
$[5.84 , 10.96]$ & CPL&  &  & $-1.10_{-0.18}^{+0.18}$ & $126_{-22}^{+52}$ &  & $12.75_{-0.76}^{+0.87}$ & $261.8 (230)$ & B,G \\ 
 
$[10.96 , 16.08]$ & CPL&  &  & $-1.27_{-0.24}^{+0.25}$ & $93_{-16}^{+51}$  &  & $6.33_{-0.73}^{+0.92}$  & $208.1 (214)$  & B,G \\   
 
$[85.71 , 96.98]$ & BPL& & & $-1.14_{-0.16}^{+0.20}$ & $3.76_{-0.82}^{+0.77}$  & $-2.14_{-0.09}^{+0.10}$  &    $0.69_{-0.09}^{+0.09}$  & $218.9 (226)$ & X,B,G \\ 
 
$[96.98 , 105.17]$ & BPL& & &  $-0.44_{-0.40}^{+0.22}$ & $3.86_{-0.48}^{+2.11}$  & $-2.10_{-0.14}^{+0.05}$  & $3.14_{-0.50}^{+0.51}$  & $204.7 (206)$ & X,B,G \\ 
 
$[105.17 , 125.65]$ & BPL & & & $-0.63_{-0.15}^{+0.14}$ & $4.52_{-0.40}^{+0.52}$  & $-2.65_{-0.07}^{+0.06}$  &  $1.85_{-0.23}^{+0.24}$ & $202.2 (193)$ & X,B,G \\
 
\\
\multicolumn{10}{c}{${ \bf GRB~110102A}$ } \\    [0.1 cm]
$\bf{[125.54 , 156.26]}$ & CPL &  &  & $-1.39_{-0.09}^{+0.10}$ & $283_{-94}^{+278}$ & & $3.55_{-0.25}^{+0.34}$ & $217.5 (205)$ & B,G\\

$\bf{[195.17 , 290.40]}$ & BCPL & $-0.85_{-0.05}^{+0.06}$ & $4.02_{-0.54}^{+0.56}$  & $-1.49_{-0.03}^{+0.03}$ & $686_{-274}^{+731}$ & & $3.83_{-0.15}^{+0.15}$ & $352.8 (315)$ & X,B,G \\

$[125.54 , 132.71]$ & CPL&  &  & $-1.24_{-0.22}^{+0.25}$ & $108_{-25}^{+69}$ & &  $1.63_{-0.34}^{+0.53}$ & $217.7 (206)$ & B,G \\ 

$[132.71 , 137.83]$ & CPL&  &  & $-1.17_{-0.06}^{+0.07}$ & $344_{-69}^{+111}$ & & $10.15_{-0.47}^{+0.54}$ & $225.3 (205)$ & B,G \\ 

$[137.83 , 142.95]$ & CPL&  &  & $-1.19_{-0.12}^{+0.14}$ & $194_{-45}^{+89}$ & & $4.23_{-0.42}^{+0.56}$ & $200.2 (189)$ & B,G \\

$[142.95 , 156.26]$ & PL&  &  & $-1.75_{-0.08}^{+0.08}$ & &  & $[2.75-10.21]$ & $274.2 (247)$ & B,G  \\

$[195.17 , 200.29]$ & BCPL & $-0.68_{-0.16}^{+0.27}$ & $4.76_{-1.96}^{+2.48}$  & $-1.18_{-0.07}^{+0.08}$ & $679_{-305}^{+840}$ & &  $5.49_{-0.94}^{+0.92}$  & $165.1 (205)$ & X,B,G \\ 

$[200.29 , 206.44]$ & BCPL & $-0.59_{-0.15}^{+0.15}$ & $5.10_{-1.19}^{+2.99}$  & $-1.13_{-0.07}^{+0.05}$ & $391_{-88}^{+161}$ &  & $9.21_{-1.29}^{+1.38}$  & $260.2 (248)$ & X,B,G  \\ 

$[206.44 , 209.51]$ & BCPL & $-0.37_{-0.40}^{+0.74}$ & $3.17_{-1.14}^{+4.65}$  & $-1.09_{-0.06}^{+0.05}$ & $554_{-121}^{+236}$ &  & $22.44_{-6.63}^{+6.64}$ & $198.3 (185)$ & X,B,G \\ 

$[209.51 , 212.58]$ & BCPL & $-0.43_{-0.19}^{+0.19}$ & $5.15_{-0.85}^{+1.55}$  & $-1.24_{-0.06}^{+0.05}$ & $509_{-142}^{+270}$ &  & $15.99_{-3.22}^{+3.40}$ & $211.1 (183)$ & X,B,G \\ 

$[212.58 , 218.73]$ & BCPL & $-0.67_{-0.14}^{+0.18}$ & $4.59_{-1.37}^{+1.52}$  & $-1.47_{-0.10}^{+0.09}$ & $220_{-86}^{+285}$  & &  $2.79_{-0.37}^{+0.38}$ & $228.2 (224)$ & X,B,G \\ 

$[218.73 , 229.99]$ & BCPL & $-0.10_{-0.32}^{+0.45}$ & $1.97_{-0.27}^{+0.36}$  & $-1.44_{-0.07}^{+0.08}$ & $70_{-23}^{+48}$ & &  $0.78_{-0.11}^{+0.10}$ & $102.7 (113)$ & X,B,G \\

$[241.25 , 252.52]$ & BPL& $-1.03_{-0.09}^{+0.09}$ & $5.74_{-1.11}^{+1.95}$  & $-1.85_{-0.11}^{+0.07}$  & &  & $[0.97-4.10]$ & $244.0 (233)$ & X,B,G \\

$[252.52 , 260.71]$ & BPL& $-0.67_{-0.22}^{+0.23}$ & $3.59_{-0.73}^{+1.39}$  & $-1.70_{-0.08}^{+0.05}$  & &  & $[1.68-17.17]$ & $213.2 (200)$ & X,B,G \\

$[260.71 , 270.95]$ & BPL& $-0.75_{-0.11}^{+0.13}$ & $4.52_{-0.69}^{+0.69}$  & $-1.74_{-0.04}^{+0.04}$  & &  & $[2.73-22.18]$ & $269.0 (235)$ & X,B,G \\

$[270.95 , 290.40]$ & BPL& $-0.68_{-0.19}^{+0.22}$ & $2.17_{-0.20}^{+0.27}$  & $-1.91_{-0.04}^{+0.04}$  & & &  $[0.60-2.64]$ & $132.5 (128)$ & X,B,G \\ 

\\
\multicolumn{10}{c}{${ \bf GRB~110205A,\,z=2.22 }$ } \\    [0.1 cm]       
 $\bf{[0.00 , 160.00]}$ & CPL &  &  & $-1.27_{-0.28}^{+0.29}$ & $72_{-10}^{+23}$ & & $0.65_{-0.17}^{+0.24}$ & $48.5 (55)$ & B \\

 $\bf{[160.00 , 350.00]}$ & BPL & $-0.88_{-0.03}^{+0.04}$ & $5.79_{-0.74}^{+0.68}$  & $-1.78_{-0.04}^{+0.04}$  & & & $[0.64-7.11]$ & $272.2 (281)$ & X,B \\

 $[0.00 , 94.00]$ & PL &  &  & $-1.63_{-0.13}^{+0.13}$ &  &    & $[0.27-11.72]$ & $52.5 (56)$ & B \\

 $[94.00 , 120.00]$ & PL&  &  & $-1.87_{-0.09}^{+0.08}$ & & &  $[0.51-5.29]$ & $61.8 (56)$ & B \\

 $[120.00 , 160.00]$ & CPL&  &  & $-1.46_{-0.23}^{+0.24}$ & $65_{-8}^{+16}$ &  & $1.23_{-0.26}^{+0.35}$ & $61.1 (55)$ & B \\

 $[160.00 , 193.00]$ & BPL& $-0.63_{-0.05}^{+0.05}$ & $5.89_{-0.46}^{+0.60}$  & $-1.85_{-0.04}^{+0.04}$  & & &  $[0.95-6.80]$ & $209.7 (190)$ & X,B \\

 $[193.00 , 210.00]$ & BPL& $-0.74_{-0.07}^{+0.08}$ & $5.82_{-0.90}^{+0.78}$  & $-1.64_{-0.05}^{+0.05}$  & & &  $[1.16-35.35]$ & $112.3 (126)$ & X,B \\

 $[210.00 , 240.00]$ & BCPL & $-0.57_{-0.08}^{+0.15}$ & $3.85_{-0.80}^{+0.66}$  & $-1.37_{-0.15}^{+0.07}$ & $108_{-28}^{+99}$ &  & $1.52_{-0.12}^{+0.37}$  & $174.3 (168)$ & X,B \\

 $[240.00 , 350.00]$ & BPL& $-1.15_{-0.05}^{+0.04}$ & $6.19_{-0.71}^{+1.79}$  & $-1.86_{-0.05}^{+0.08}$  & & &  $[0.30-1.91]$ & $235.9 (225)$ & X,B \\   

\\       
\multicolumn{10}{c}{${ \bf GRB~121123A}$ } \\    [0.1 cm] 

$\bf{[193.15 , 299.65]}$ & CPL &  &  & $-0.86_{-0.03}^{+0.03}$ & $75_{-4}^{+4}$ & & $1.11_{-0.06}^{+0.07}$ & $148.1 (164)$ & X,B \\

$[193.15 , 214.65]$ & CPL&  &  & $-0.73_{-0.05}^{+0.05}$ & $121_{-16}^{+22}$  &  &  $1.33_{-0.14}^{+0.15}$  & $109.2 (127)$ & X,B \\ 

$[214.65 , 231.04]$ & CPL&  &  & $-0.54_{-0.04}^{+0.05}$ & $99_{-8}^{+10}$ & &  $1.86_{-0.16}^{+0.10}$ & $108.7 (120)$ & X,B \\ 

$[231.04 , 239.23]$ & CPL&  &  & $-0.84_{-0.06}^{+0.06}$ & $87_{-9}^{+12}$  &  &  $2.02_{-0.22}^{+0.25}$  & $87.6 (88)$ & X,B \\

$[239.23 , 247.42]$ & BCPL & $-0.19_{-0.26}^{+0.46}$ & $2.63_{-0.68}^{+1.13}$  & $-1.04_{-0.17}^{+0.09}$ & $61_{-10}^{+20}$ &  &  $1.69_{-0.29}^{+0.51}$ & $74.4 (93)$ & X,B \\

$[247.42 , 267.90]$ & BCPL & $-0.59_{-0.16}^{+0.35}$ & $2.18_{-0.83}^{+4.10}$  & $-0.93_{-0.12}^{+0.07}$ & $47_{-6}^{+9}$ & &  $0.91_{-0.07}^{+0.17}$ & $162.7 (157)$ & X,B \\ 

$[267.90 , 299.65]$ & CPL&  &  & $-1.10_{-0.06}^{+0.06}$ & $44_{-4}^{+5}$ &  & $0.57_{-0.07}^{+0.08}$ & $185.4 (185)$ & X,B \\   

\\
\multicolumn{10}{c}{${ \bf GRB~130907A,\,z =1.238 }$ } \\    [0.1 cm]
 $\bf{[-80.00 , 71.00]}$ & CPL &  &  & $-0.93_{-0.08}^{+0.08}$ & $284_{-50}^{+91}$ & & $19.56_{-1.24}^{+1.33}$ & $22.5 (55)$ & B \\ 
 
 $\bf{[71.00 , 550.00]}$ & BPL & $-1.37_{-0.07}^{+0.10}$ & $2.30_{-0.60}^{+0.62}$  & $-1.67_{-0.01}^{+0.01}$  & & & $[0.78-17.44]$ & $307.1 (281)$ & X,B \\
     
  $[-80.00 , -65.00]$ & PL&  &  & $-1.32_{-0.31}^{+0.32}$ & & &   $[0.60-310.90]$ & $62.4 (56)$ & B \\ 
 
 $[-65.00 , -44.00]$ & PL&  &  & $-1.22_{-0.06}^{+0.06}$ & &  & $[1.38-1550.40]$ & $43.2 (56)$ & B \\ 
 
 $[-44.00 , -30.00]$ & PL&  &  & $-1.27_{-0.03}^{+0.03}$ & & &  $[3.43-2648.80]$ & $36.8 (56)$ & B \\
 
 $[-30.00 , 20.00]$ & CPL&  &  & $-0.95_{-0.08}^{+0.08}$ & $365_{-85}^{+193}$ &  & $32.84_{-2.09}^{+2.25}$ & $21.6 (55)$ & B  \\ 
 
 $[20.00 , 40.00]$ & CPL&  &  & $-0.84_{-0.09}^{+0.09}$ & $275_{-50}^{+94}$ &  & $28.26_{-2.07}^{+2.26}$ & $24.4 (55)$ & B \\ 
 
 $[40.00 , 52.00]$ & CPL&  &  & $-1.02_{-0.11}^{+0.11}$ & $288_{-76}^{+217}$ &  &  $18.28_{-1.69}^{+1.90}$ & $33.6 (55)$ & B  \\ 
 
 $[52.00 , 71.00]$ & CPL&  &  & $-0.95_{-0.09}^{+0.09}$ & $249_{-45}^{+86}$ &  & $26.22_{-1.96}^{+2.14}$ & $25.0 (55)$ & B \\
 
 $[71.00 , 79.00]$ & BPL& $-0.58_{-0.33}^{+0.90}$ & $2.75_{-0.93}^{+2.01}$  & $-1.29_{-0.03}^{+0.02}$  & & &  $[5.47-2978.90]$ & $71.9 (81)$ & X,B \\
 
 $[79.00 , 87.00]$ & PL&  &  & $-1.12_{-0.01}^{+0.02}$ & & &  $[7.28-18644.00]$ & $77.1 (89)$ & X,B \\
 
 $[87.00 , 110.00]$ & BPL& $-1.03_{-0.21}^{+0.23}$ & $2.58_{-0.51}^{+0.95}$  & $-1.70_{-0.04}^{+0.04}$  & &  & $[1.50-26.50]$ & $142.6 (124)$  & X,B \\
 
 $[200.00 , 220.00]$ & PL&  &  & $-1.50_{-0.07}^{+0.03}$ & & &   $[0.88-77.48]$ & $68.1 (79)$ & X,B \\
 
 $[220.00 , 250.00]$ & BPL& $-1.08_{-0.13}^{+0.25}$ & $4.54_{-1.66}^{+1.61}$  & $-1.75_{-0.05}^{+0.05}$  & &  & $[1.60-20.06]$ & $120.4 (133)$ & X,B \\
 
 $[250.00 , 350.00]$ & BPL& $-1.46_{-0.07}^{+0.11}$ & $4.08_{-1.34}^{+1.03}$  & $-1.91_{-0.04}^{+0.05}$  & & &  $[0.69-3.02]$ & $215.1 (232)$ & X,B \\ 
 
 $[350.00 , 550.00]$ & BPL& $-1.59_{-0.03}^{+0.03}$ & $5.01_{-0.82}^{+0.69}$  & $-2.04_{-0.03}^{+0.04}$  & & &  $[0.42-1.02]$ & $341.6 (297)$  & X,B \\
 
\\
\\
\multicolumn{10}{c}{${ \bf GRB~140108A}$ } \\    [0.1 cm]
$\bf{[-7.21 , 16.34]}$  & CPL &  &  & $-1.43_{-0.13}^{+0.14}$ & $143_{-36}^{+94}$ & & $2.33_{-0.27}^{+0.37}$ & $80.4 (97)$ & B,G \\ 

$\bf{[76.76 , 101.33]}$ & BCPL & $0.35_{-0.55}^{+0.37}$ & $2.54_{-0.22}^{+0.82}$  & $-1.33_{-0.05}^{+0.03}$ & $844_{-310}^{+1548}$ & &  $7.24_{-1.43}^{+2.41}$ & $136.1 (137)$ & X,B,G \\

$[-3.11 , 2.01]$ & CPL&  &  & $-1.11_{-0.32}^{+0.40}$ & $105_{-28}^{+89}$ &  & $1.34_{-0.37}^{+0.72}$  & $104.3 (135)$ & B,G \\

$[2.01 , 4.05]$ & CPL&  &  & $-1.34_{-0.20}^{+0.23}$ & $116_{-30}^{+82}$  &  & $3.72_{-0.67}^{+1.05}$ & $105.2 (98)$ & B,G \\ 

$[4.05 , 7.13]$ & CPL&  &  & $-1.27_{-0.12}^{+0.14}$ & $172_{-40}^{+83}$  &  & $6.45_{-0.69}^{+0.92}$  & $134.7 (129)$ & B,G \\

$[7.13 , 11.22]$ & PL&  &  & $-1.70_{-0.05}^{+0.05}$ & &  & $[4.29-50.00]$ & $96.3 (99)$ & B,G \\ 

$[76.76 , 81.88]$ & BPL& $-0.63_{-0.32}^{+0.40}$ & $7.12_{-1.58}^{+4.75}$  & $-1.37_{-0.05}^{+0.05}$  & & & $[4.03-358.53]$ & $134.0 (165)$ & X,B,G \\

$[81.88 , 83.92]$ & BPL & $-0.61_{-0.43}^{+0.40}$ & $7.54_{-1.69}^{+14.21}$  & $-1.34_{-0.04}^{+0.04}$  & & & $[9.87-1137.50]$  & $129.1 (124)$ &  X,B,G \\

$[83.92 , 85.97]$ & BPL& $-0.17_{-0.48}^{+0.62}$ & $7.14_{-1.55}^{+3.92}$  & $-1.37_{-0.04}^{+0.04}$  & & & $[11.18-1055.20]$ & $138.2 (121)$ & X,B,G \\

$[85.97 , 88.02]$ & CPL &  &  & $-1.12_{-0.05}^{+0.06}$ & $314_{-81}^{+150}$ & & $10.74_{-0.75}^{+0.89}$ & $117.7 (131)$ & X,B,G \\

$[88.02 , 93.14]$ & PL&  &  & $-1.42_{-0.03}^{+0.03}$ &  &  & $[4.99-325.48]$ & $192.4 (182)$ & X,B,G \\

$[93.14 , 101.33]$  & PL&  &  & $-1.58_{-0.05}^{+0.05}$ &  & &  $[1.17-24.19]$ & $178.5 (202)$ & X,B,G \\

\\       
\multicolumn{10}{c}{${ \bf GRB~140206A,\,z =2.73 }$ } \\    [0.1 cm]

 $\bf{[-0.50 , 11.00]}$ & CPL &  &  & $-0.98_{-0.18}^{+0.19}$ & $145_{-28}^{+69}$ & &  $4.37_{-0.71}^{+0.88}$ & $39.1 (55)$ & B \\
 
 $\bf{[50.25 , 100.00]}$ & BCPL & $-0.70_{-0.07}^{+0.11}$ & $5.42_{-2.34}^{+1.96}$  & $-1.05_{-0.08}^{+0.10}$ & $102_{-13}^{+18}$ & &  $3.53_{-0.26}^{+0.38}$ & $78.6 (99)$ & X,B  \\
        
  $[-0.50 , 4.30]$ & CPL&  &  & $-0.92_{-0.31}^{+0.33}$ & $98_{-18}^{+53}$  & &  $2.51_{-0.72}^{+1.09}$  & $48.4 (55)$ & B \\  
 
 $[4.30 , 8.60]$ & PL&  &  & $-1.33_{-0.05}^{+0.05}$ & &  & $[2.95-1401.90]$ & $73.4 (56)$  & B \\ 
 
  $[8.60 , 11.00]$ & PL&  &  & $-1.30_{-0.07}^{+0.07}$ & &  & $[3.00-1782.70]$ & $58.3 (56)$ & B \\ 
 
 $[50.25 , 55.00]$ & BPL& $-0.86_{-0.15}^{+0.10}$ & $8.01_{-2.81}^{+4.11}$  & $-1.58_{-0.09}^{+0.08}$  & &  & $[1.87-90.47]$ & $48.9 (74)$ & X,B \\
 
 $[55.00 , 58.00]$ & BCPL & $0.16_{-0.55}^{+0.16}$ & $2.26_{-0.90}^{+1.64}$  & $-0.87_{-0.12}^{+0.10}$ & $128_{-23}^{+39}$ &  & $12.31_{-5.75}^{+4.89}$ & $52.8 (67)$ & X,B \\ 
 
 $[58.00 , 60.00]$ & BCPL & $-0.14_{-0.29}^{+0.40}$ & $5.24_{-1.74}^{+1.90}$  & $-0.83_{-0.17}^{+0.10}$ & $112_{-18}^{+45}$ &  & $15.36_{-3.89}^{+9.88}$ & $55.2 (73)$ & X,B \\ 
 
 $[60.00 , 62.00]$ & CPL&  &  & $-0.59_{-0.08}^{+0.06}$ & $95_{-6}^{+5}$ & &  $15.75_{-1.07}^{+0.91}$ & $88.1 (78)$ & X,B \\ 
 
 $[62.00 , 64.00]$ & BCPL & $0.39_{-0.83}^{+1.54}$ & $2.14_{-0.57}^{+2.75}$  & $-0.86_{-0.14}^{+0.08}$ & $87_{-10}^{+23}$ &  & $11.44_{-4.17}^{+6.21}$ & $60.1 (77)$ & X,B \\ 
 
 $[64.00 , 70.00]$ & BCPL & $-0.40_{-0.17}^{+0.22}$ & $4.26_{-1.09}^{+2.05}$  & $-1.36_{-0.19}^{+0.12}$ & $81_{-23}^{+85}$ & &  $3.44_{-0.53}^{+0.67}$ & $66.7 (84)$ & X,B  \\
 
 $[70.00 , 80.00]$ & CPL&  &  & $-1.13_{-0.06}^{+0.06}$ & $41_{-5}^{+6}$ &  & $0.90_{-0.15}^{+0.19}$ & $94.1 (84)$ & X,B \\ 
 
 $[80.00 , 100.00]$ & CPL&  &  & $-1.34_{-0.10}^{+0.07}$ & $47_{-13}^{+22}$ &  & $0.28_{-0.02}^{+0.09}$ & $135.1 (122)$ & X,B \\
 
\\
 \multicolumn{10}{c}{${ \bf GRB~140512A,\,z =0.725 }$ } \\    [0.1 cm]
 
$\bf{[-21.05 , 10.70]}$ & CPL &  &  & $-1.09_{-0.11}^{+0.12}$ & $439_{-134}^{+293}$ & &  $3.32_{-1.12}^{+1.50}$ & $196.9 (317)$ & B,G \\
   
$\bf{[102.86 , 158.16]}$ & BCPL & $-0.76_{-0.04}^{+0.05}$ & $7.18_{-1.00}^{+1.12}$  & $-1.26_{-0.04}^{+0.04}$ & $532_{-123}^{+190}$ & & $5.52_{-0.27}^{+0.27}$ & $442.8 (478)$ & X,B,G \\
                 
$[-21.05 , 0.46]$ & CPL&  &  & $-1.20_{-0.12}^{+0.14}$ & $598_{-259}^{+1030}$ &  & $3.03_{-0.27}^{+0.33}$ & $223.8 (299)$ & B,G \\ 
   
$[0.46 , 10.70]$ & CPL&  &  & $-1.01_{-0.15}^{+0.17}$ & $306_{-86}^{+190}$ &  & $3.28_{-0.34}^{+0.43}$ & $224.6 (250)$ & B,G  \\ 
   
$[102.86 , 107.98]$ & BCPL & $-0.59_{-0.18}^{+0.18}$ & $7.77_{-2.69}^{+6.25}$  & $-1.19_{-0.10}^{+0.10}$ & $580_{-259}^{+1150}$ &  & $5.26_{-1.04}^{+1.11}$ & $190.7 (213)$ & X,B,G \\ 
   
$[107.98 , 113.10]$ & BCPL & $-0.40_{-0.15}^{+0.19}$ & $6.67_{-2.02}^{+2.99}$  & $-1.06_{-0.07}^{+0.07}$ & $513_{-135}^{+220}$ & &  $9.68_{-1.78}^{+1.79}$ & $246.3 (228)$ & X,B,G \\ 
   
$[113.10 , 118.22]$  & BCPL & $-0.58_{-0.14}^{+0.15}$ & $7.96_{-2.14}^{+3.63}$  & $-1.25_{-0.10}^{+0.10}$ & $328_{-123}^{+287}$ &  & $4.56_{-0.75}^{+0.76}$ & $196.9 (225)$ & X,B,G \\ 
   
$[118.22 , 123.34]$ & BCPL & $-0.56_{-0.07}^{+0.10}$ & $19.52_{-5.08}^{+4.40}$  & $-1.23_{-0.05}^{+0.05}$ & $942_{-292}^{+484}$ &  & $18.52_{-2.49}^{+2.47}$ & $239.3 (246)$ & X,B,G \\ 
   
$[123.34 , 128.46]$ & BCPL & $-0.35_{-0.14}^{+0.16}$ & $7.24_{-1.38}^{+1.74}$  & $-1.29_{-0.06}^{+0.06}$ & $529_{-176}^{+366}$ &  & $10.40_{-1.70}^{+1.75}$ & $263.3 (241)$ & X,B,G \\ 
   
$[128.46 , 133.58]$ & BPL& $-0.76_{-0.14}^{+0.18}$ & $6.08_{-1.60}^{+2.18}$  & $-1.45_{-0.04}^{+0.04}$  & &  & $[20.82-275.19]$ & $223.8 (228)$ & X,B,G \\ 
   
$[133.58 , 138.70]$  & CPL&  &  & $-1.16_{-0.04}^{+0.04}$ & $170_{-37}^{+60}$ & &  $1.88_{-0.19}^{+0.22}$ & $214.5 (226)$ & X,B,G \\ 
   
$[138.70 , 143.82]$ & CPL&  &  & $-1.18_{-0.05}^{+0.05}$ & $213_{-67}^{+146}$ &  & $1.63_{-0.20}^{+0.25}$ & $220.3 (216)$ & X,B,G \\
   
$[143.82 , 158.16]$ & BPL& $-0.61_{-0.19}^{+0.34}$ & $2.07_{-0.38}^{+0.33}$  & $-1.45_{-0.03}^{+0.03}$  & & &  $[5.17-70.95]$ & $367.6 (340)$ & X,B,G\\[0.1cm]
\end{longtable}

\newpage

\section{Selection of the Best-fit Model}

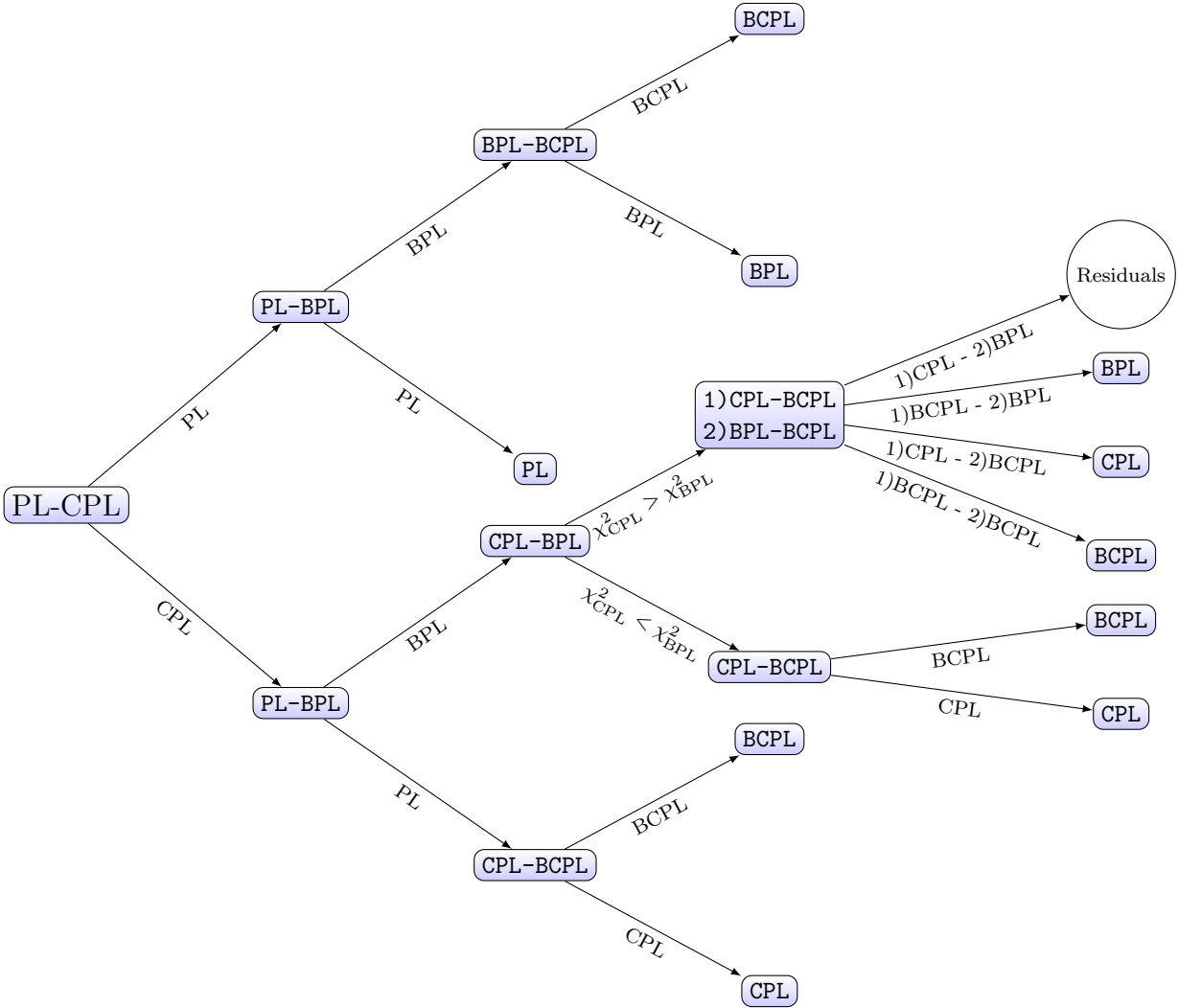
\begin{figure*}[!hbt]
\begin{tikzpicture}
  [
    grow                    = right,
    sibling distance        = 10em,
    level distance          = 10em,
    edge from parent/.style = {draw, -latex},
    every node/.style       = {font=\footnotesize},
    sloped
  ]
 \tikzstyle{level 1}=[sibling distance = 55 mm]
  \tikzstyle{level 2}=[sibling distance = 45 mm]
    \tikzstyle{level 3}=[sibling distance = 35 mm]
    \tikzstyle{level 4}=[sibling distance = 13 mm, level distance = 15em]

  \node [root] {PL-CPL}
    child { node [env] {PL-BPL}
      child { node [env] {CPL-BCPL}
      child { node [env] {CPL}
      edge from parent node [below] {CPL} }
      child { node [env] {BCPL}
      edge from parent node [below] {BCPL} }
      edge from parent node [below] {PL} }
      child { node [env] {CPL-BPL}
      child { node [env] {CPL-BCPL}
      child { node [env] {CPL}
      edge from parent node [below] {CPL} }
      child { node [env] {BCPL}
      edge from parent node [below] {BCPL} }
      edge from parent node [below] {$\chi_{\rm CPL}^{2}<\chi_{\rm BPL}^{2}$} }
      child { node [env] {1)CPL-BCPL \\ 2)BPL-BCPL}
      child { node [env] {BCPL}
      edge from parent node [below] {1)BCPL  -  2)BCPL} }
      child { node [env] {CPL}
      edge from parent node [below] {1)CPL  -  2)BCPL} }
      child { node [env] {BPL}
      edge from parent node [below] {1)BCPL  -  2)BPL} }
      child { node [dummy] {Residuals}
      edge from parent node [below] {1)CPL  -  2)BPL} }
      edge from parent node [below] {~~$\chi_{\rm CPL}^{2}>\chi_{\rm BPL}^{2}$} }
      edge from parent node [below] {BPL} }
      edge from parent node [below] {CPL} }
     child { node [env] {PL-BPL}
         child { node [env] {PL}
          edge from parent node [below] {PL} }
          child { node [env] {BPL-BCPL}
          child { node [env] {BPL}
      edge from parent node [below] {BPL} }
         child { node [env] {BCPL}
      edge from parent node [below] {BCPL} }
      edge from parent node [below] {BPL} }
      edge from parent node [below] {PL} };

\end{tikzpicture}
\caption{\label{fig:scheme}Flow chart summarizing the procedure adopted to select the best-fit model among four models: power law (PL), cutoff power law (CPL),
broken power law (BPL), and broken power law with high-energy cutoff (BCPL). The selection proceeds from left to right. 
The nodes represent the models that are compared using the $F-test$ (except for the case CPL-BPL, where models are not nested and the total chi-square is compared).
Next to the arrows it is reported the model chosen as a result of the comparison is reported. There is a special case in the scheme where there is no possibility 
to find statistical difference between CPL and BPL models and select the best-fit model on the basis of the chi-square or $F-test$. In this case (which occurred six times) visual inspection of the residuals is adopted.}
\end{figure*}
 
\newpage

%---------------------------------------------------------------------
\section{Time-resolved Spectra of GRB~140512A}

\begin{figure*}[ht!]
\begin{center}
\includegraphics[scale = 0.098]{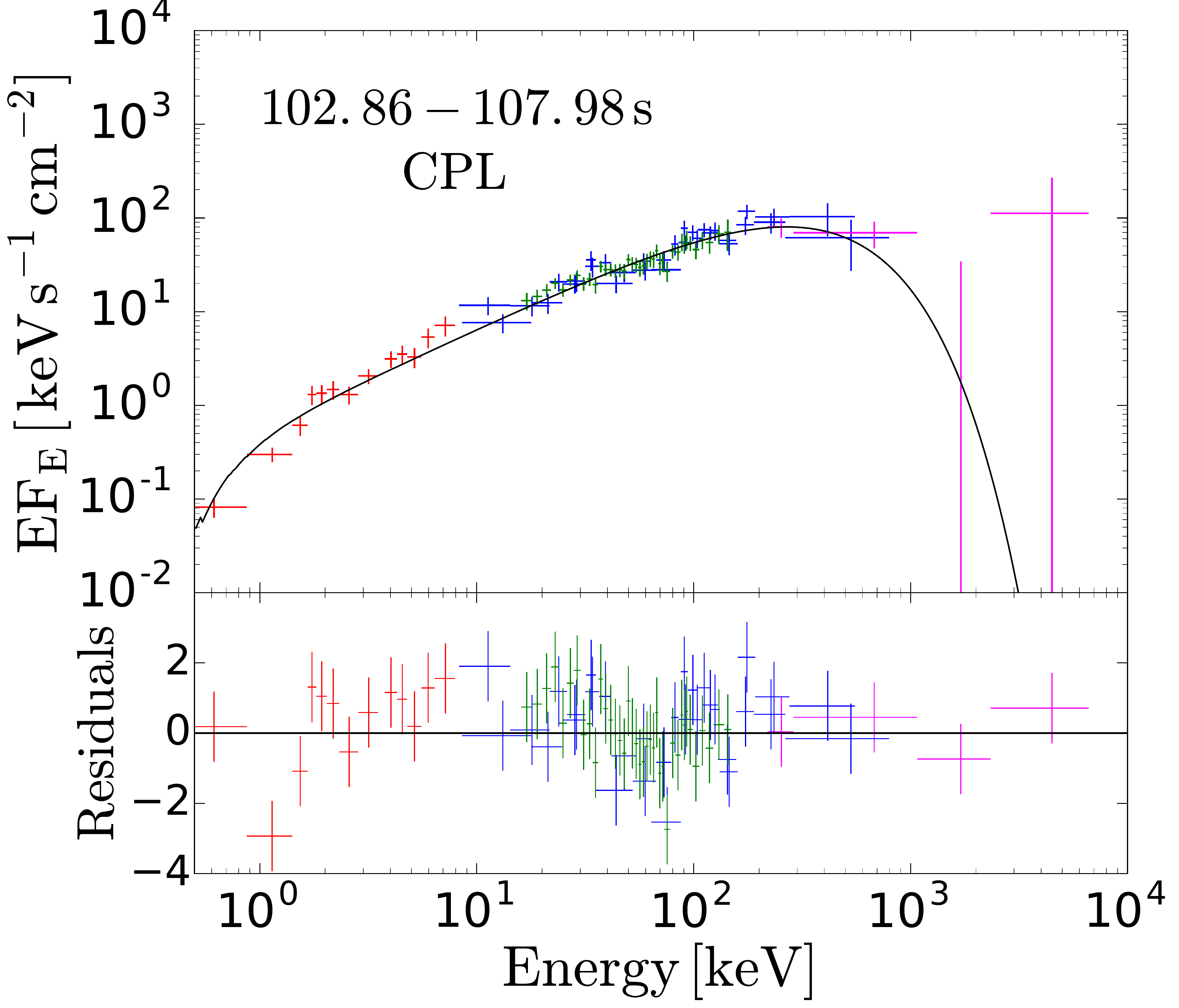} \hspace{0.4em}
\includegraphics[scale = 0.098]{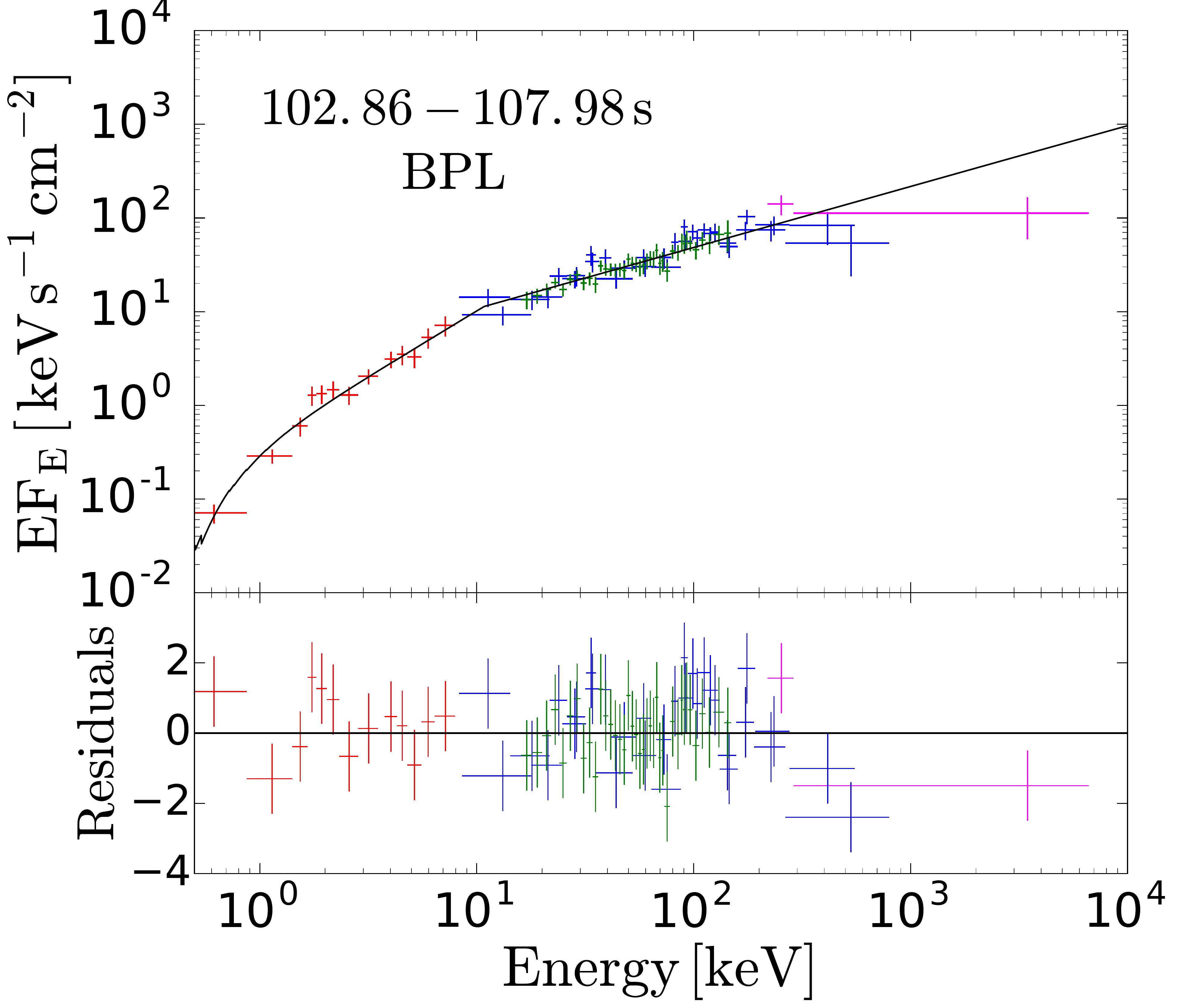} \hspace{0.4em}
\includegraphics[scale = 0.098]{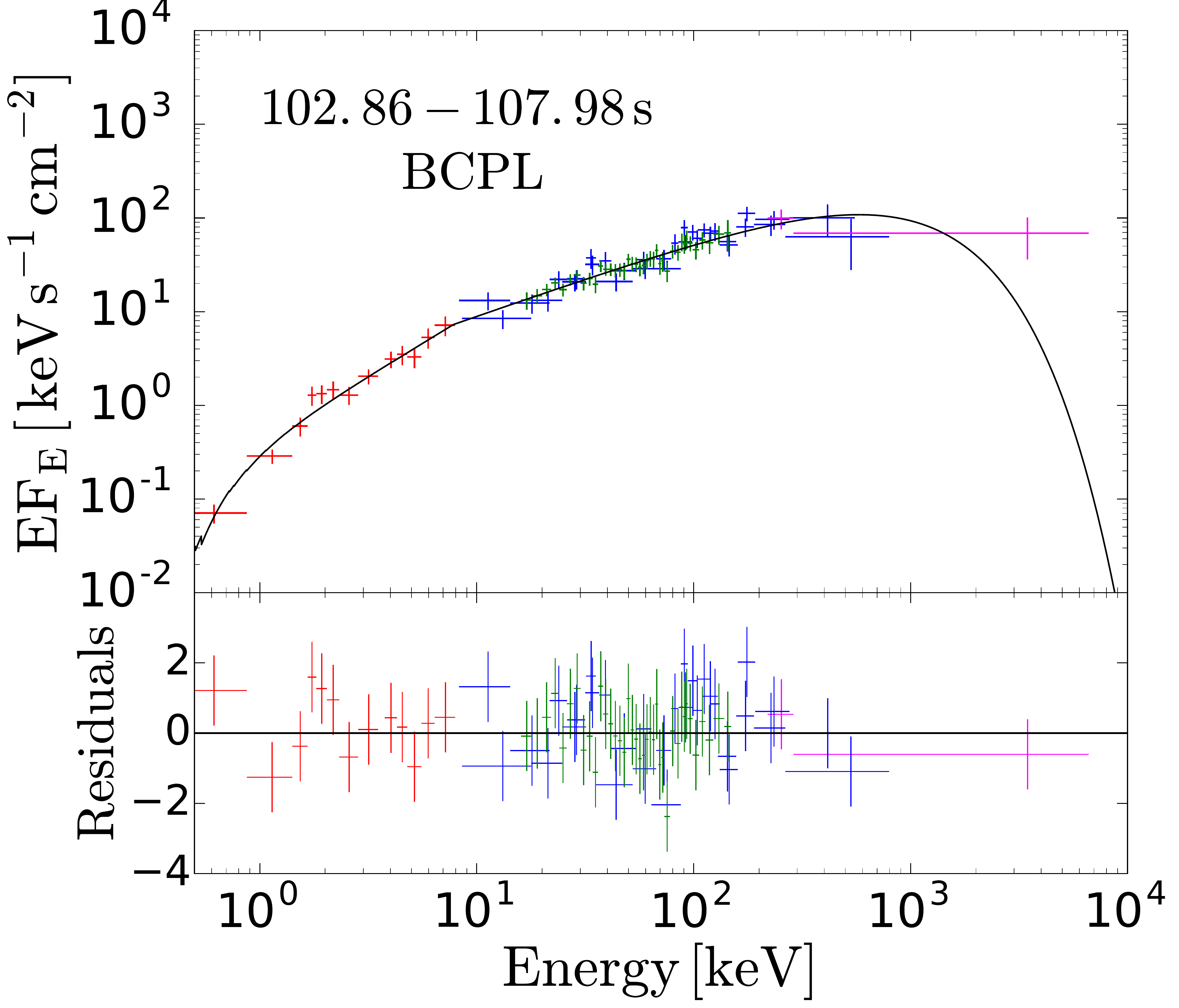} \\
\includegraphics[scale = 0.098]{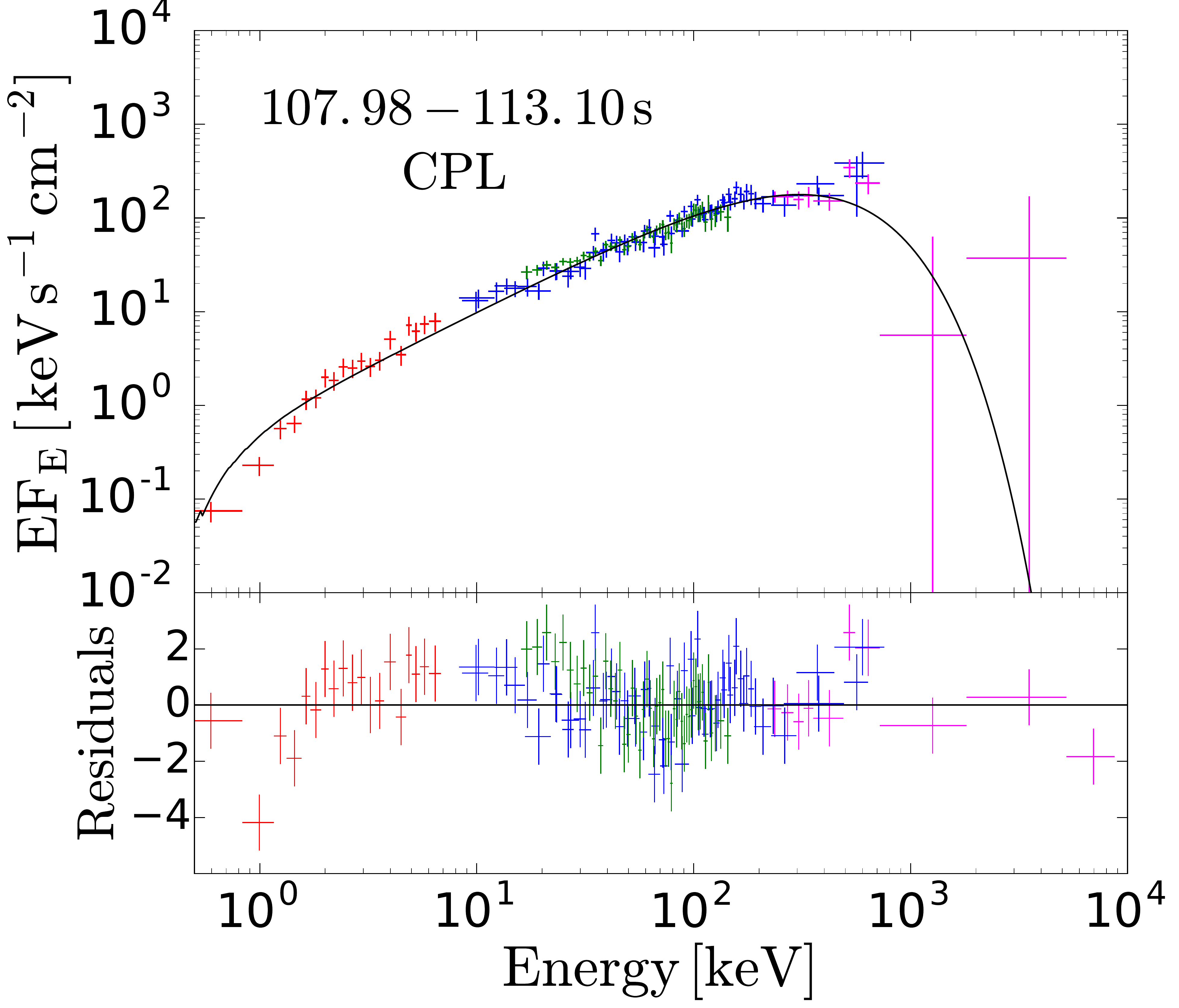} \hspace{0.4em}
\includegraphics[scale = 0.098]{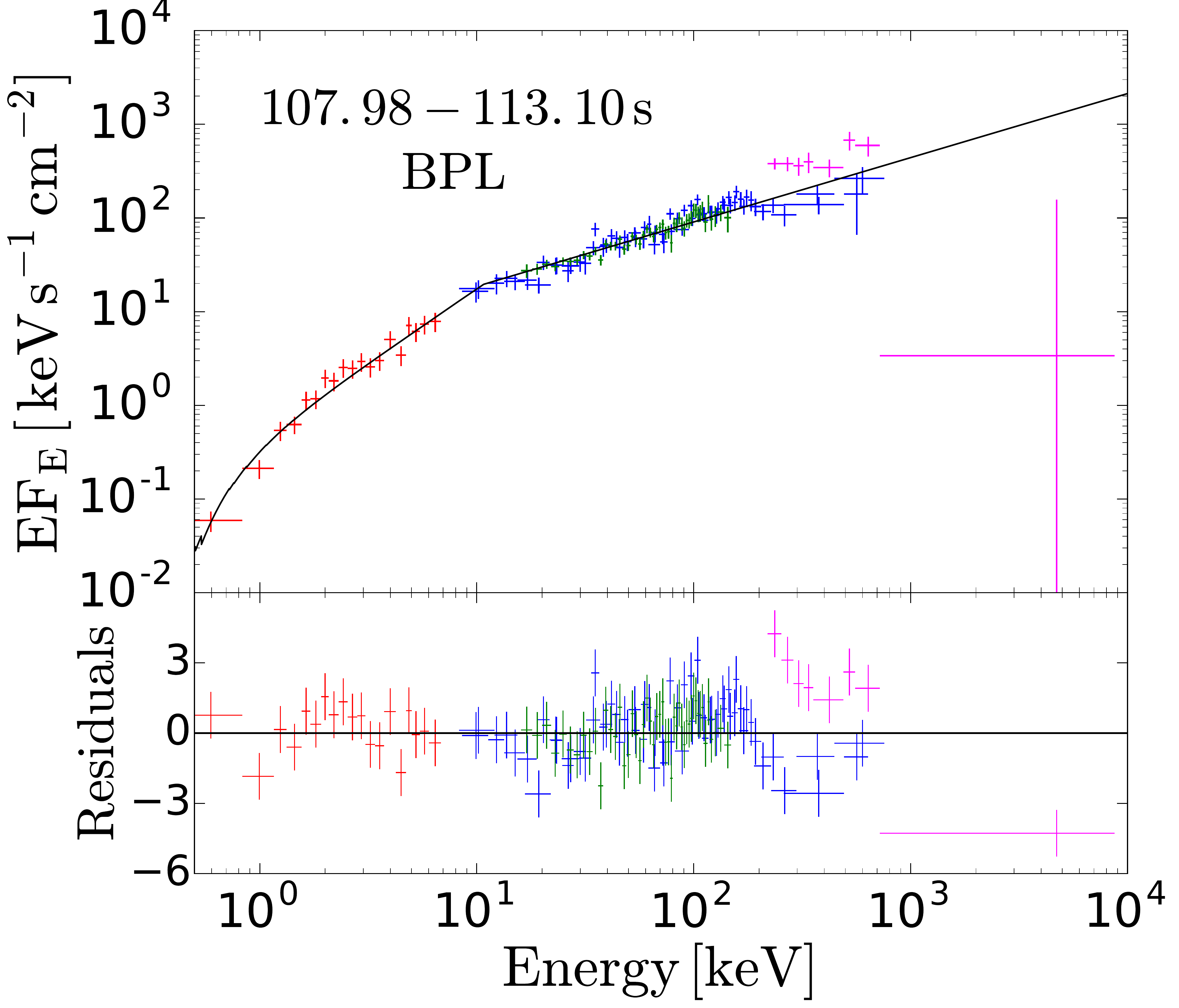} \hspace{0.4em}
\includegraphics[scale = 0.098]{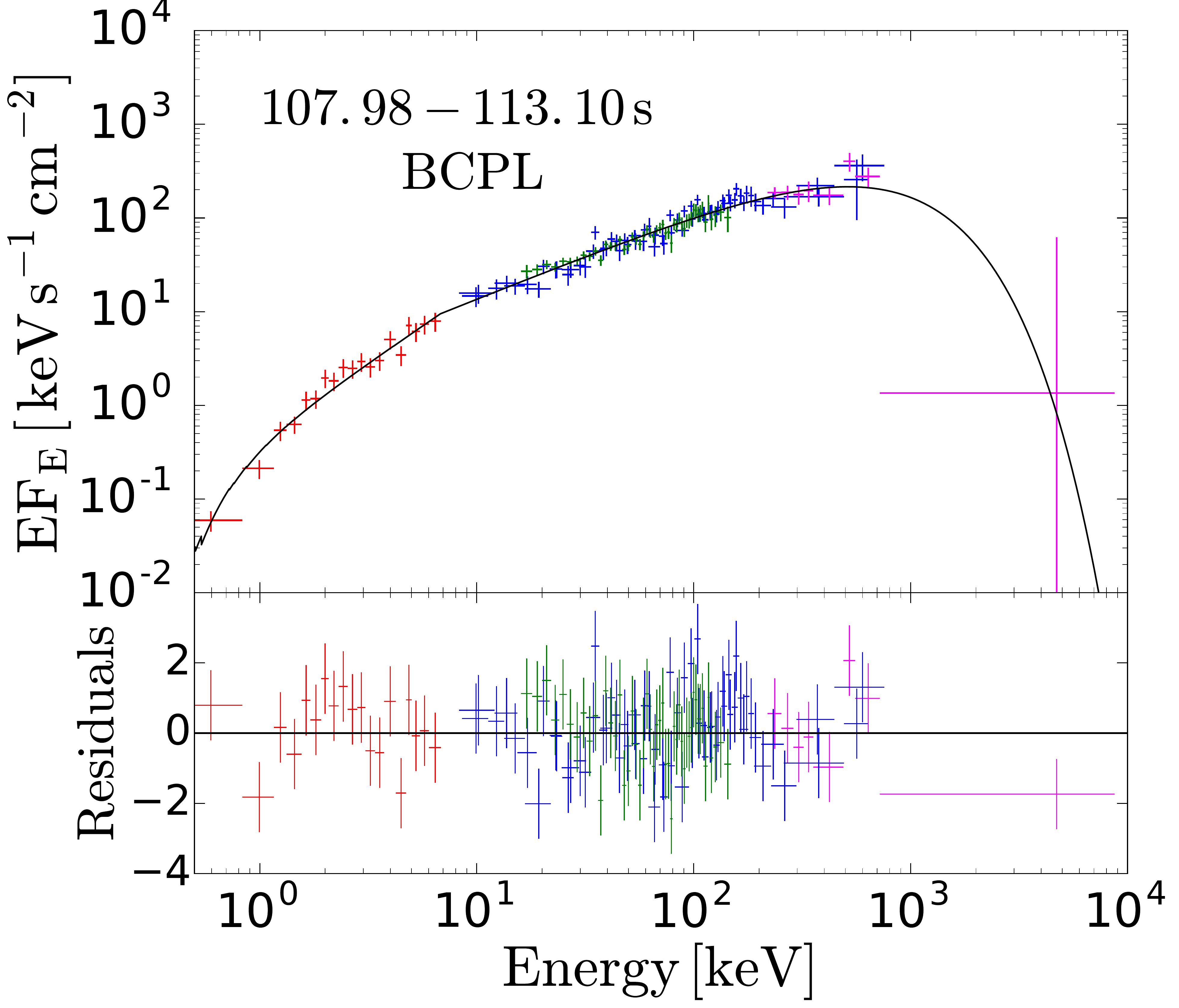} \\
\includegraphics[scale = 0.098]{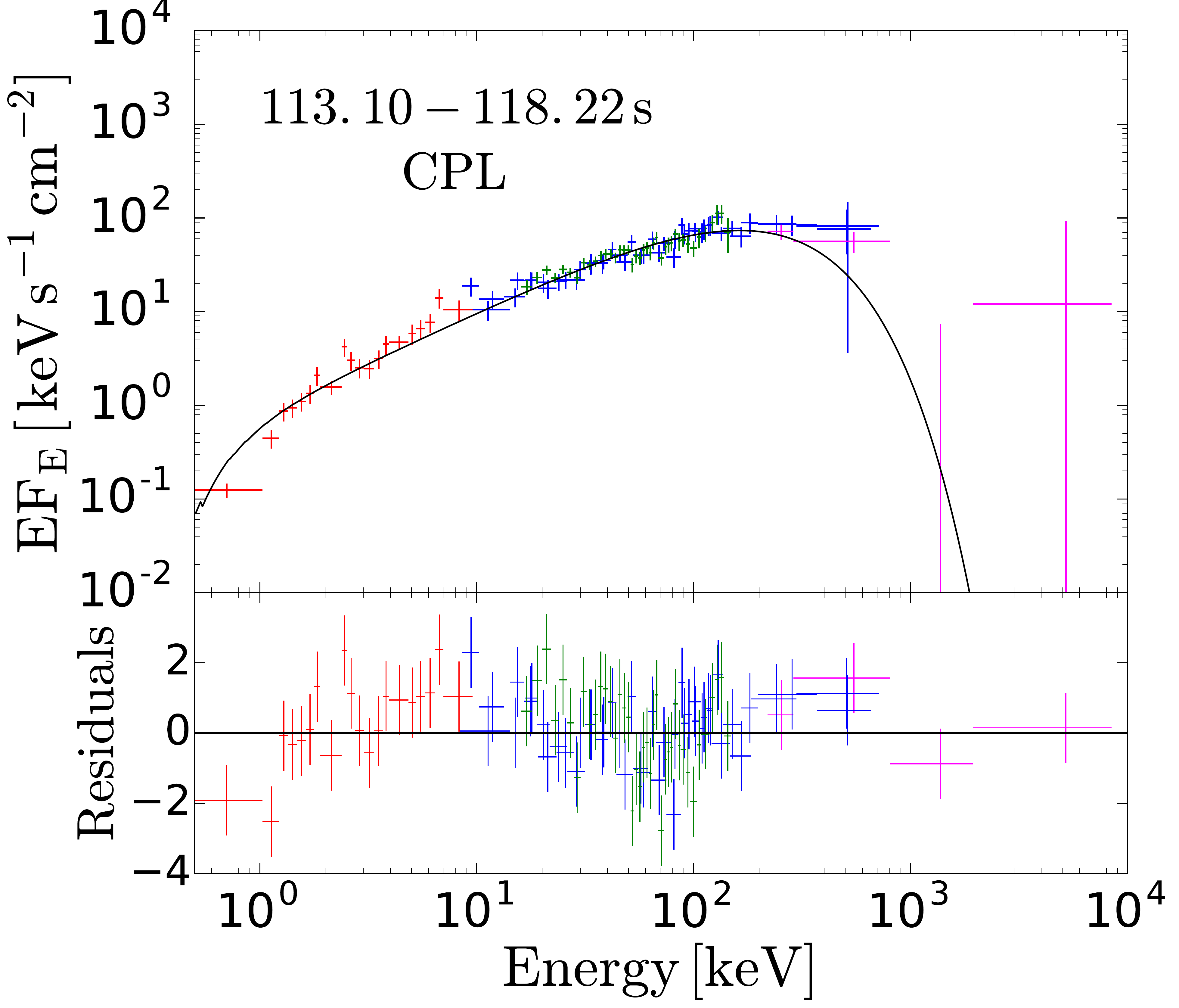} \hspace{0.4em}
\includegraphics[scale = 0.098]{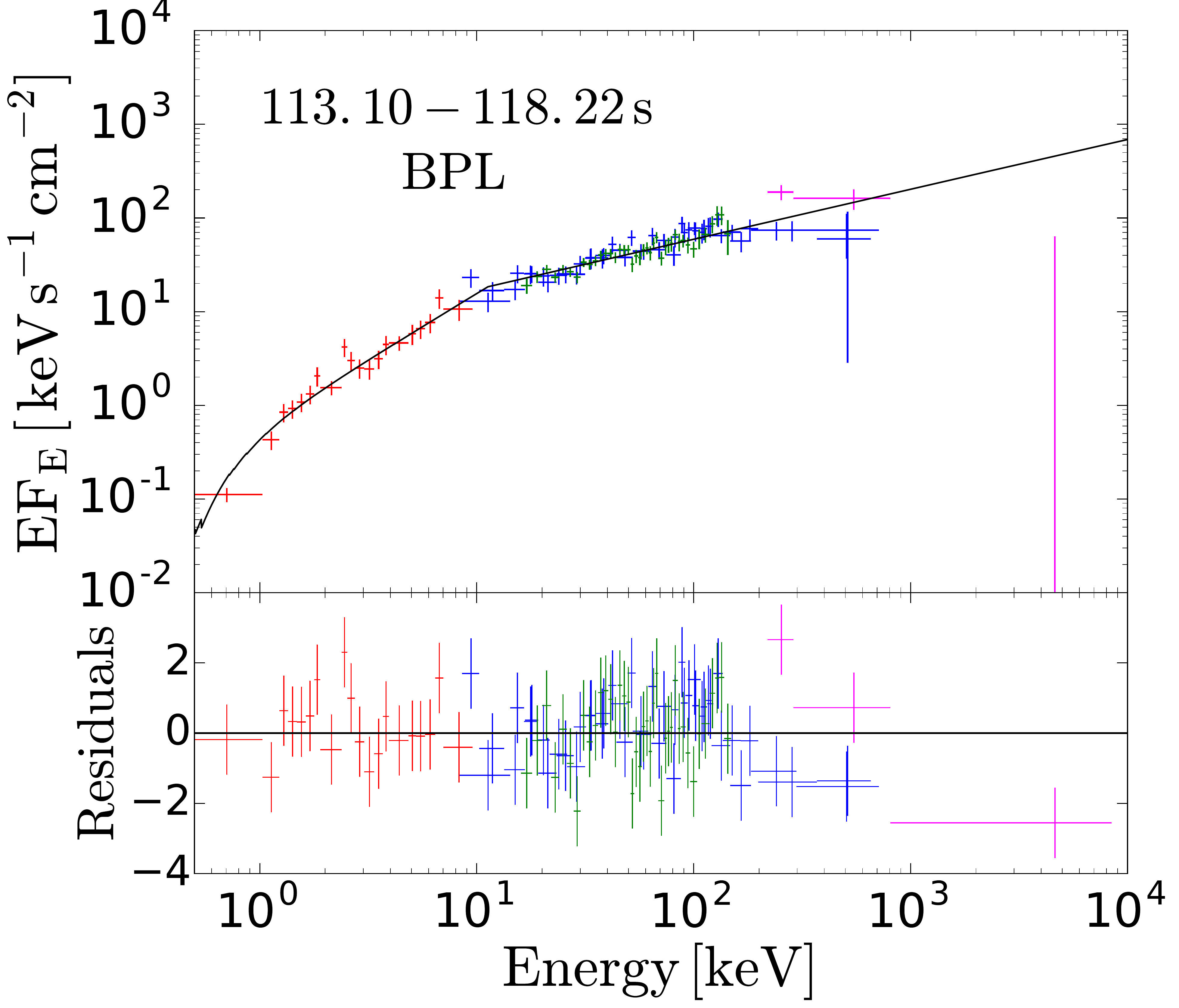} \hspace{0.4em}
\includegraphics[scale = 0.098]{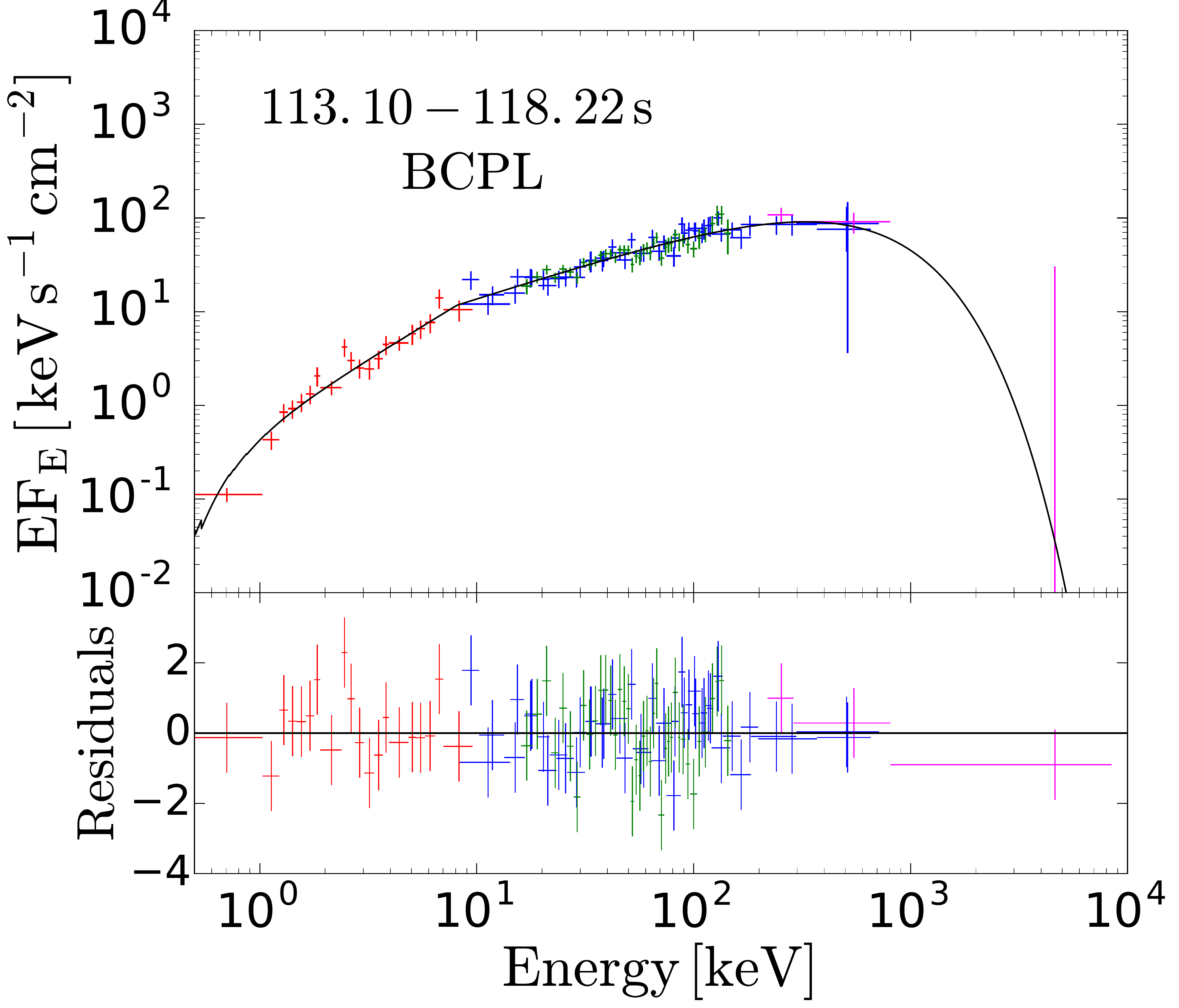} \\
\includegraphics[scale = 0.098]{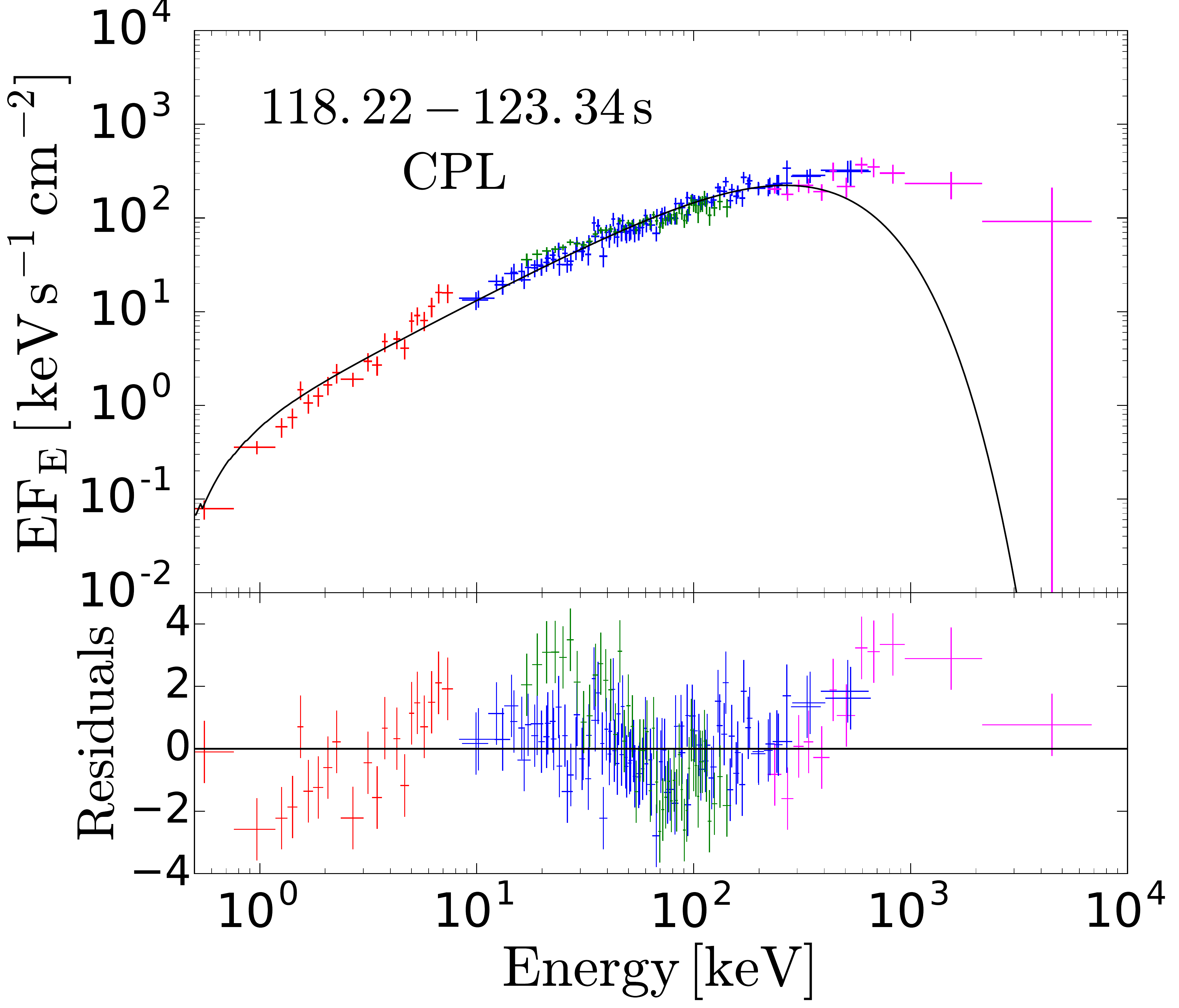} \hspace{0.4em}
\includegraphics[scale = 0.098]{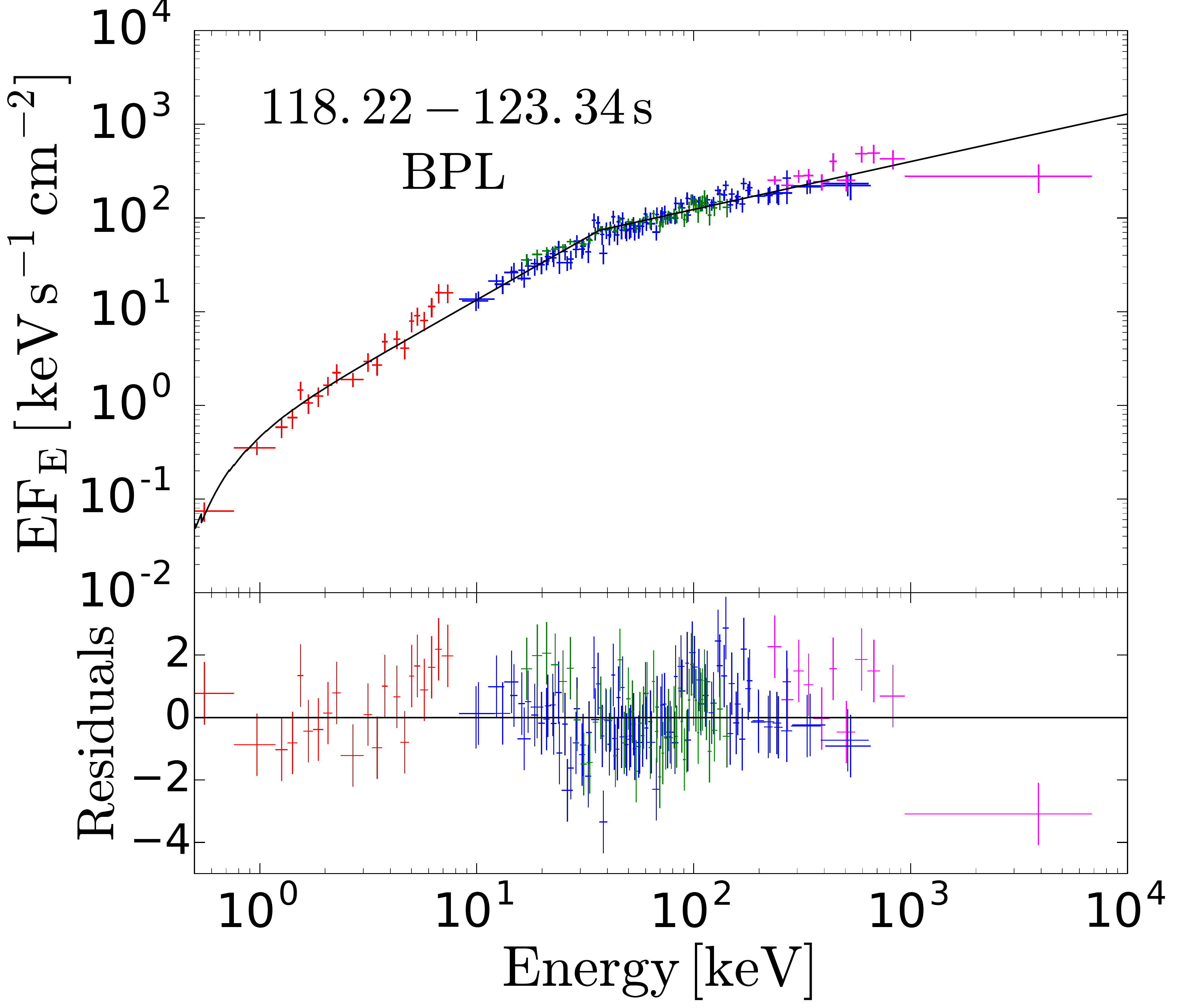} \hspace{0.4em}
\includegraphics[scale = 0.098]{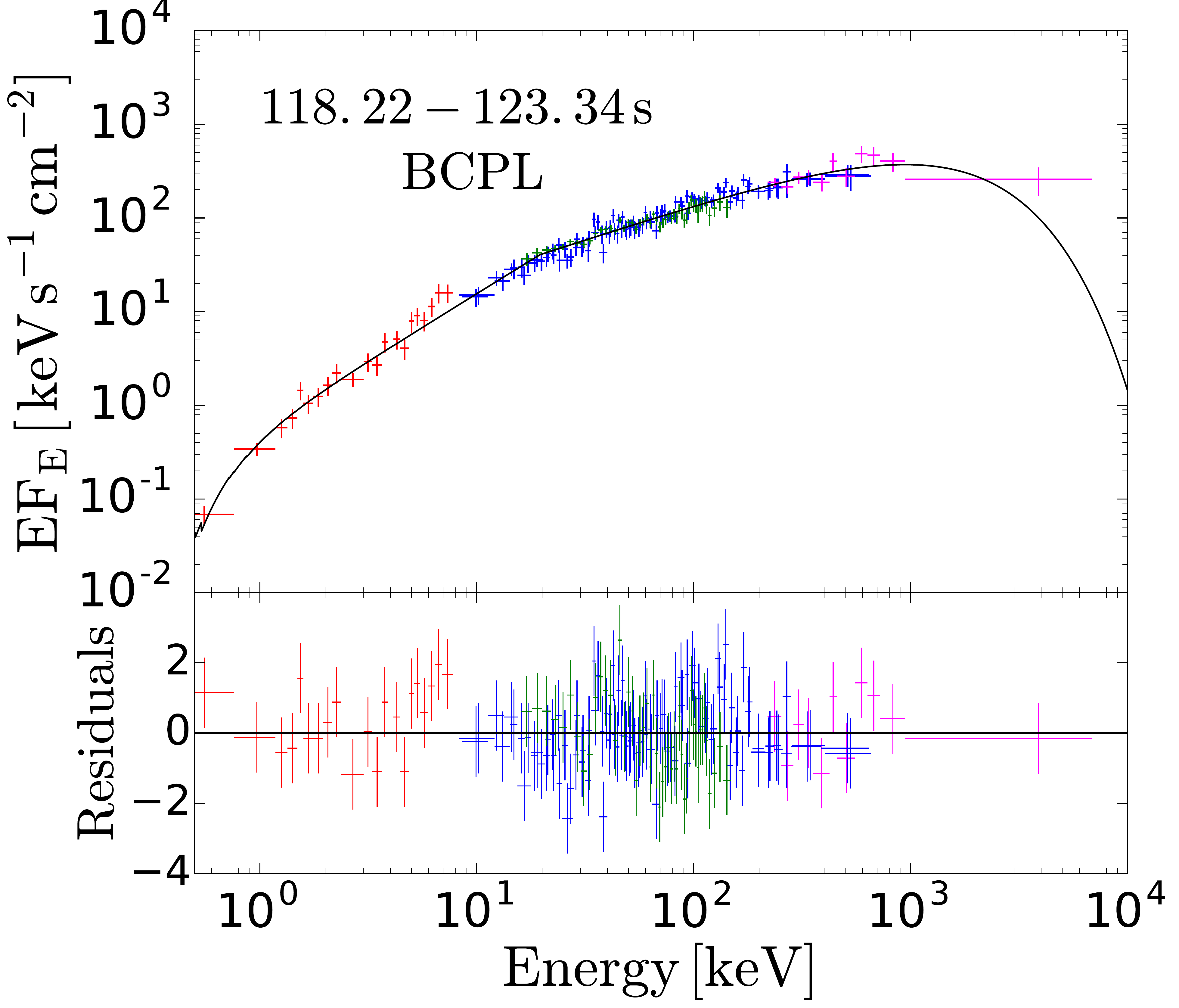} \\
\includegraphics[scale = 0.098]{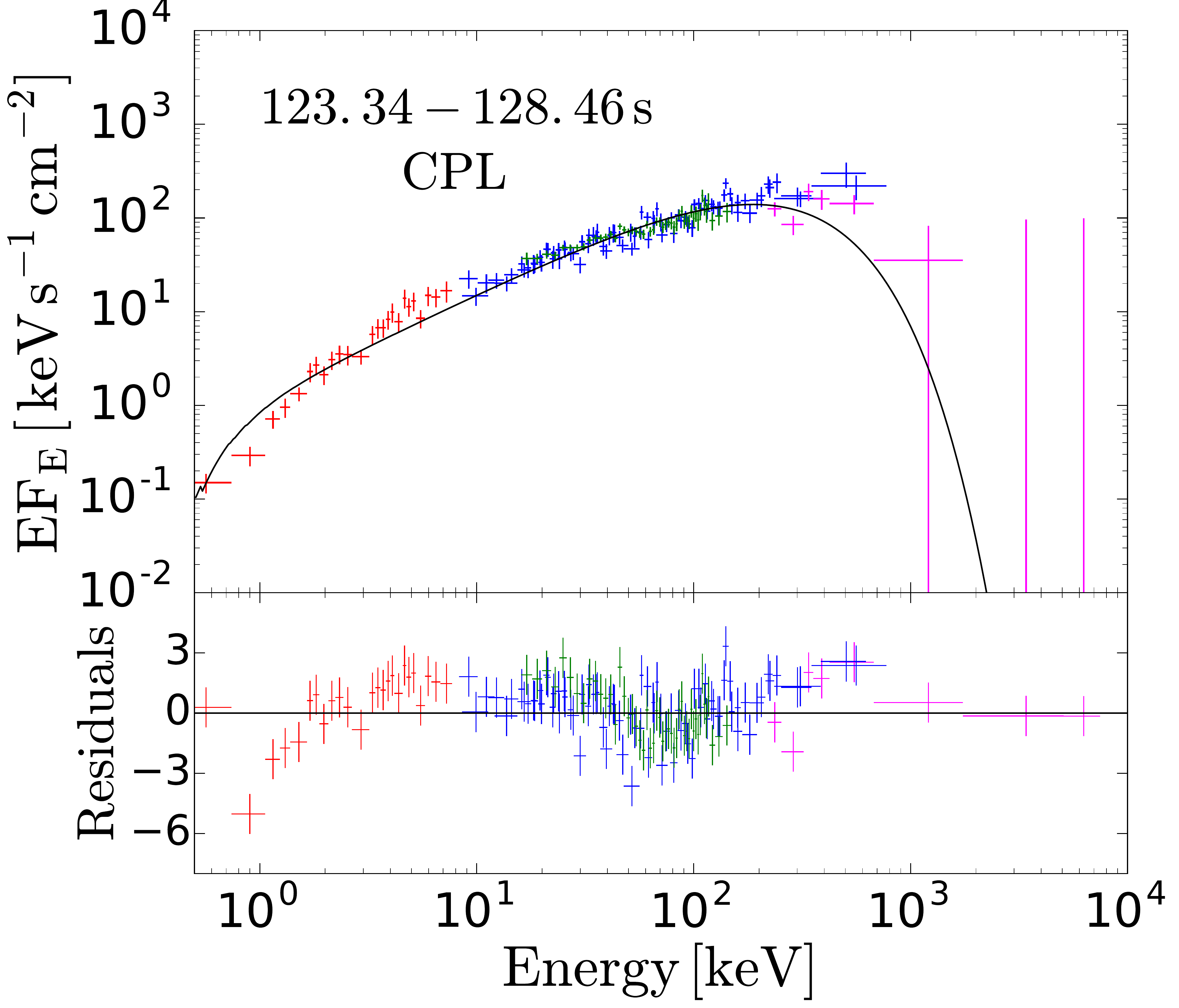} \hspace{0.4em}
\includegraphics[scale = 0.098]{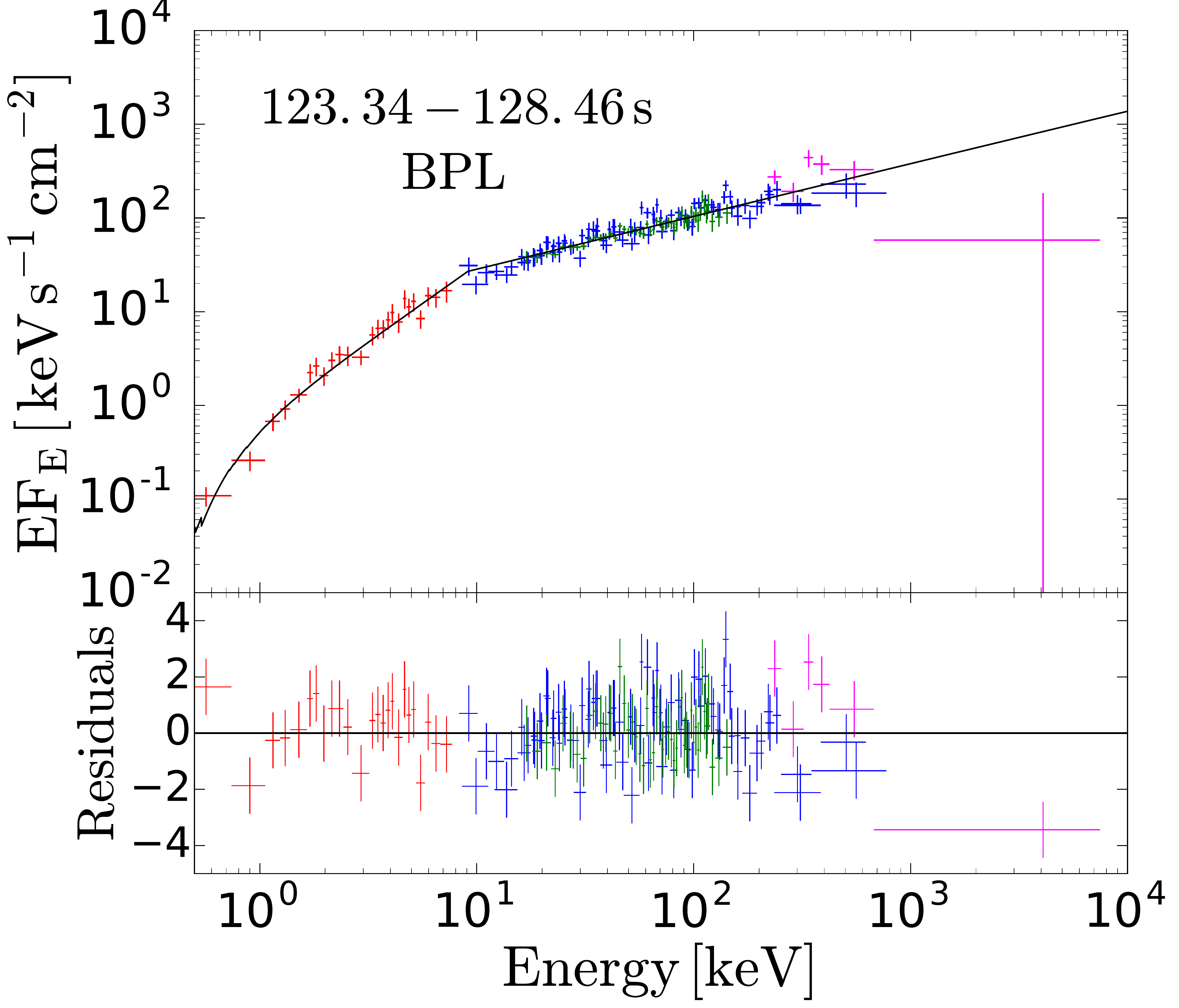} \hspace{0.4em}
\includegraphics[scale = 0.098]{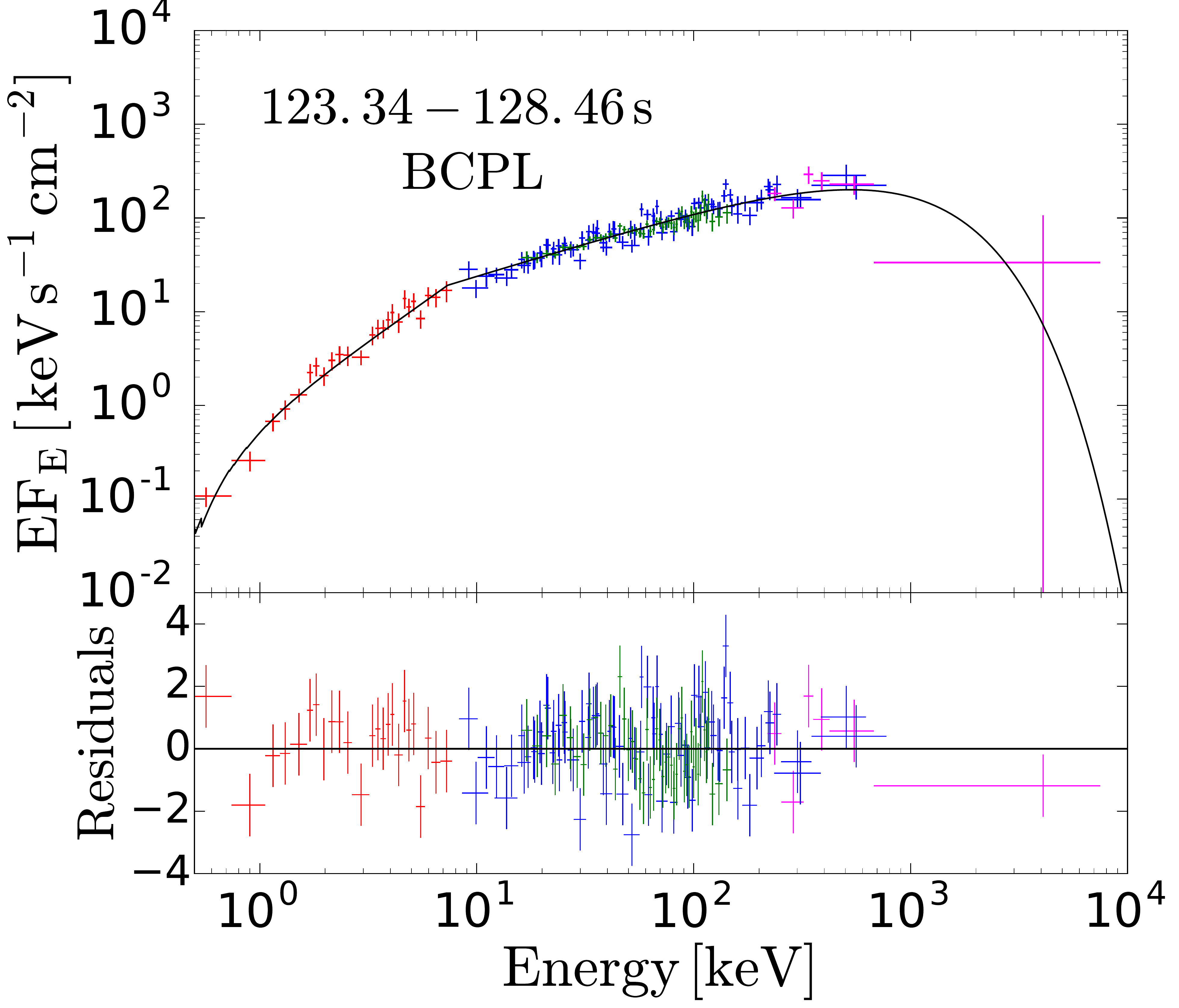} \\
\includegraphics[scale = 0.098]{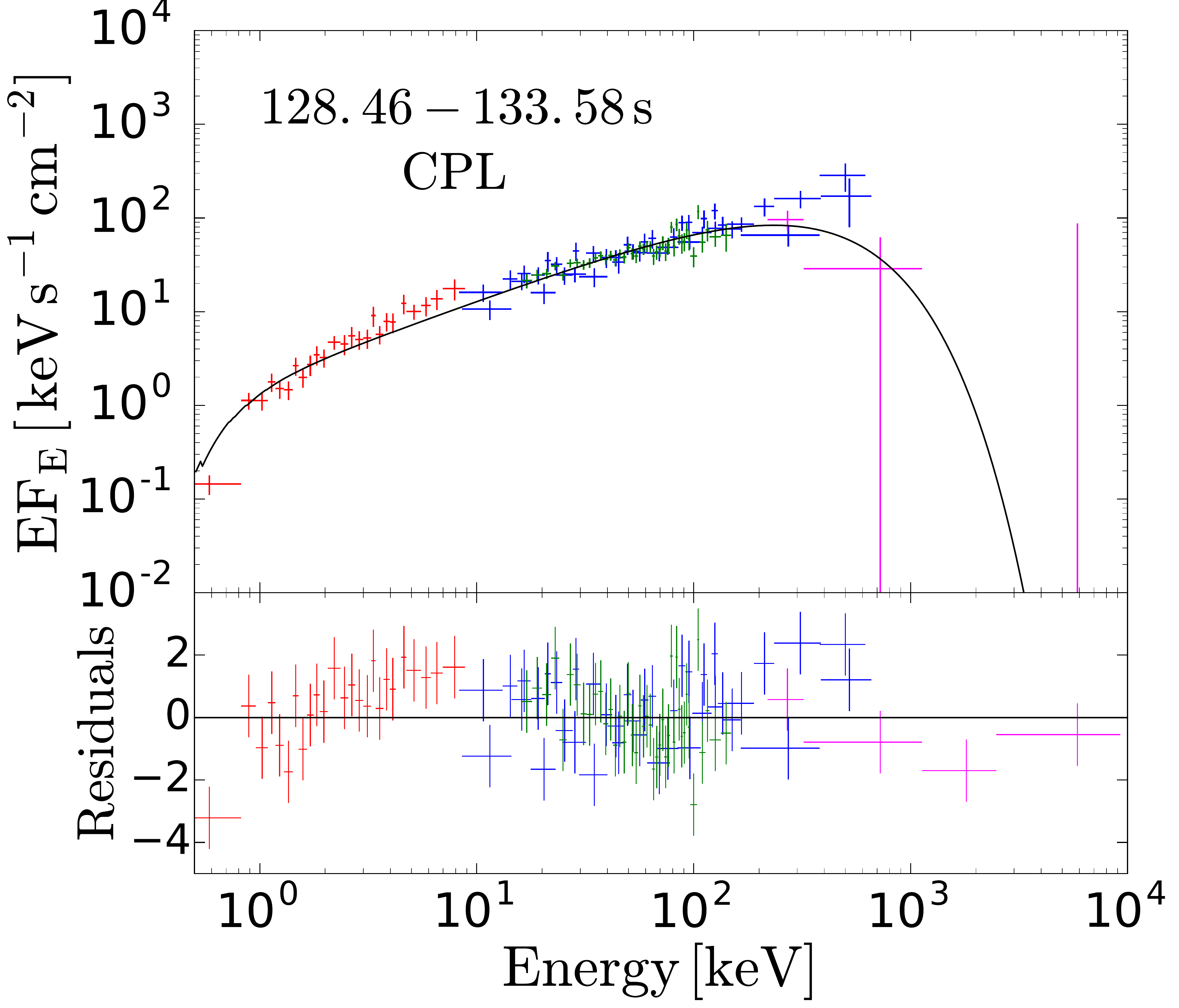} \hspace{0.4em}
\includegraphics[scale = 0.098]{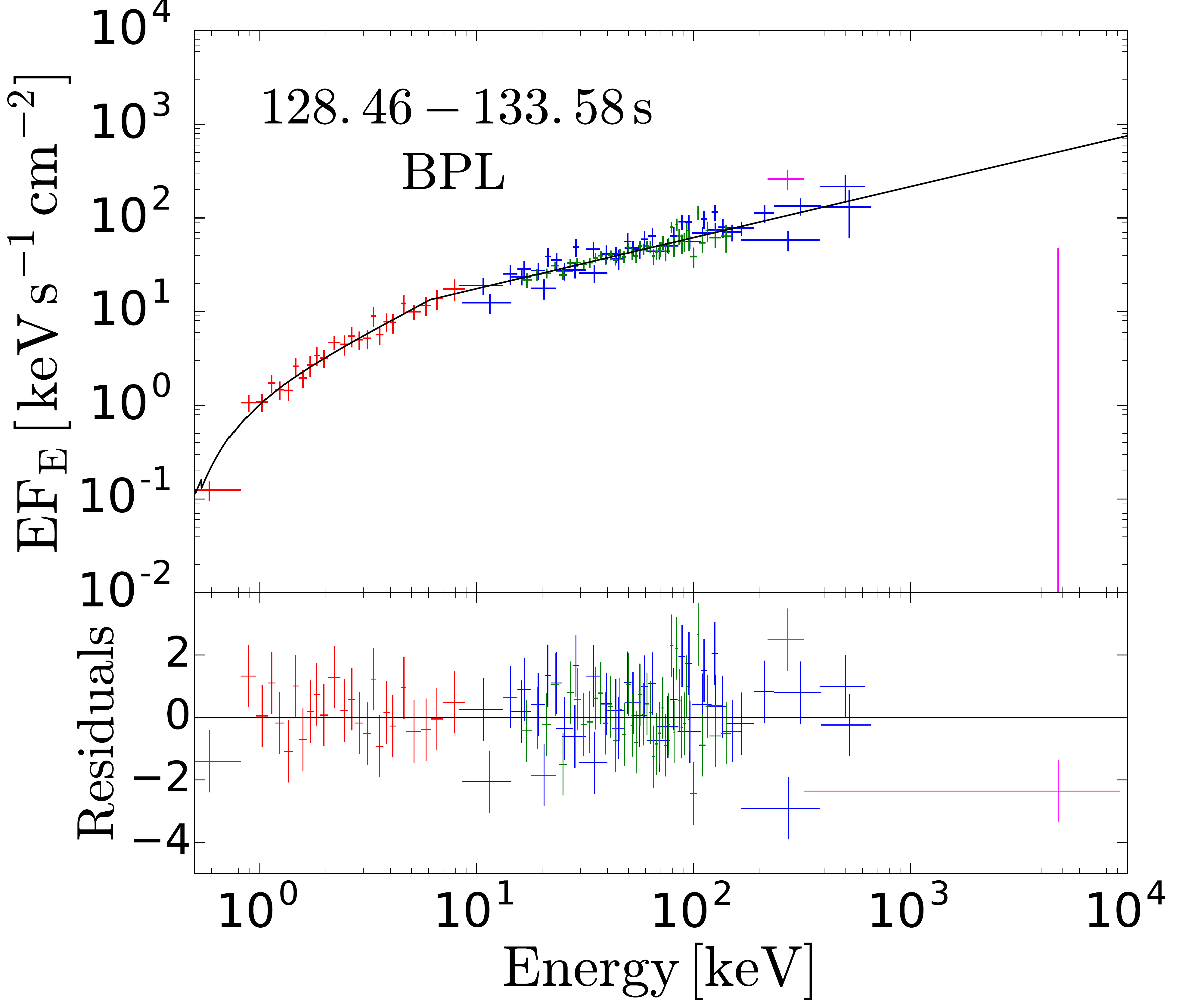} \hspace{0.4em}
\includegraphics[scale = 0.098]{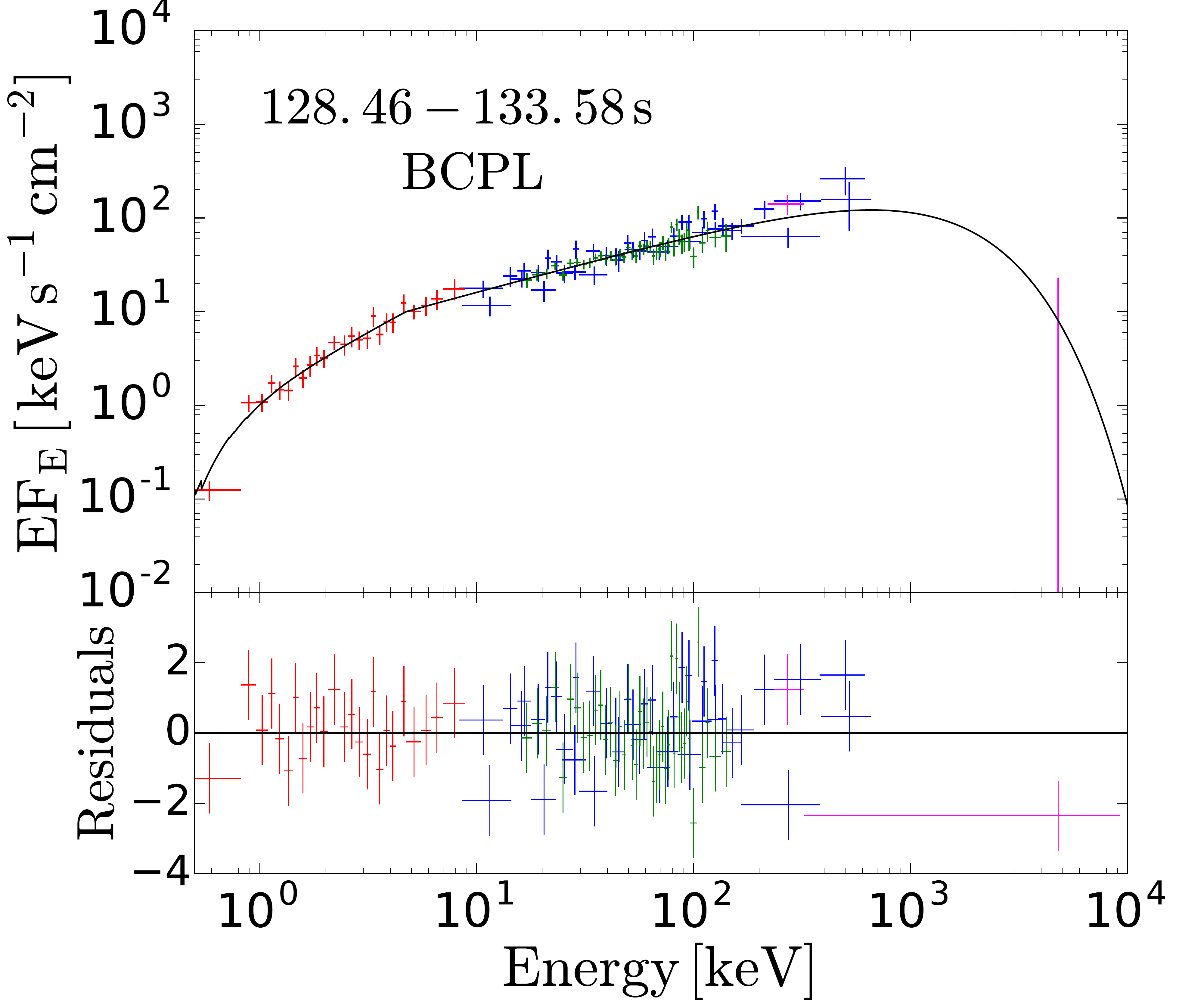} \\
\end{center}
\vskip -0.5 truecm
\caption{\label{fig:TR_spectra_140512A} Time-resolved spectra of GRB~140512A for six different time bins, including XRT (red data points), BAT (green), and GBM (blue and purple) data. Each row refers to a different time bin (the time interval is reported in each panel). For each time bin, the modeling with three different models and the residuals are shown: cutoff power law (CPL; left panel), broken power law (middle panel), and broken power law with an exponential cutoff (right panel).}
\end{figure*}
%------------------------------------------------------------------------

\newpage
\section{Testing the Effects of $N_{\rm h}$ and Pileup on the Results}\label{sec:tests}
In this section, we test how pileup effects and intrinsic absorption in the soft X-ray energy band can affect the results on the presence of the low-energy break. 
We have performed systematics tests on the spectra of three GRBs: GRB~140206A, GRB~110102A, and GRB~140512A.
The tests performed showed that the results are robust: the corrections adopted for pileup are sufficient to remove any spurious effect on the spectral shape at low energy, and the estimates of $N_{\rm H}$ are not responsible for the need of an intrinsically curved spectrum in the XRT band.
In the following, we explain in detail the tests applied and we show,
as an example, the results obtained on GRB~140512A.

\subsection{Pileup}
The spectra of bright X-ray sources, like the ones in our sample, observed by the Swift/XRT instrument in WT mode might be heavily piledup and very accurate corrections are needed in order to extract clean spectral files.
For the analysis presented in this work, we adopted the following method.
We excluded the central region of the X-ray images, in order to have a maximum count rate smaller than 150\,counts s$^{-1}$. 
To check whether this is enough to avoid contamination from pileup effects, one possibility is to further reduce the maximum count rate (i.e. excluding an even larger region), repeat the spectral analysis, and verify whether the results are affected.
In Table~\ref{tab:pileup} we show the results of this analysis applied to one time-resolved spectrum taken from GRB~140512A (128.46-133.58\,s). With a maximum count rate of 150\,counts s$^{-1}$, we found that the best-fit model is a BPL with $\alpha_1=-0.76^{+0.18}_{0.14}$, $\alpha_2=-1.45^{+0.04}_{0.04}$, and $E_{\rm break}=6.1^{+2.2}_{-1.6}$.
We progressively decreased the maximum count rate and refitted the spectrum with all four spectral models. We performed the $F-test$ to compare models with and without a low-energy break and verified that even when the count rate is reduced to 70\,counts s$^{-1}$ (where pileup is completely negligible) the presence of a break is still significant at more than 3$\sigma$. 

\floattable
\begin{deluxetable}{ccccccc}[ht!]
\tablecaption{  \label{tab:pileup} Results of the test performed to verify the possible effects of pileup on the presence of a break energy in the XRT energy range. The test is applied to one time-resolved spectrum of GRB~140512A (from 128.46 to 133.58s). The first column reports the maximum 
rate of the light curve after the central region of the source has been excluded. Columns 2-5 list the $\rm \chi^{2}$ (d.o.f.) of the four different spectral models. 
Models that differ from each other for the presence of a break (i.e. PL vs BPL and CPL vs BCPL) are compared in the last two columns, where the significance of the F-test is reported.}
\startdata \\[-0.1cm]
Rate [cts/s] & PL & CPL & BPL & BCPL &  $\rm F_{PL-BPL}$ & $\rm F_{CPL-BCPL}$  \\[0.1cm]
\hline
$120 $ &412.97 (228)& 243.89 (227) &217.09 (226) &210.95 (225)  & 1.11e-16 (8.4)& 8.14e-08 (5.4) \\ 
$90 $ &264.89 (220)& 224.17 (219) &216.90 (218) &211.05 (217)  & 3.45e-10 (6.3)& 1.44e-03 (3.2) \\ 
$70 $ &253.40 (218)& 218.80 (217) &213.14 (216) &207.45 (215)  & 7.67e-09 (5.8)& 3.26e-03 (2.9) \\ 
\enddata
\end{deluxetable}

\subsection{Intrinsic Absorption}
\begin{figure*}
\includegraphics[width=0.50\textwidth]{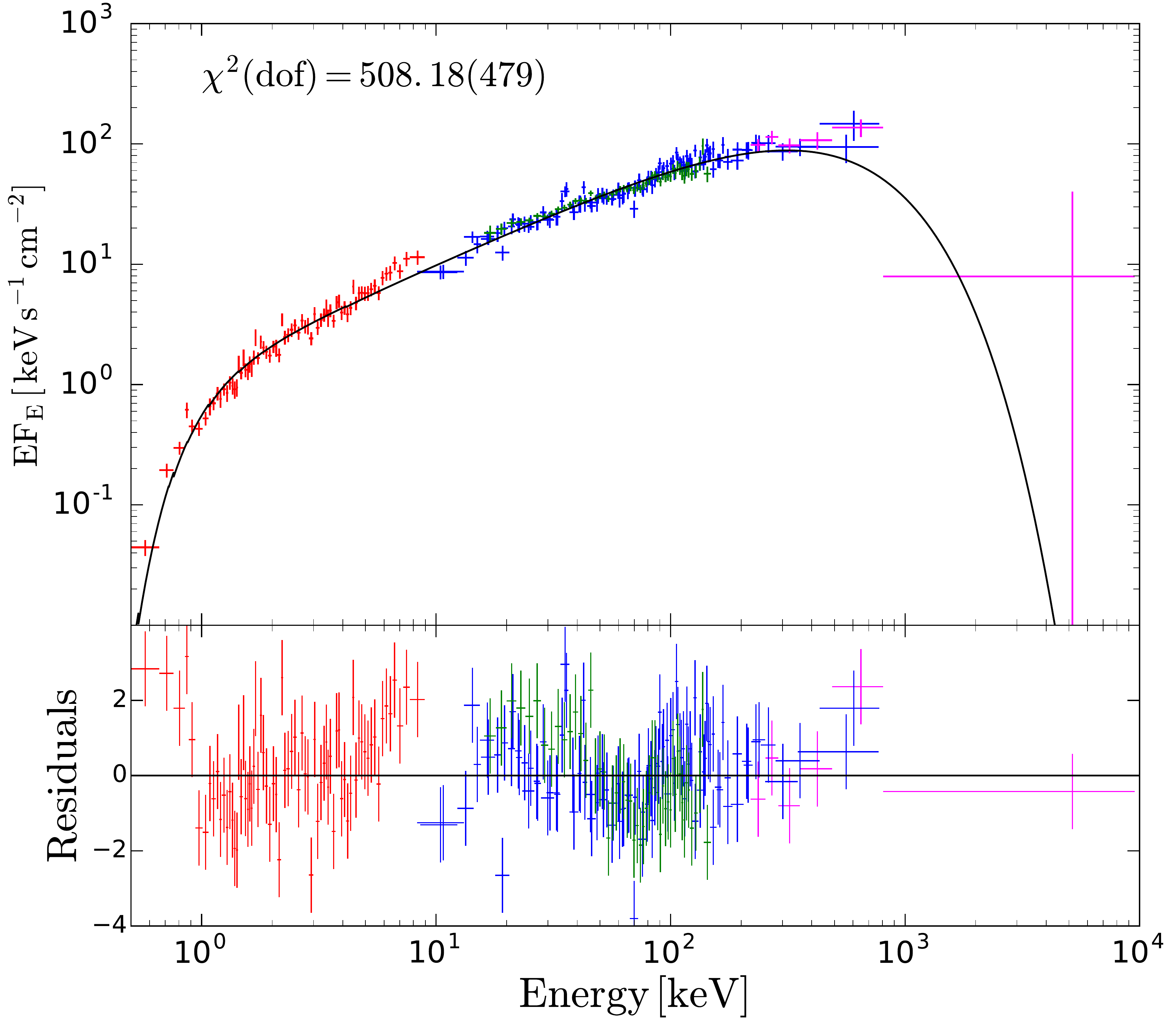}
\includegraphics[width=0.50\textwidth]{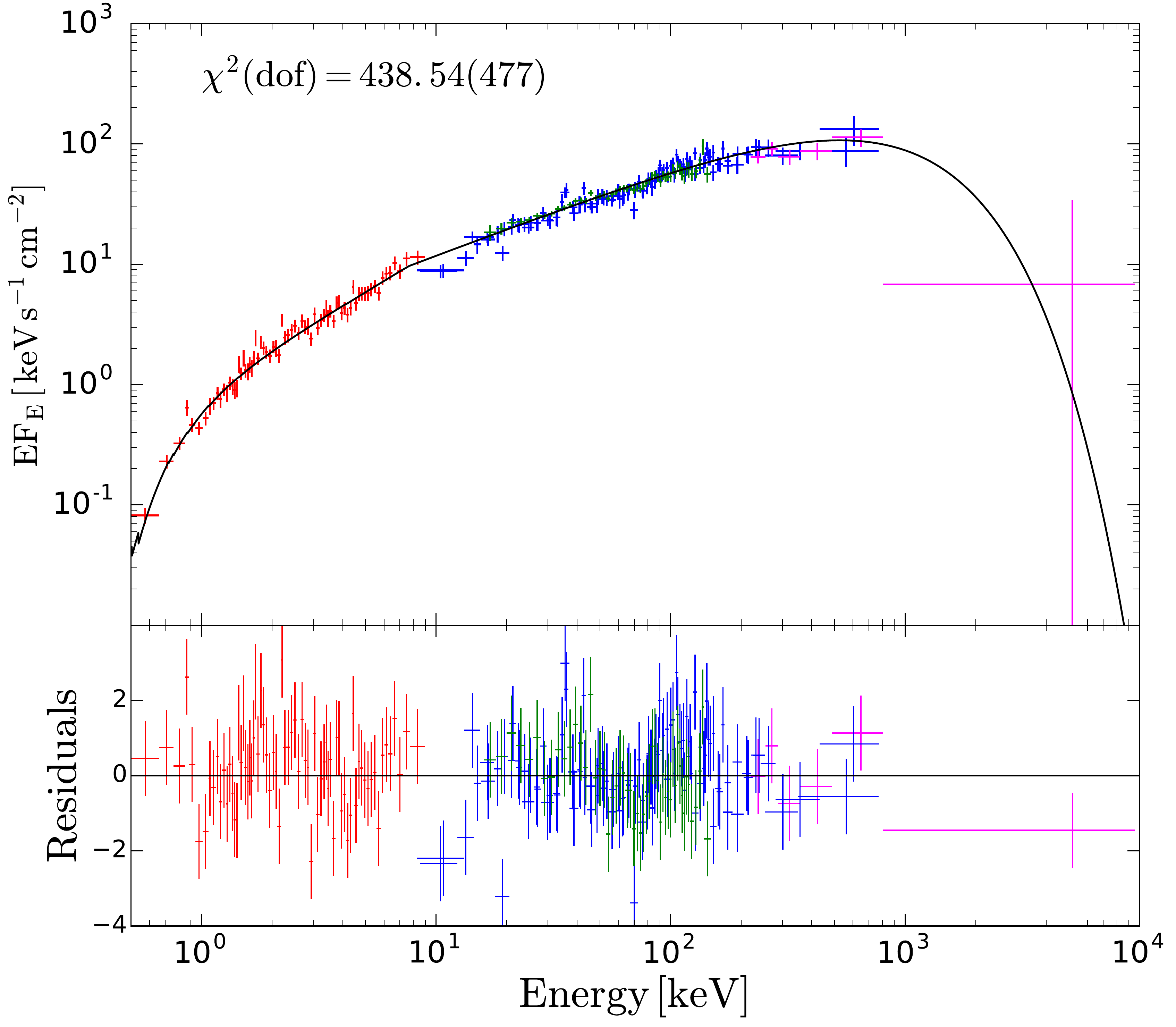}
   
\caption{\label{fig:Nh_test1} Time-integrated spectrum of GRB~140512A during the second pulse, fitted by CPL (left) and BCPL (right) models with intrinsic $N_{\rm H}$ as a free parameter. The value of the chi-square is reported in the upper left corner of each panel. The addition of the low-energy break improves the fit with a significance level corresponding to $8.1\sigma$.}
\end{figure*}

In order to exclude a possible influence of intrinsic absorption on the low-energy breaks found in this work, we perform two different tests. 
In the first test, we consider the intrinsic absorption a free parameter and refit the data with a CPL and a BCPL models. 
We then perform the $F-test$ to compare the two different fits and verify the significance of the improvement obtained thanks to the addition of the break.
An example is proposed in Figure~\ref{fig:140512A} and refers to the time-averaged spectrum of the second emission episode of GRB~140512A.
Even when the intrinsic absorption is a free parameter, the addition of a break improves the fit at more than $8\sigma$ (see Figure~\ref{fig:Nh_test1}). 

\begin{figure*}
\includegraphics[width=0.50\textwidth]{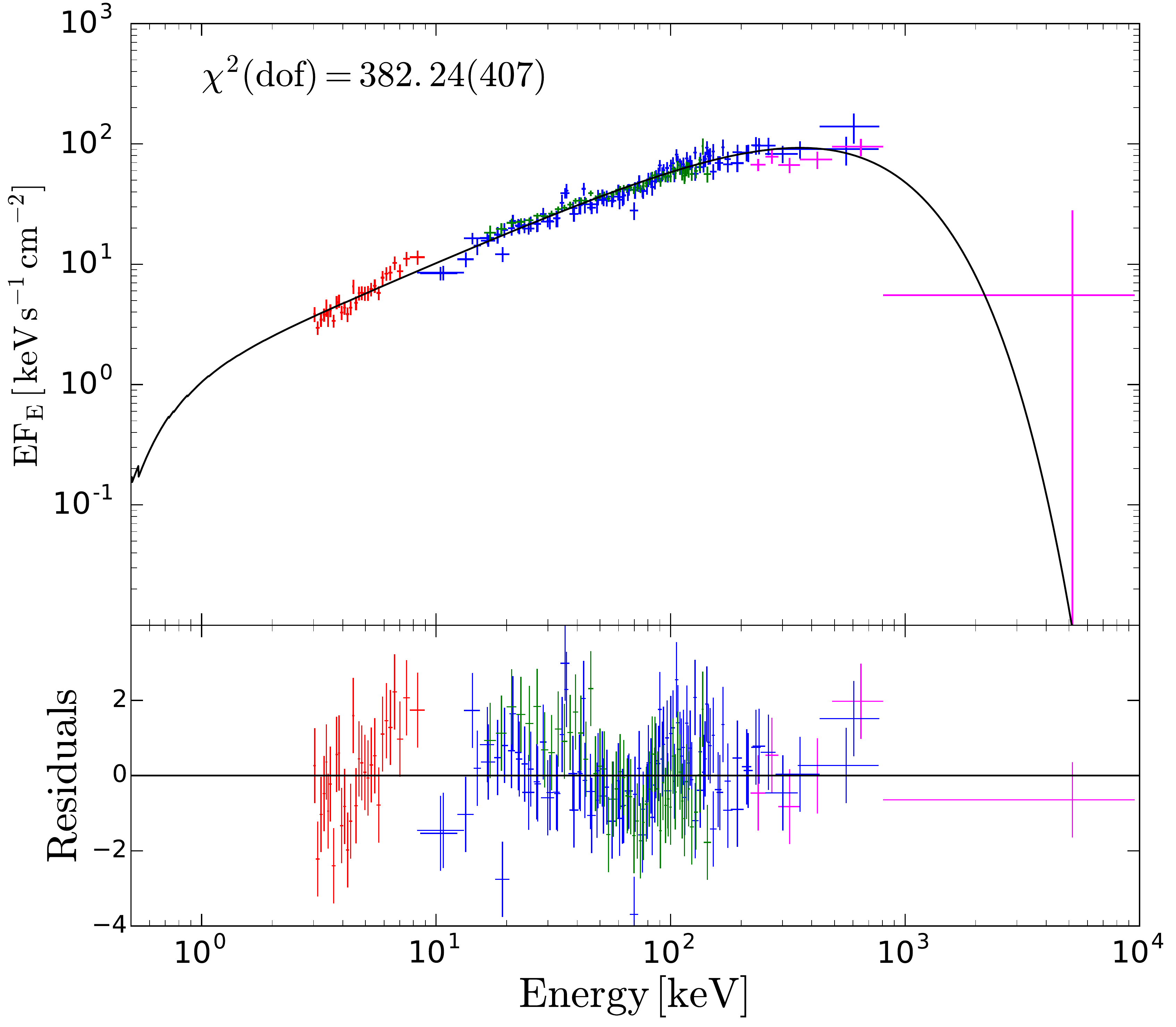}
\includegraphics[width=0.50\textwidth]{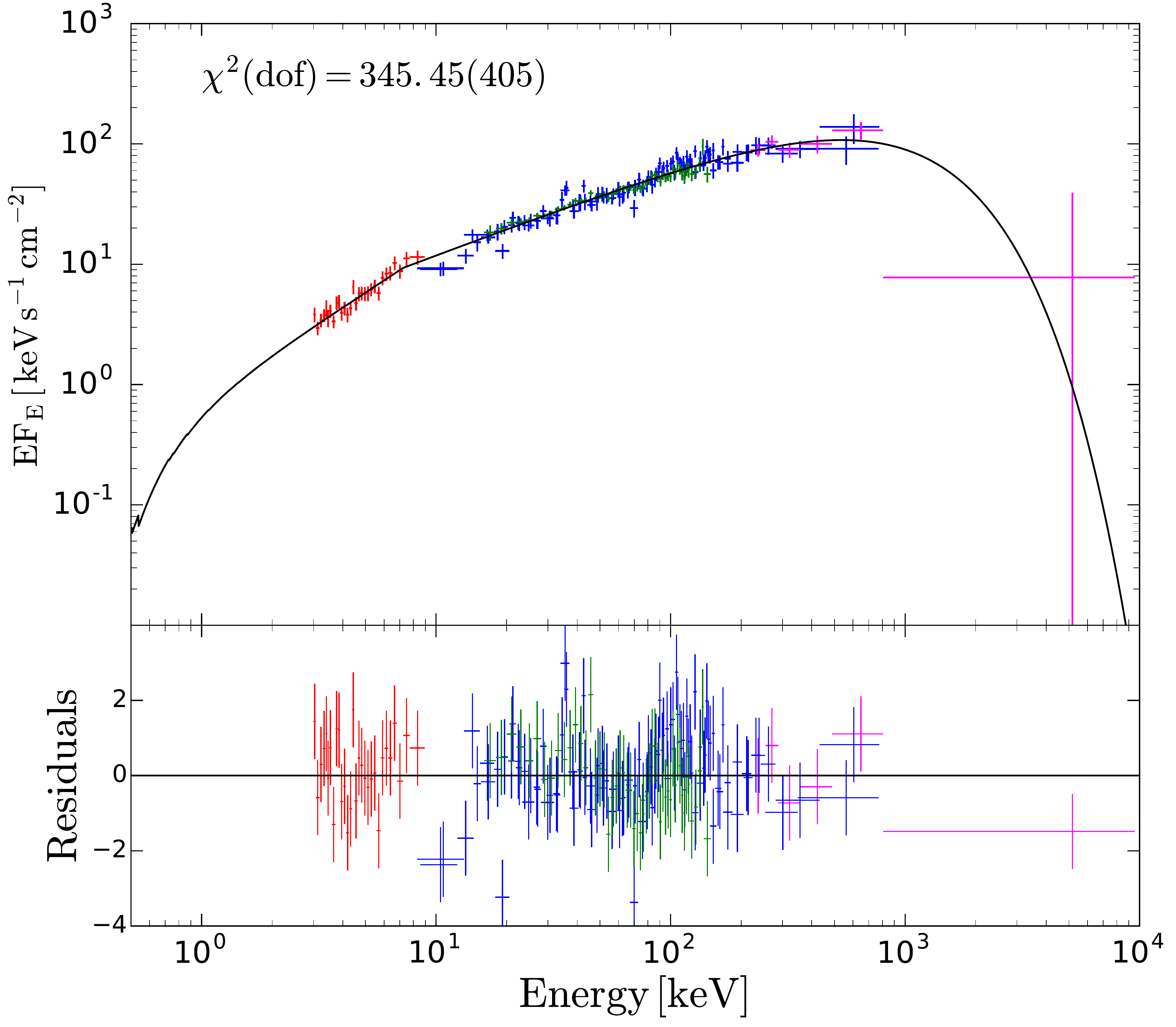}
\caption{\label{fig:Nh_test2} Time-integrated spectrum of GRB~140512A during the second pulse, fitted by CPL (left) and BCPL (right) models when data below 3\,keV are excluded. The improvement of the chi-square (reported in the upper left corner of each panel) when the low-energy break is added is significant at $6.1\sigma$, according to the $F-test.$}
\end{figure*}

In the second test, we exclude XRT data below 3\,keV, and refit the data.
Also in this case, a break in the spectrum is still required by the data.
Taking again the second emission episode of GRB~140512A as an example, we find that a BCPL improves the fit as compared to the CPL at more than $6\sigma$ (see Figure~\ref{fig:Nh_test2}).

%------------------------------------------------------------------
%------------------------------------------------------------------

%------------------------------------------------------------------
%------------------------------------------------------------------

\section{Comparison with previous studies}\label{app:comparison}

\subsection{GRBs in Our Sample}
For most of the GRBs included in our sample, the analysis of XRT+BAT spectral data has already been published in the literature. 
In this section, we discuss, case by case, the modeling proposed by different authors, as compared to those proposed in this work.
 
\begin{figure*}
{\includegraphics[scale=0.26]{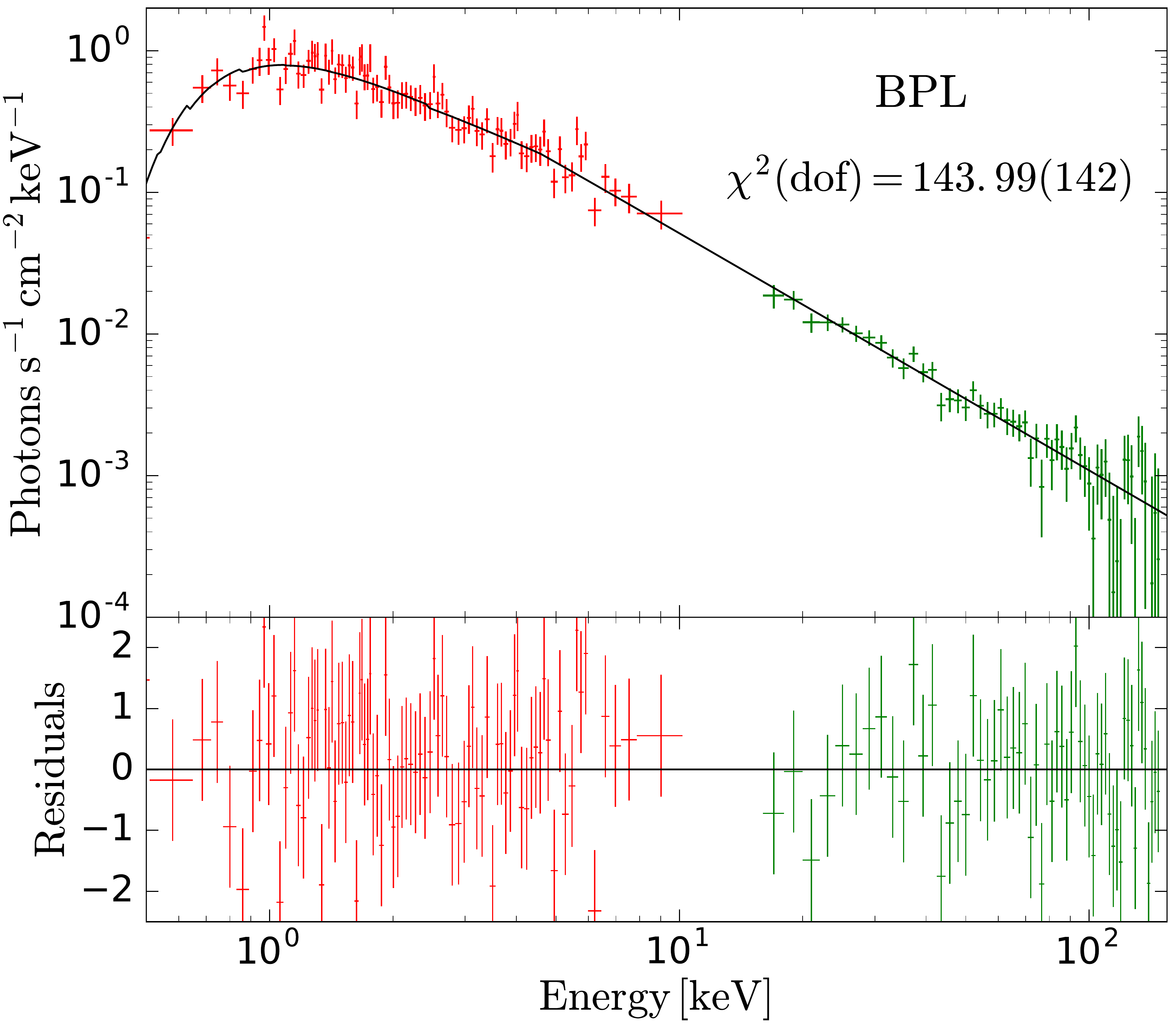}
\includegraphics[scale=0.26]{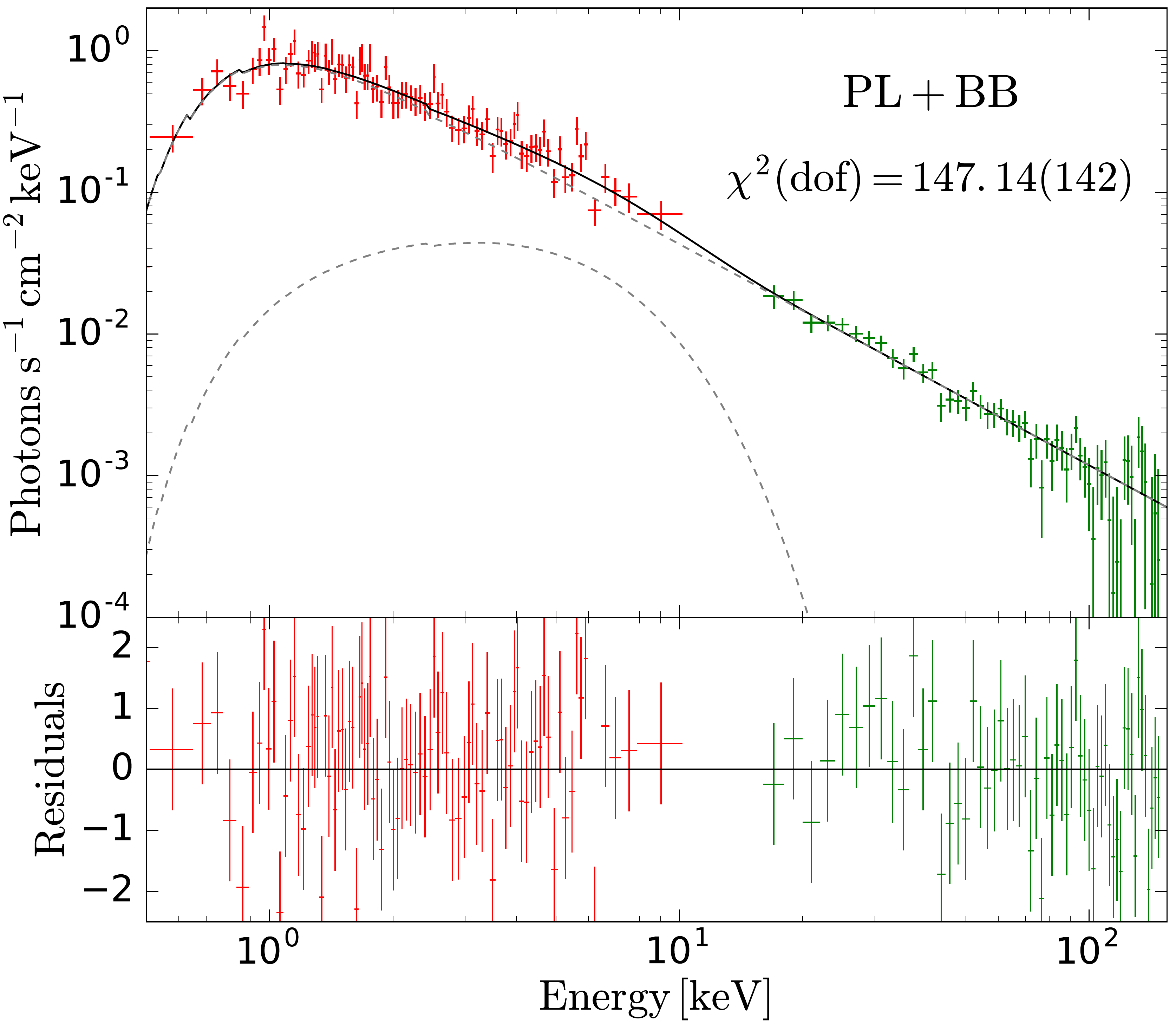}}
\caption{Spectrum of GRB~060814A integrated from 121 to 151\,s. XRT data points are shown in red, and BAT data points are shown in green. Two different spectral modelings are compared: a BPL with a break at $\sim5$\,keV, and a PL+BB, where the BB temperature is $kT\sim2$\,keV.}
\label{fig:peng}
\end{figure*}

%Peng method
A systematic analysis of GRBs with prompt XRT+BAT observations has been performed by \cite{Peng_14} (hereafter P14). 
A comparison with our findings is not straightforward, since the methods for data extraction and modeling are quite different. 
First, P14 considered intrinsic absorption as a free parameter. Moreover, they never discuss correction for the pileup effect, and it is not clear whether and how pileup has been treated. Time bins chosen for the analysis also differ from those chosen in our work. 
The spectral models tested by P14 are a single PL, a blackbody plus a PL (BB+PL), and the Band model. Sometimes, fits are performed by fixing to -1 the value of the low-energy spectral index.
%Peng results in comparison with our results
With these differences in mind, we report in the following a comparison between our modeling and the modeling proposed by P14 for the 10 GRBs common to both studies. 
In P14, the spectra of GRB~060814, GRB~061121, and GRB~100725A are fitted by BB+PL. The PL dominates at low and high energies, and the BB contributes to the flux at intermediate energies. In our analysis we proposed that the best-fit model for these three GRBs is a BPL.
We choose one of these GRBs (GRB~060814) as an example, to understand how two apparently completely different interpretations (a BPL and a BB+PL) can both give a satisfactory description of the same data, and perform spectral analysis with the two different models: a BPL and the combination of a BB plus a PL (Figure~\ref{fig:peng}). 
To be consistent with the method applied by P14, we leave the intrinsic $N_{\rm H}$ free to vary and choose the same time interval analyzed by P14 (from 121 to 151\,s). 
Both modelings return an acceptable fit: the reduced chi-square values for the BPL and BB+PL are $\chi^2_{\rm BPL}=1.01$ and $\chi^2_{\rm BB+PL}=1.04$, for the same number of degrees of freedom.
First, we note that, even though the value of the intrinsic $N_{\rm H}$ is a free parameter, a BPL model returns a well-constrained break energy $E_{\rm break}=4.54_{-1.56}^{+3.48}$\,keV. The BB+PL fit returns a BB temperature $kT=1.80_{-0.60}^{+1.00}$.
The role of the BB is to contribute to the emission at intermediate energies, producing a deviation from a single PL behavior between 4 and 8\,keV.
The overall shape of the BB+PL model mimics than a BPL behavior. We verified that the same explanation applies to the other two GRBs in our sample for which P14 claim the presence of a BB.

%-------------------------------------------------------------
%-------------------------------------------------------------

For GRB~100619A and GRB~110102, the best fit proposed by P14 is a Band function with $\beta>-2$, and $E_{\rm peak}$ around 10\,keV. Since $\beta>-2$ the characteristic energy cannot be properly identified with the spectral peak energy and must be more properly identified with what we called in this work the break energy, making their analysis of these two GRBs consistent with the one proposed in this work.
Also, the analysis of GRB~100906A is consistent, since for this GRB we also find a peak energy but no evidence of a break energy.
For GRB~100728A and GRB~121123 the differences can also be easily understood: the break energy is very small (2\,keV) and can be hardly constrained (see also \citealt{Abdo}), especially in the time interval studied by P14. There is agreement instead on the measure of $E_{\rm peak}$, which is large in the first case, and can be constrained only thanks to the inclusion of GBM data, and is around 50\,keV in the second GRB.
Similar considerations hold for GRB~140206A: the small value of $E_{\rm break}$ during the temporal window studied by P14 makes it difficult to recognize the presence of a break, while the peak energy, inside the BAT range, is constrained in both their and our analysis to be around 100\,keV.
Finally, a strong break around 7-8\,keV is found in this work in the spectrum of GRB~140108A, while in P14 it is claimed that the best model is a single PL. However, their spectrum is mainly accumulated over a time where we also find that the best fit is a PL, with the very same slope reported by P14 (-1.4).

We conclude that the analysis either is consistent or differs owing to the interpretation of the X-ray hardening as the result of a combination of two different components, one of which is assumed to be a BB in P14. Comparison between these two different interpretations is discussed in more detail in Appendix~\ref{app:BB}.

\subsubsection{GRB~061121}
In \cite{Page_07}, time-resolved spectra (from 62 to 90~s) are fitted by a broken power-law model with photon indices below and above the break 
$\rm \Gamma_{1}=-0.69_{-0.13}^{+0.07}$ and $\rm \Gamma_{2}=-1.61_{-0.14}^{+0.13}$, and a break energy varying in time in the $\rm 1-6\,$keV range (see their Figure~5), in agreement with the analysis reported in this work. 
The XRT+BAT time-averaged spectrum has been considered also by \cite{Peng_14} and \cite{Friis_13}.
They proposed a model composed of a single PL plus a BB with $\rm kT \sim3\,$keV. 
As discussed before, 
we then believe that the same change of spectral slope is found also in these studies, but is interpreted as the result of the composition of two different spectral components (see below for further details).

\subsubsection{GRB~070616}
In our analysis, this GRB is best fit by a BCPL, with $E_{\rm break}$ ranging between 8 and 3\,keV and $E_{\rm peak}$ evolving from 170 to 16\,keV.
The joint XRT+BAT time-resolved spectral analysis of this GRB has been performed also by \cite{starling08}. 
They tested both a BPL and a Band model and found that they are both acceptable, though the $\chi^2$ of the Band model is systematically higher (see their Table 2).
Their BPL fit identifies a break in the range 4-8\,keV, in agreement with our findings.
Their Band fit identifies a peak energy in the range 135 to 14\,keV, also in agreement with our findings. A model including both features (i.e. a low-energy break and a high-energy peak) is never tested by these authors.

\subsubsection{GRB~110205A}
We find a break energy around 4-6 keV, and a peak energy at $\sim100\,$keV. The peak energy is constrained only in two time-resolved spectra (GBM observations are not available for this GRB). In \cite{zheng12} joint Swift/XRT+BAT and Suzaku/WAM time-resolved spectra are best fitted by a Band function with a high-energy exponential cutoff. The photon indices below and 
above the break energy vary in the ranges $\rm -0.8<\alpha_1<-0.1$ and $\rm -1.8<\alpha_2<-1.2$ (within $\rm 90\%$ confidence level). The break energy is found to be located at $\rm \sim5\,keV$. These best-fit parameters estimated in \cite{zheng12} with an inclusion of Suzaku/WAM observations are consistent with our spectral fit.
An alternative modeling has been proposed by \cite{guiriec16b}, who included also data from Suzaku/WAM. Their modeling is composed 
of the superposition of three spectral components: a modified blackbody and two CPL. 
The reason why two completely different modelings can both give a good fit to the data is clear from Figure 2 in \cite{guiriec16b}: their best-fit model (black line), which in their interpretation is the sum of three different components, can be alternatively seen as a single component from X-rays to MeV energies: a broken power law with a high-energy cutoff (BCPL). A change of slope around 5 keV is clearly visible also in their data. The difference then is not in a different extraction/analysis of the data, but in a different interpretation of the same spectral features. However, a BCPL model does not reproduce the optical emission and would require an additional component at low energy. The three-component model proposed in \cite{guiriec16b} explains both the optical and the gamma-ray emission. \\

\subsection{Thermal Components}\label{app:BB}
\begin{figure}
\begin{center}
\includegraphics[scale=0.27]{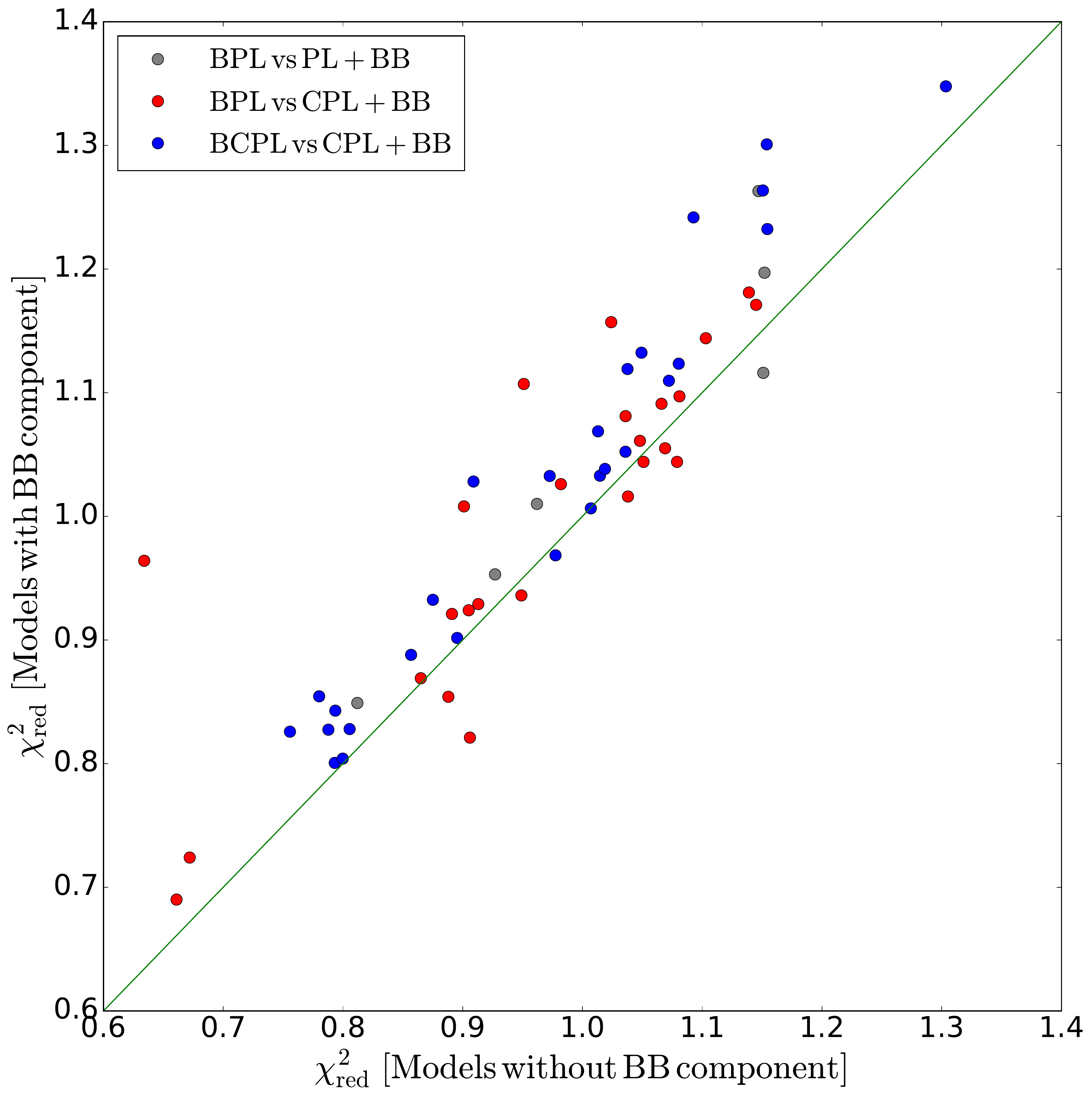}
\caption{Comparison between the reduced chi-square of models adopted in this work (labeled as {\it Models without BB component}; $x$-axis) and models invoking the presence of a thermal component plus an unbroken nonthermal component (labeled as {\it Models with BB component}; $y$-axis). Each point represents one of the time-resolved spectra for which we claim the presence of a break in the $\sim$keV range. Gray filled circles refer to cases where, according to our analysis, the best-fit model is a BPL, while if a BB is included, the best-fit model is a BB+PL. The number of dof for the two different modelings in this case is the same. Red filled circles show cases for which the best-fit models are a BPL and a BB+CPL for models without and with a thermal component, respectively. In this case the model including a BB has one more free parameter. Blue filled circles show cases where the best-fit models are a BCPL and a BB+CPL (same number of dof), for fits without and with a thermal component,respectively.}
\label{fig:BB}
\end{center}
\end{figure}

A two-component model, including a BB and a nonthermal component, has been often suggested to describe XRT+BAT spectral data.
To compare this interpretation with the one-component models proposed in this work, we considered all time-resolved spectra for which we claim the presence of a keV spectral break and refit them with a BB+PL or BB+CPL model. We chose the best fit among BB+PL and BB+CPL by adopting an $F-test$ and requiring a significance level of at least 3$\sigma$.
The results of this analysis and comparison with our one-component models are shown in Figure~\ref{fig:BB}.
We plot the reduced chi-square of models with a BB component ($y$-axis) versus the reduced chi-square of models without a BB component (i.e., either a BPL or a BCPL). 
Gray filled circles identify those cases where the best fits for models without and with a BB component are a BPL and a BB+PL, respectively. In this case, the number of dof for the two different fits is the same.
Red filled circles refer to cases where the best-fit models are a BPL and a BB+CPL: in these cases the models with a BB component have one more free parameter.
Blue filled circles refer to cases where the best-fit models are a BCPL and a BB+CPL (same dof).
The comparison shows that both modelings return acceptable fits in terms of reduced chi-square, with a tendency of single-component models to give a smaller value.
%-------------------------------------------------------------
%-------------------------------------------------------------
We note that when the best-fit model is a BPL, in most cases the alternative model invoking a BB component also requires the addition of a high-energy cutoff, i.e., the nonthermal component is not a simple PL, but a CPL (red filled circles in Figure~\ref{fig:BB}). The high-energy cutoff is required because a simple PL would be too hard at high energies, overpredicting the flux around 100-150 keV. A cutoff is then required to suppress the predicted flux. The actual presence of the peak energy identified by the BB+CPL fits can be tested with data at higher energies ($>150$\,keV) when Konus-{\it Wind} and/or Suzaku/WAM data are available (GBM data are not available for these GRBs).

We found that, for GRB~061121, BB+CPL time-resolved fits between 62 and 90\,s require peak energies in the range of 143-423\,keV, while, according to Konus-{\it Wind} data, the time-averaged spectrum from 61.9 to 83.4\,s peaks at $E_{\rm peak}=606^{+90}_{-72}$\,keV \citep{Page_07}.
For GRB~070616, spectra between 138 and 210\,s can be fitted by BB+CPL with peak energies
between 91 and 139\,keV, while the addition of Suzaku-WAM data shows that the spectrum integrated between 133 and 159\,s peaks at $E_{\rm peak}=356\pm78$\,keV \citep{starling08}.
For GRB~110205A, three time-resolved spectra at 160-193\,s, 193-210\,s, and 240-350\,s can be fitted by the BB+CPL
model with peak energies at 58-98\,keV. The time-integrated spectrum observed by Konus-Wind (up to 330 s)
is fitted by CPL with peak energy $222\pm74$\,keV \citep{2011GCN..11659...1G}. The spectrum observed by
Suzaku/WAM jointly with Swift/BAT from 20.2 to 318.2\,s is fitted by CPL with $E_{\rm peak}=230_{-65}^{+135}$\,keV \citep{2011GCN..11692...1S}.
In the context of models including a thermal component, XRT+BAT+WAM data for this GRB have been fitted by \cite{guiriec16b}. They find that a third nonthermal component is necessary in order to explain the peak at $\sim200$\,keV.
Finally, for GRB~130907A two spectra at 71-79\,s and 220-250\,s can be fitted with BB+CPL with peak energies at 323 and at 95 keV. The time-averaged spectrum observed by the Konus-{\it Wind} experiment up to 206 s shows has a peak energy of $394\pm11$\,keV \citep{2013GCN..15203...1G}.
A proper comparison would require spectral analysis on the same temporal bin. 
However, at least in some cases, it seems evident that the BB+CPL fits predict a peak energy that is in conflict with spectral data available at higher energies.

%%%%%%%%%%%%%%%%%%%%%%%  BIBLIOGRAPHY %%%%%%%%%%%%%%%%%%%%%

\end{document}